%% file: report.tex
\documentclass[whitelogo]{TUD-report2020}

\usepackage{biblatex}
\usepackage{enumitem}
\usepackage{changes}

\usepackage{glossaries}
\input{content/glossary}

\usepackage{pdfpages}
\usepackage{setspace}
\usepackage[utf8]{inputenc}  % for arxiv

\usepackage{placeins}
\makeatletter
\AtBeginDocument{%
  \expandafter\renewcommand\expandafter\subsection\expandafter
    {\expandafter\@fb@secFB\subsection}%
  \newcommand\@fb@secFB{\FloatBarrier
    \gdef\@fb@afterHHook{\@fb@topbarrier \gdef\@fb@afterHHook{}}}%
  \g@addto@macro\@afterheading{\@fb@afterHHook}%
  \gdef\@fb@afterHHook{}%
}
\makeatother

\begin{document}

%% Use Roman numerals for the page numbers of the title pages and table of
%% contents.
\frontmatter

%% Uncomment following 16 lines for a cover with a picture on the lower half only
\author{Nicolas~F.~Chaves-de-Plaza |\space Prerak~Mody |\space Klaus~Hildebrandt |\space Marius~Staring |\space Eleftheria~Astreinidou |\space Mischa~de~Ridder |\space Huib~de~Ridder |\space Rene~van~Egmond}
\title[tudelft-white]{Report on AI-Infused Contouring Workflows for Adaptive Proton Therapy in the Head and Neck}
%\subtitle[tudelft-black]{For Adaptive Proton Therapy in the Head and Neck}

%\affiliation{Technische Universiteit Delft}
%\coverimage{images/hptc-gantry}
% \covertext[tudelft-white]{
%     \textbf{Cover Text} \\
%     possibly \\
%     spanning 
%     multiple 
%     lines
%     \vfill
%     ISBN 000-00-0000-000-0
% }
%\setpagecolor{tudelft-cyan}
%\makecover[split]

\includepdf[pages={1}]{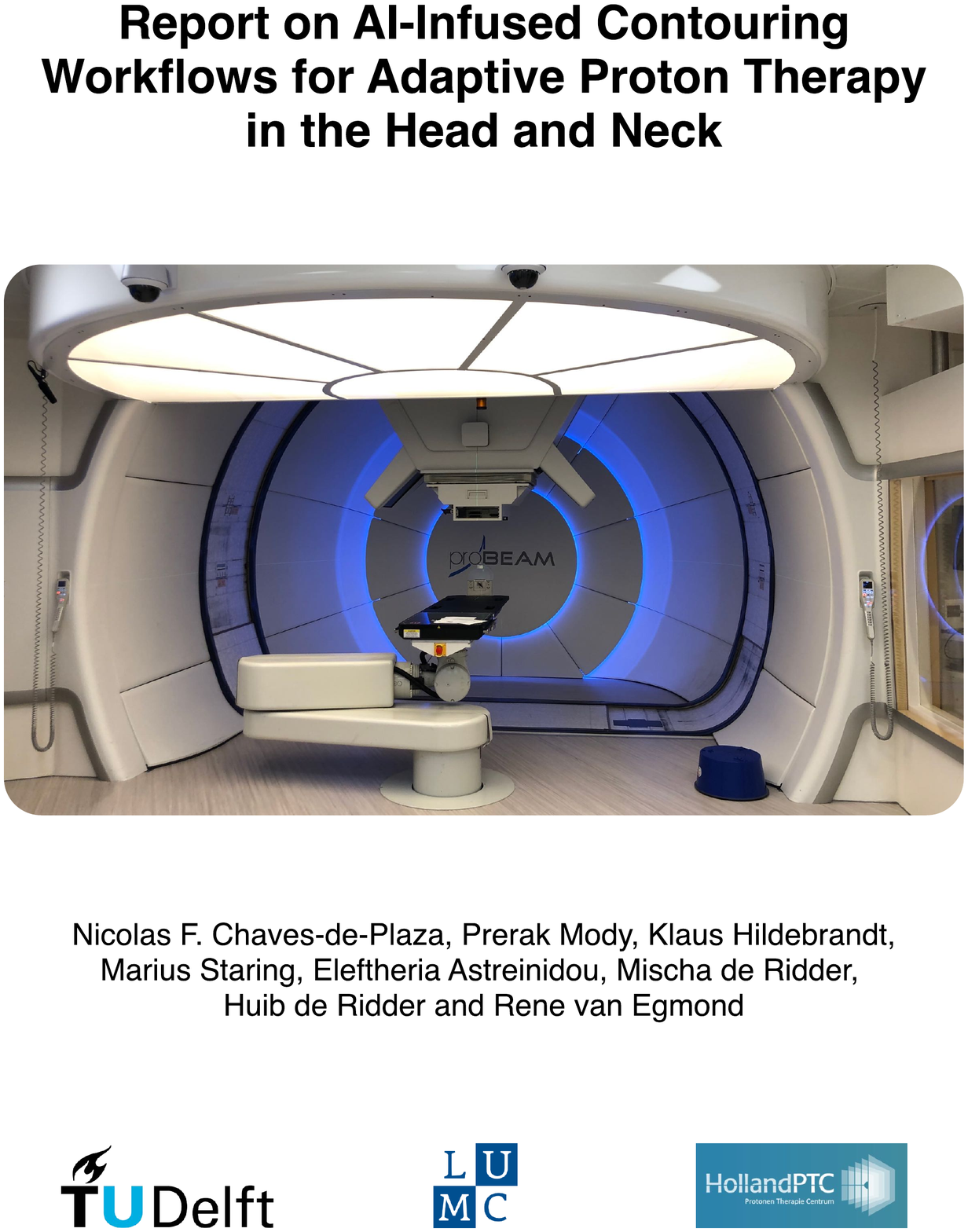}

%% Include an optional title page.
\input{title}

\input{preface}

\tableofcontents

\clearpage
\printglossary

%% Use Arabic numerals for the page numbers of the chapters.
\mainmatter

\input{content/introduction}

\input{content/adaptive-proton-therapy}

\input{content/current-contouring-wf}

\input{content/towards-apt-ready-contouring-wf}

\input{content/conclusion}

%% Use letters for the chapter numbers of the appendices.
\appendix

\input{content/appendix-process-specification}

\input{content/appendix-study-apt-methods}

\printbibliography

\end{document}

%% file: content/glossary.tex
\makeglossaries

\newglossaryentry{gtv}
{
    name=gross tumor volume,
    description={Visible extension of the tumor.}
}

\newglossaryentry{ctv}
{
    name=clinical target volume,
    description={Microscopic, and therefore invisible, part of the tumor confirmed based on histopathological studies.}
}

\newglossaryentry{itv}
{
    name=internal target volume,
    description={}
}

\newglossaryentry{ptv}
{
    name=planning target volume,
    description={Margin surrounding the ITV or CTV if the former was not considered. It accounts for setup errors.}
}

%% file: title.tex
\begin{titlepage}

\begin{center}

%% Insert the TU Delft logo at the bottom of the page.

%% Print the title in cyan.
{\makeatletter
\begin{spacing}{1.2} 
    \Huge\@title
\end{spacing} 
%\largetitlestyle\fontsize{64}{94}\selectfont\@title
%\largetitlestyle\color{tudelft-cyan}\Huge\@title
\makeatother}

%% Print the optional subtitle in black.
{\makeatletter
\ifx\@subtitle\undefined\else
    \bigskip
   %{\tudsffamily\fontsize{22}{32}\selectfont\@subtitle}    
    %\titlefont\titleshape\LARGE\@subtitle
    \begin{spacing}{1.2} 
        \titlefont\titleshape\huge\@subtitle
    \end{spacing} 
\fi
\makeatother}

\bigskip
\bigskip

by
%door

\bigskip
\bigskip

%% Print the name of the author.
{\makeatletter
%\largetitlefont\Large\bfseries\@author
\begin{spacing}{1.2} 
    %\largetitlestyle\fontsize{26}{26}\selectfont\@author
    \large\@author
\end{spacing} 
\makeatother}

\bigskip
\bigskip

\vfill

\begin{tabular}{lll}
    Project number: & IGF62H\\
    Work package: & 2 \\
    Deliverable: & D2.1 \\
\end{tabular}
%% Only include the following lines if confidentiality is applicable.

\bigskip
\bigskip
%\emph{This report is confidential and cannot be made public until December 31, 2013.}
%\emph{Op dit verslag is geheimhouding van toepassing tot en met 31 december 2013.}

\bigskip
\bigskip
%An electronic version of this thesis is available at \url{http://repository.tudelft.nl/}.
%\\[1cm]

%\centering{\includegraphics{cover/logo_black}}

\end{center}

\begin{tikzpicture}[remember picture, overlay]
    \node at (current page.south)[anchor=south,inner sep=0pt]{
        \includegraphics{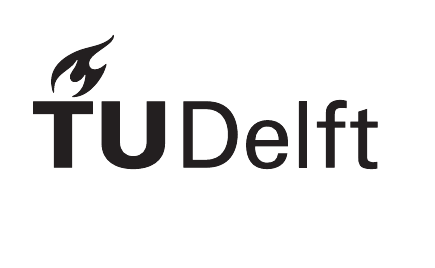}
    };
\end{tikzpicture}

\end{titlepage}

%% file: preface.tex
\chapter{Disclaimer}
%\setheader{Disclaimer}
This document has been prepared in good faith considering the information available at the publication date without any independent verification. The authors do not guarantee or warrant the accuracy, reliability, completeness, or currency of the information of this report nor its usefulness in achieving any purpose. Readers are responsible for assessing the relevance and accuracy of the content of this report. The authors will not be liable for any loss, damage, cost or expense incurred or arising because of any person using or relying on information in this publication.

\chapter{Acknowledgment}
%\setheader{Acknowledgment}
 The authors of this work are grateful for the assistance and collaboration of the personnel at both Holland Proton Therapy Center and Leiden University Medical Center. On the one hand, for the participation of radiation oncologists and radiotherapy technologists in the observational study. On the other, to other members of the staff of both centers who helped coordinate the meetings and who provided valuable feedback that improved this work. The research for this work was funded by Varian, a Siemens Healthineers Company, through the HollandPTC-Varian Consortium (grant id 2019022) and partly financed by the Surcharge for Top Consortia for Knowledge and Innovation (TKIs) from the Ministry of Economic Affairs and Climate.

\chapter{Abstract}
%\setheader{Abstract}

Delineation of tumors and organs-at-risk permits detecting and correcting changes in the patients' anatomy throughout the treatment, making it a core step of adaptive proton therapy (APT). Although AI-based auto-contouring technologies have sped up this process, the time needed to perform the quality assessment (QA) of the generated contours remains a bottleneck, taking clinicians between several minutes up to an hour to complete. This paper introduces a fast contouring workflow suitable for time-critical APT, enabling detection of anatomical changes in shorter time frames and with a lower demand of clinical resources. The proposed AI-infused workflow follows two principles uncovered after reviewing the APT literature and conducting several interviews and an observational study in two radiotherapy centers in the Netherlands. First, enable targeted inspection of the generated contours by leveraging AI uncertainty and clinically-relevant features such as the proximity of the organs-at-risk to the tumor. Second, minimize the number of interactions needed to edit faulty delineations with redundancy-aware editing tools that provide the user a sense of predictability and control. We use a proof of concept that we validated with clinicians to demonstrate how current and upcoming AI capabilities support the workflow and how it would fit into clinical practice.

%% file: content/introduction.tex
\chapter{Introduction}

\section{Motivation}

Proton therapy (PT) is a type of treatment against cancer in which a beam of protons is directed at the patient’s tumors to destroy the malignant cells \cite{blanchard_proton_2018}. In contrast with previous photon-based external beam radiotherapy (EBRT) technologies like intensity-modulated radiotherapy (IMRT) and volumetric arc therapy (VMAT), the use of protons allows targeting cancerous regions more precisely, which translates into less risk of toxicity to surrounding healthy tissue \cite{kamran_proton_2019}. 

The extra precision that using proton-based modern dose delivery technologies grants makes PT more suitable for treating cancer in areas of the body like the head and neck, which house critical organs that can produce severe side effects if damaged. For instance, the usage of PT has been linked to a reduction in the incidence of xerostomia (dry mouth) and dysphagia (difficulty swallowing) \cite{lundkvist_proton_2005, langendijk_selection_2013}. Also, it has shown potential in reducing growth-related side effects in pediatric cancers \cite{merchant_clinical_2013}.

\begin{figure}[h]
  \centering
  \includegraphics[width=\linewidth]{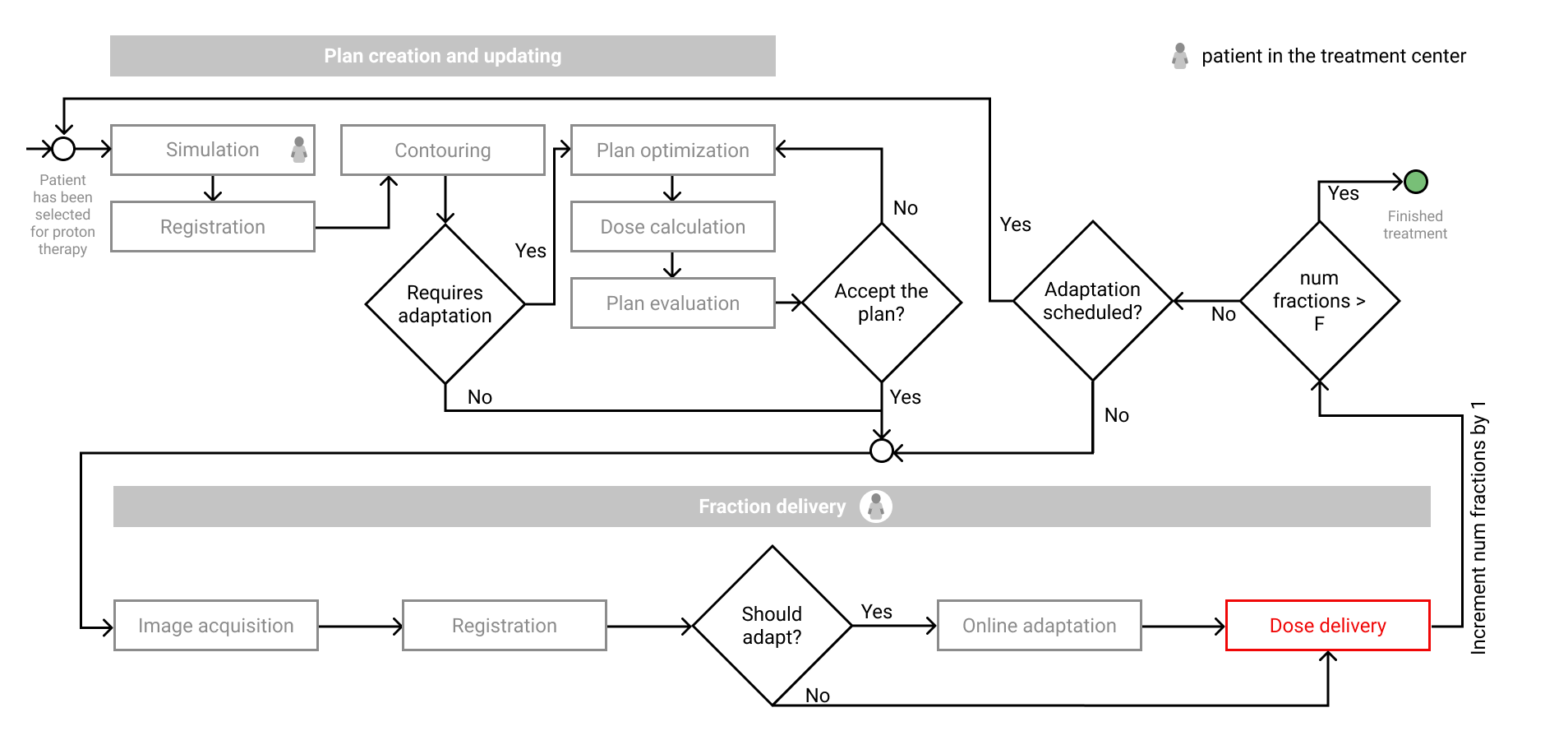}
  \caption{
  Simplified representation of the dose delivery pipeline of Holland Proton Therapy Center (HollandPTC), a cancer treatment clinic in the Netherlands. Gray boxes represent processes and diamonds decisions. Processes in the plan creation and updating phase change depending on whether it is the first time they are executed (treatment plan creation) or not (treatment plan updating). The goal of the pipeline is to deliver the planned dose to the patient, which corresponds to the process highlighted in red. In HollandPTC's case, treatment plan updates are scheduled weekly. Therefore, this workflow is an example of an offline adaptive proton therapy one. Although this kind of scheme can deal with changes that occur over multiple days, it would not be able to account for anatomical variations that happen in-between or within fractions.
  }
  \label{fig:simplified-dose-delivery-pipeline-hptc}
\end{figure}

The main obstacle for implementing PT in the clinic is that it requires resource-intensive adaptive workflows for realizing its organ-sparing capabilities \cite{albertini_online_2019}. The traditional EBRT dose delivery pipeline is composed of two phases, which Figure \ref{fig:simplified-dose-delivery-pipeline-hptc} depicts. The first one consists of planning the patient’s treatment plan, which involves acquiring three-dimensional images, delineating the relevant structures such as the tumors and organs at risk, and finding a configuration of the treatment machine that can optimally deliver the planned dose. In the second stage, the patient goes to the treatment center, usually daily, to receive a fraction of the radiation dose. 

Uncertainties in multiple steps of the pipeline can cause the delivered dose to deviate from the planned one \cite{van_herk_errors_2004}. For example, if the patient loses weight and the tumor position shifts, the amount of radiation it receives will reduce, leading to decreased probability of eradicating it. To avoid this from happening, safety margins around the relevant structures are added, which ensure that the treatment is robust against variability. Margins are often computed based on population estimates and consider uncertainties due to, for example, image quality issues (image acquisition), organ motion and delineation variability (delineation), and dose calculation uncertainty (treatment plan creation) \cite{veiga_toward_2015}.

Margins are an effective uncertainty management tool, ensuring that the tumor receives the planned dose. Nevertheless, they can lead to more toxicity of the organs at risk, defeating the chief benefit of PT. Therefore, several complementary strategies that can reduce uncertainty in the pipeline have been proposed. An example of such a strategy, known as Image-guided Radiotherapy (IGRT), adds imaging devices to the treatment room, allowing detection and correction of deviations in the patient’s positioning before the delivery of the dose. By adjusting the machine parameters based on the setup deviation, IGRT can lead to a reduction of the margins of between 3 to 5 mm, resulting in improved sparing of the organs at risk \cite{sterzing_image-guided_2011, mackie_image_2003}.

A significant portion of the safety margins used in PT corresponds to geometric uncertainties, which can arise due to changes in the setup, organ motion, and variability in their delineations. While the former, which results in translations and rotations of the couch, has been mitigated by IGRT, the latter two usually require adapting the treatment plan to the new anatomy, which requires re-executing most of the steps of the treatment creation process \cite{castadot_adaptive_2010}.

The substantial increase of time and resources associated with plan adaptation has made it hard to implement adaptive schemes that respond to the changes in the patient's anatomy as they occur. Instead, current adaptive proton therapy (APT) workflows address geometric changes weekly, scheduling an offline adaptation session. It is not clear whether this is enough for realizing the benefits of PT given the uncertainties that arise between the adaptation sessions. 

Having said this, future APT workflows require two components to maximize the effect of adaptations and reduce the additional resource load associated with unnecessary executions of the treatment creation process. On the one hand, it should be possible to detect and predict anatomical changes that might require plan updating, which would allow for efficient allocation of adaptation sessions. On the other, the resource footprint of the treatment creation process should be reduced to allow for on-demand offline adaptations. Figure \ref{fig:online-adaptive-template} in Chapter \ref{ch:towards-apt} presents a schematic of an APT workflow that includes these components.

% figure: APT workflow with new building blocks

\section{Contouring in APT}

Contouring is an essential process in EBRT, given that delineations of the anatomical structures are the input of the plan optimization. In adaptive treatments like APT, contouring plays a more significant role because new delineations are required every time the shape of the patient's anatomy changes. Furthermore, daily monitoring of the contours could enable detection of the patient's anatomical changes as they occur, yielding responsive treatments. 

Although there have been several works that analyze and optimize the contouring workflow \cite{ramkumar_user_2016, aselmaa_using_2017, aselmaa_medical_2014, aselmaa_influence_2017, ramkumar_exploring_2013, ramkumar_using_2017, ramkumar_design_2016, ramkumar_hci_2017}, most of the time they focus on the clinical scenario when time is not a constraint. Namely, producing the initial set of patient's delineations. For instance, in \cite{aselmaa_influence_2017} the authors compared the performance gains of using between-slice interpolation, which is one of the preferred tools when performing contouring from scratch. Similarly, in \cite{ramkumar_user_2016} the authors picked an interactive method, region growing,  and analyzed aspects of the interactions like performance and cognitive load. 

Nevertheless, to implement time-critical resource-intensive APT in the clinic, it is necessary to rely on automation (AI) for contour generation. Current auto-contouring tools like deformable image registration (DIR) and deep learning-based (DLBS) ones have shown expert-level performance in several delineation tasks \cite{kumarasiri_deformable_2014, nikolov_deep_2020}, making them good candidates for the adaptive paradigm. Nevertheless, AI's results are not infallible and usually require attention from the user before feeding them to downstream processes. 

Future contouring workflows require adapting to increasingly shorter time frames while maintaining the patient's treatment quality. This report proposes recommendations for clinical tool developers and researchers interested in implementing delineation in an APT setting.  It leans on state-of-the-art auto-contouring tools such as DIR and DLBS and bases its results on the literature and empirical data gathered from observational sessions and interviews at several radiotherapy centers in the Netherlands.

\section{Report Goals and Structure}

This report follows the line of research traced by works that have described the RT and PT workflows \cite{aselmaa_workflow_2013, green_practical_2019, albertini_online_2019} and also those that have analyzed the performance of the contouring process and how to optimize it. Nevertheless, it advances them by focusing on contouring's role in adaptive therapies, which promise to be the key to fully realizing the organ-sparing benefits of technologies like proton therapy.

The report departs from the previous analysis of the contouring process in that it does not follow the bottom-up approach of choosing a tool currently being used in the clinic and optimizing its performance to make it fit in the adaptive paradigm. Instead, it first establishes the context of APT and how delineation currently works in the clinic. Then, it leverages this knowledge to propose avenues of work for developing contouring tools that consider the scenario and its constraints. 

\subsubsection{Summarizing, the main goal of the report is:}
\begin{itemize}[noitemsep,nolistsep]
    \item To understand what is the role that contouring will play in future APT workflows and how to evolve the process with new tools and clinical protocols. 
\end{itemize}

\subsubsection{This is done by breaking it into three parts:}
\begin{itemize}[noitemsep,nolistsep]
    \item Chapter \ref{ch:adaptive-proton-therapy} introduces the reader to the concept of adaptation in PT. The goal is to make clear why it is needed and how it can impact the PT workflow.
    \item Chapter \ref{ch:current-contouring-wf} presents the results of the study of the Offline-APT Contouring Workflow. The study's goal was to get a picture of the role that contouring plays in the adaptive RT workflows that treatment centers in the Netherlands currently implement. The study yielded factors that affect delineation performance and requirements for the users to adopt new tools. 
    \item Chapter \ref{ch:towards-apt} puts the previous two in the context of the more responsive future APT that leverages advances of AI auto-contouring. It proposes a proof of concept of a system that could accelerate contouring by leveraging the risk of uncertain contours in downstream tasks and exploiting redundancy in the delineations to reduce the number of interactions required. 
\end{itemize}

%% file: content/adaptive-proton-therapy.tex
\chapter{Adaptive Proton Therapy (APT)}
\label{ch:adaptive-proton-therapy}

The goal of this chapter is to present Adaptive Proton Therapy (APT) and the impact implementing this type of treatment has on the clinical workflow. The explanations are based on a review of the external beam radiotherapy literature, focusing on adaptive treatments and proton therapy. Also, it was complemented with the first round of interviews in the Study of the Offline-APT Contouring Workflow, which took place at two cancer therapy centers in the Netherlands: Holland Proton Therapy Center (HollandPTC) and the radiotherapy department of Leiden University Medical Hospital (LUMC). The next chapter presents the complete results of the study.

In radiotherapy research, there are different streams of literature, depending on the region of the body in which cancer resides. The reason is that, depending on the tumor site, there might be different considerations, such as the amount of movement, the type of changes that can occur or the number of structures. Although most of the concepts in this chapter apply regardless of the location of cancer, Section \ref{sec:head-and-neck-considerations} presents specific considerations for the head and neck (HN) area, which is the focus of the project.

\section{What is PT?}
\label{sec:what-is-pt}

\begin{figure}[h]
  \centering
  \includegraphics[width=10cm]{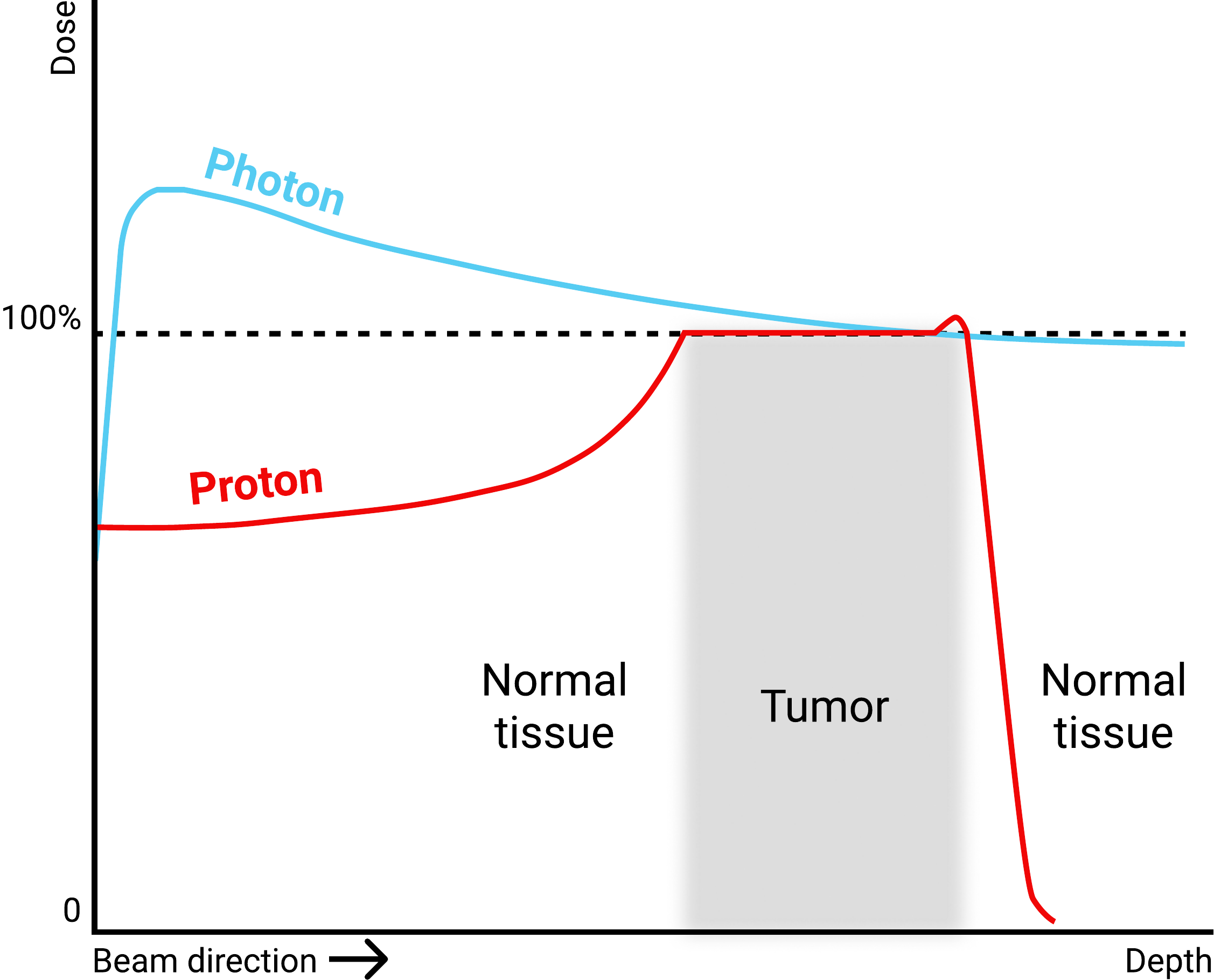}
  \caption{
  Depth-dose distributions for photons and protons. Note how the protons' dose distribution is concentrated at the tumor, rapidly falling off afterward. Image adapted from \cite{kamran_proton_2019}.}
  \label{fig:bragg-photon-vs-proton}
\end{figure}

Proton therapy (PT) is a type of external beam radiotherapy treatment (EBRT) in which a ray of charged particles, protons, in this case, is directed at the patient's tumor \cite{blanchard_proton_2018}. What differentiates protons from the more commonly used photons is that it is possible to shape the rays of the former in a way that most of the radiation is deposited in the tumor. As can be observed in Figure \ref{fig:bragg-photon-vs-proton}, this is possible because their Bragg peak can be precisely aligned with the tumor and quickly decays afterward, preventing dose residues \cite{kamran_proton_2019}. In contrast, photon rays (blue line) deposit radiation along the path, unnecessarily irradiating healthy tissue.

\begin{figure}[h]
  \centering
  \includegraphics[width=\linewidth]{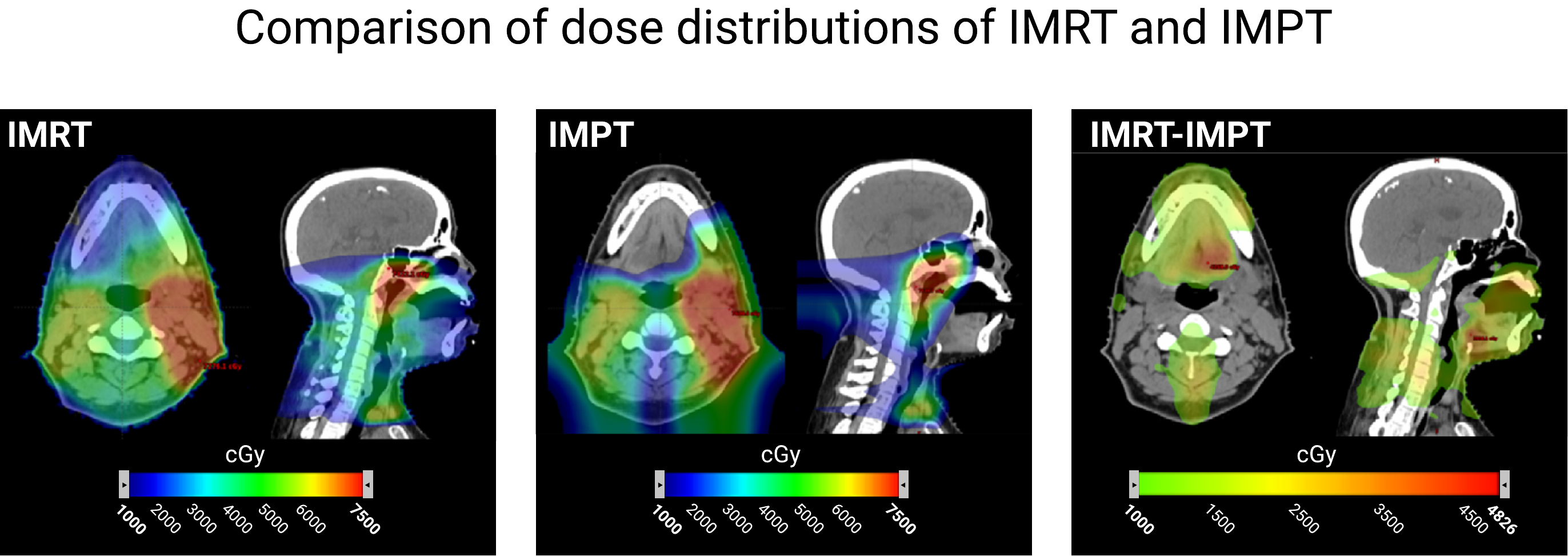}
  \caption{
  Comparison of the dose distribution generated by a photon-based (left, IMRT) and a proton-based(center, IMPT) treatment plan of a 56-year old woman with T1N1 left-sided nasopharyngeal carcinoma (4 nodes in left levels IIa, IIb, III, and Va). The right pane shows the difference in dose between the two plans. Observe how IMRT delivers significantly more radiation in the surrounding healthy tissue. Image adapted from \cite{blanchard_proton_2018}.}
  \label{fig:imrt-vs-impt}
\end{figure}

Compared to photon-based EBRT methods, PT can achieve more conformal dose distributions that adapt to the tumor's shape, sparing more healthy surrounding tissue. Figure \ref{fig:imrt-vs-impt} shows the dose distributions delivered by photon and proton-based EBRT technologies. In the right pane of the figure, it is possible to observe how a significant portion of the patient's head and neck area can be spared if protons are used. In turn, this can result, for this example, in a reduction of the risk of suffering side effects like difficulty swallowing (dysphagia) and dry mouth (xerostomia) \cite{langendijk_selection_2013, lundkvist_proton_2005}.

\section{Dose Delivery Pipeline}
\label{sec:dose-delivery-pipeline}

\begin{figure}[h]
  \centering
  \includegraphics[width=\linewidth]{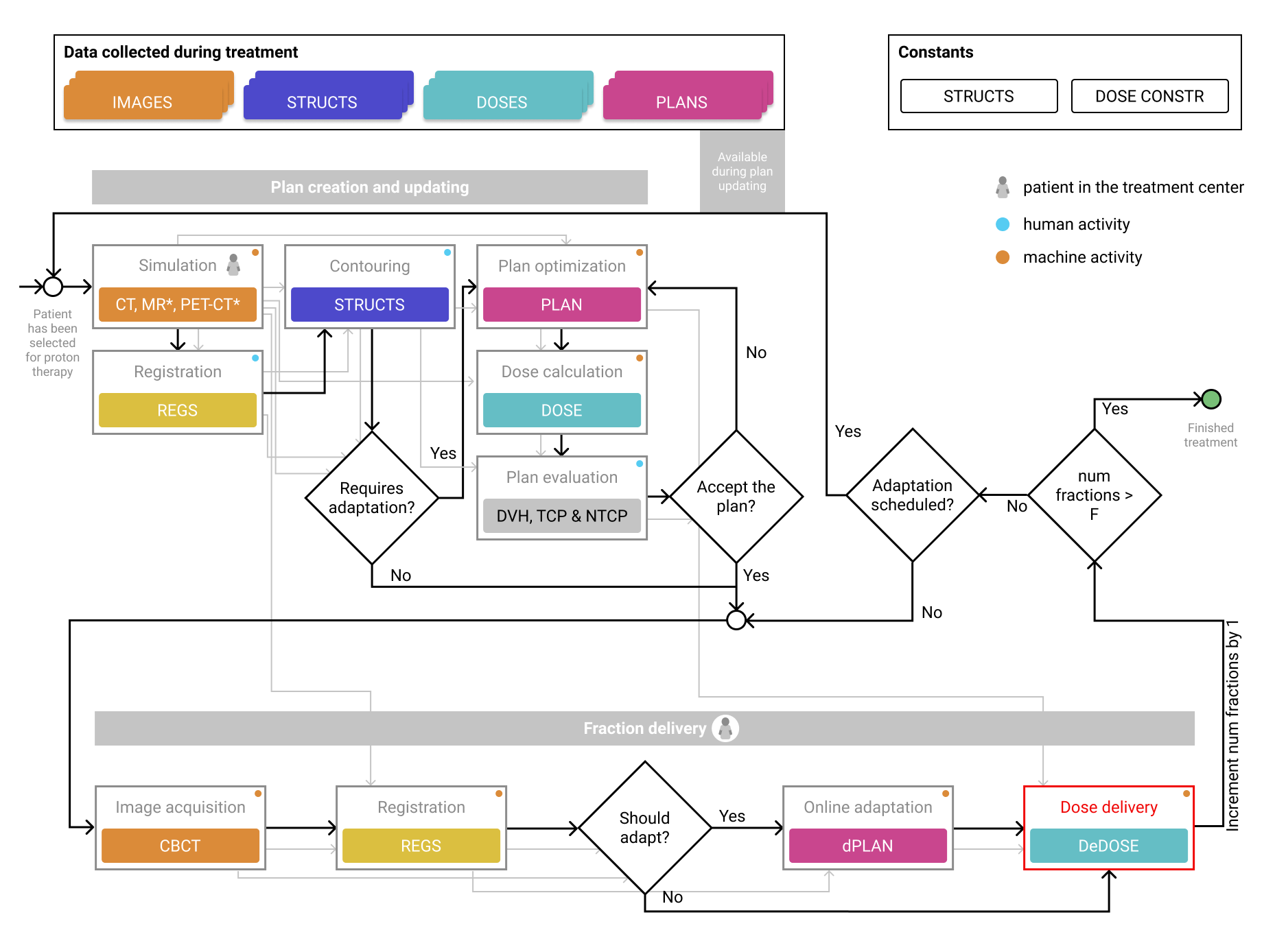}
  \caption{
  Schematic of HollandPTC's dose delivery pipeline, which aims at delivering the planned dose to the patient (box highlighted in red) as precisely as possible. Gray boxes represent processes and diamonds decisions. Processes in the plan creation and updating phase change depending on whether it is the first time they are executed (treatment plan creation) or not (treatment plan updating). Every process can require inputs (gray lines) and produces outputs (colored boxes). As the treatment progresses, information such as contours (STRUCTS) and dose distributions (DOSE) can be saved and used later when updating the treatment plan for comparison purposes. In HollandPTC's case, treatment plan updates are scheduled weekly. Therefore, this workflow is an example of an offline adaptive proton therapy one. Although this kind of scheme can deal with changes that occur over multiple days, it would not be able to account for anatomical variations that happen in-between or within fractions.
  }
  \label{fig:detailed-dose-delivery-pipeline-hptc}
\end{figure}

The end goal of PT is to eliminate the tumor from the patient's body by irradiating it. In clinical practice, this process has two stages. The first one consists of creating a treatment plan that delivers the optimal dose distribution to the patient. After that, the planned dose needs to be delivered to the patient throughout multiple fractions (patient visits to the clinic). Figure \ref{fig:detailed-dose-delivery-pipeline-hptc} presents an overview of HollandPTC's dose delivery pipeline, including its processes and their outputs and the decisions that need to be made. The following sections explain these elements in more detail. 

\subsection{Treatment Plan Creation}
\label{sec:treatment-plan-creation}

The goal of the plan creation phase is to find a treatment machine configuration that can deliver the dose optimally. An optimal dose distribution satisfies clinical goals and constraints like ensuring that the tumor receives the radiation dose (usually 70 Grays (Gy)), while the toxicity of the surrounding organs stays below a given threshold, respectively. Table \ref{tab:dosimetric-info-hn} presents an example of the dosimetric goals and objectives currently used for Head and Neck cancers in HollandPTC.

% Table: link to table in section of head and neck considerations

As can be observed in the plan creation portion of Figure \ref{fig:detailed-dose-delivery-pipeline-hptc}, several processes and their outputs contribute to the end goal of this stage, which is to find an acceptable plan (see decision point "Accept the plan?"). Concretely, the plan optimization process, which produces the treatment plan, requires the contours of the anatomical structures, their dosimetric constraints, and a CT image. Then, the plan evaluation consists of assessing whether the planned dose can satisfy the dosimetric constraints, which are expressed in terms of Tumor Control Probability (TCP) and Normal Tissue Complication Probability (NTCP) (see Section \ref{sec:assessing-plan-quality}). Appendix \ref{ch:appendix-process-specification} provides the full specification of the processes involved in the initial creation of the treatment plan.

In the current clinical workflow at HollandPTC and LUMC, several of the processes of the plan creation stage can be re-executed if the plan needs to be adapted. Although the structure of the workflow remains largely the same, the processes' context changes. Section \ref{sec:workflow-implications-apt} will provide an in-depth discussion of how the plan creation stage changes in the context of APT. 

\subsection{Fractionated Dose Delivery}

\begin{figure}[h]
  \centering
  \includegraphics[width=10cm]{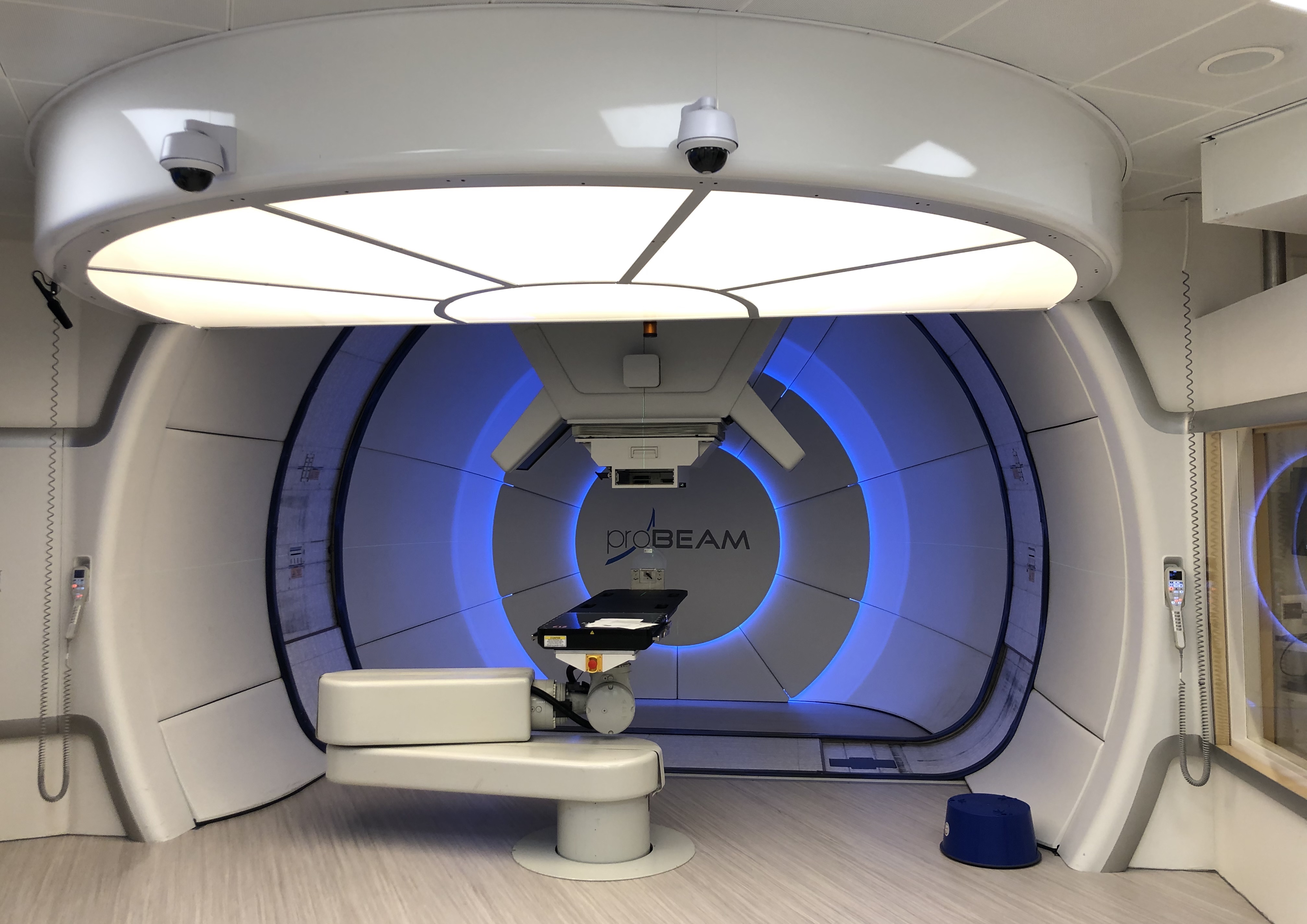}
  \caption{
  HollandPTC treatment room. In a regular head and neck fractionation schedule, patients come 35 days to receive a portion of the dose. The aim is to maximize tumor damage while allowing healthy cells to recover.
  }
  \label{fig:hptc-gantry}
\end{figure}

The bottom half of the pipeline in Figure \ref{fig:detailed-dose-delivery-pipeline-hptc} corresponds to the execution of the treatment plan, delivering the planned dose to the patient. In practice, the radiation dose is fractionated and administered for multiple days. Interleaving small dose batches with resting periods maximizes the damage to the tumor cells while allowing the healthy ones to recover \cite{hellevik_radiotherapy_2014}.
 
Fractionation imposes a significant resource load on the PT workflow given that, for several days, the patients need to come back to the PT center to receive the dose (Figure \ref{fig:hptc-gantry} shows HollandPTC's treatment room, also known as gantry). For HN cancer, it is common to administer a total of 70 Gy of radiation for 35 days (2 Gy per day). In the current workflow, each visit of the patient, known as a fraction, can last between 10 and 30 minutes and involves steps such as acquiring a daily image, positioning the patient on the treatment couch, and delivering the dose (which the machine does automatically). A specification of these processes can be found in Appendix \ref{ch:appendix-process-specification}.

As the following sections explore, changes in the patient's anatomy between and within fractions can lead to deviations of the delivered dose from the planned one, which is especially problematic in the highly-precise PT. Currently, setup uncertainties are addressed in-room by using image guidance to detect and correct positioning errors. When more complex anatomical changes arise, it is likely that the treatment plan needs to be updated, which would trigger the execution of several processes of the plan creation phase discussed before. This plan updating is currently done offline (in between fractions when the patient leaves the treatment center), further increasing the resource load of the treatment center.

\subsection{Assessing Treatment Plan Quality}
\label{sec:assessing-plan-quality}

An integral part of the dose delivery pipeline is to ensure that the planned dose satisfies clinical constraints and that this holds as the treatment progresses. A tool used in practice for assessing and comparing treatment plans' quality is the Dose-Volume Histogram (DVH) which presents, as can be observed in the left section of Figure \ref{fig:dvh-and-metrics-comparison-translation}, for several dose levels (horizontal axis), the percentage of the structure's volume that receives it \cite{drzymala_dose-volume_1991}. 

Qualitatively speaking, the DVH of an acceptable plan is characterized by a large gap between the curves of tumor-related structures and healthy tissue, which indicates better sparing of the latter. In quantitative terms, thresholds are set for the doses that different organs can receive. The first column of the table in Figure \ref{fig:dvh-and-metrics-comparison-translation} presents examples of constraints currently used in the clinic for head and neck cancer \cite{jensen_danish_2020}. Note how the maximum dose that any part of the spinal cord can receive is 50 Gy ($D_{max}<45$ Gy), while the parotid should receive no more than $26$ Gy on average ($D_{mean}<26$ Gy). Regarding the tumor-related structures like the GTV and CTV (more on this later), the whole volume should receive as much of the dose as possible. 

An additional set of tools that build on top of the DVH are the Tumor Control Probability (TCP) and Normal Tissue Complication Probability (NTCP) models, which relate dose and volume with patient outcomes \cite{dean_normal_2016}.  On the one hand, TCP models estimate the probability of eradicating the patient's cancer with the current treatment plan and fractionation schedule. On the other, NTCP ones aim to predict whether a patient is likely to suffer from side-effects like xerostomia based on the amount of dose that, in this case, the salivary glands receive. 

In the clinical workflow, there are two moments for assessing the plan quality. First, when creating the initial treatment plan, DVH, TCP, and NTCP, together with the clinical thresholds, will be used to select one among a set of candidates. Second, the process is repeated every time the treatment plan changes during the treatment. In this case, the goal is to ensure that the quality metrics of the updated plan match those of the initial one. A specification of plan evaluation-related processes can be found in Appendix \ref{ch:appendix-process-specification}.

\section{Errors and Uncertainties}
\label{sec:errors-and-uncertainties}

In PT, uncertainties can arise at multiple steps of the dose delivery pipeline due to setup and anatomical variations, dose calculation approximations, and biological considerations \cite{paganetti_range_2012}. Propagation of these uncertainties can lead to deviations of the delivered dose from the planned one, resulting in underdosing of the tumor and overdosing of the healthy organs. This work focuses on geometric uncertainties, which this section presents. The reason is that these are tightly connected to the delineation process. For an in-depth review of uncertainty management in EBRT refer to \cite{van_der_merwe_accuracy_2017}. Then, it introduces margins and treatment plan adaptation as mechanisms to cope and eliminate them, respectively.

\subsection{Geometric Uncertainties in PT}
\label{sec:geometric-uncertainties}

\begin{figure}[h] % TODO add correct image
  \centering
  \includegraphics[width=\linewidth]{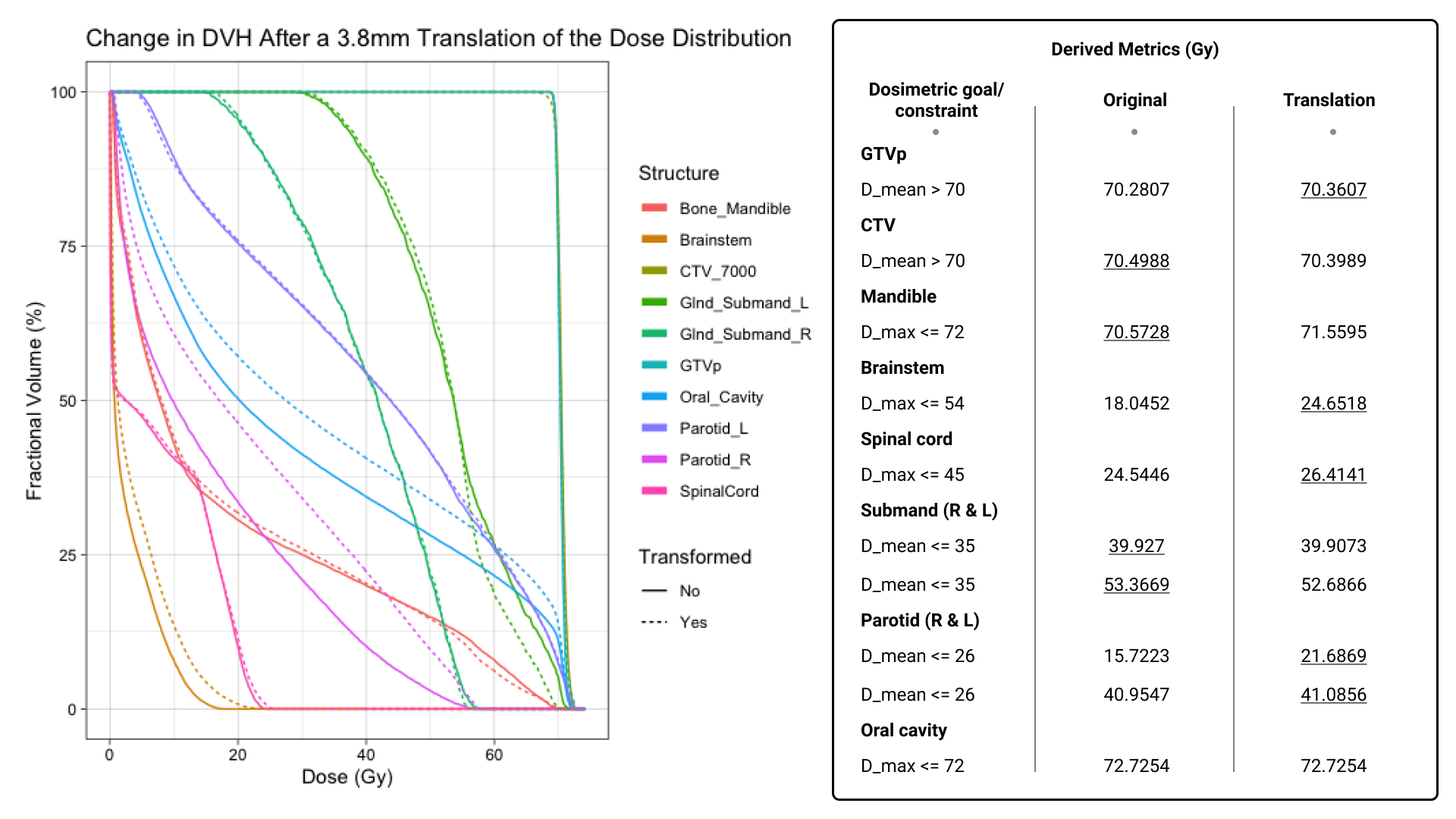}
  \caption{
  Comparison of Dose Volume Histograms (DVH) and a subset of derived metrics used for clinical decision making of two treatment plans. The solid line on the plot corresponds to the original plan. The dashed line corresponds to a simulated setup error of 3.8mm (a translation with x: 3mm; y: 1.2mm; z: 2mm) on the original plan. These errors often happen when positioning the patient on the treatment couch due to the body sliding in the fixation mask. As the table on the right shows, the translation of the patient resulted in a shift of the DVH that can compromise the plan's quality. For instance, even though the mean dose in the right parotid is still below the clinical threshold, it increased significantly, raising the probability that the patient suffers from dry mouth (xerostomia).
  }
  \label{fig:dvh-and-metrics-comparison-translation}
\end{figure}

Geometric uncertainties arise as a consequence of changes in the organ's shape or position with respect to what was recorded in the treatment delineation process. For instance, a change in the patient positioning (due to a failure to replicate the planning setup) during a fraction will result in translations or rotations of the whole anatomy. As Figure \ref{fig:dvh-and-metrics-comparison-translation} shows, this can produce a transformation in the delivered dose, with respect to the planned one, which would, in turn, affect the DVHs, potentially compromising the plan's quality and the patient's outcome.

% figure: example of the effect of shifts in patient setup
% original DVH vs DVH with shift (translation + rotation)

In addition to simple transformations like translations and rotations, the patient's anatomy can change in more complex ways if, for example, they lose weight or if the organs' volume, shape, or position change. For instance, parotid glands can shrink during treatment \cite{morgan_adaptive_2020}, which can cause an increase in the mean dose that they receive. Some other examples of anatomical changes on different time scales for HN are:
\begin{itemize}[noitemsep]
    \item Swallowing (multiple times per fraction).
    \item Filling of nasal cavity (in-between fractions).
    \item Swelling of the tongue (in-between fractions).
    \item Tumor volume change or displacement (throughout multiple fractions). 
    \item Organ volume change or displacement (throughout multiple fractions). 
    \item Weight loss (throughout multiple fractions). 
\end{itemize}

In addition to setup variation and changes in the organs' shapes and position, a different, but related, source of variation comes from the delineations of anatomical structures. Highly uncertain regions in the image are prone to high variability in the contours both between and within observers\cite{vinod_review_2016}. While the first one, known as inter-observer variability, refers to different observers producing different delineations for the same patient, the second, intra-observer variability, points towards a lack of consistency when an observer delineates the same structure multiple times. %Figure \ref{TODO} presents an example of a slice of a parotid with uncertain areas. In these regions, observers will need to complete the contour based on their experience, which will result in a lack of consistency.

% figure: image showing inter and intra-observer variability

As was mentioned, delineation uncertainties and those due to organ motion are interrelated. For example, suppose the parotid gland's pre and post-treatment volumes differ. These can happen for two reasons. On the one hand, the parotid gland shrank, reducing its volume. On the other, due to inter-observer variability, the contour of the post-treatment parotid is tighter, which results in a decreased volume even if that is not the case. In practice, both scenarios would be treated as a geometric uncertainty to address by using safety margins around the structure.

% \subsection{Random vs Systematic Errors}

% As mentioned before, uncertainties can result in errors in the delivery of the dose. 

\subsection{Coping with Uncertainty: Margins}
\label{sec:coping-with-unc-margins}

\begin{figure}[h]
  \centering
  \includegraphics[width=10cm]{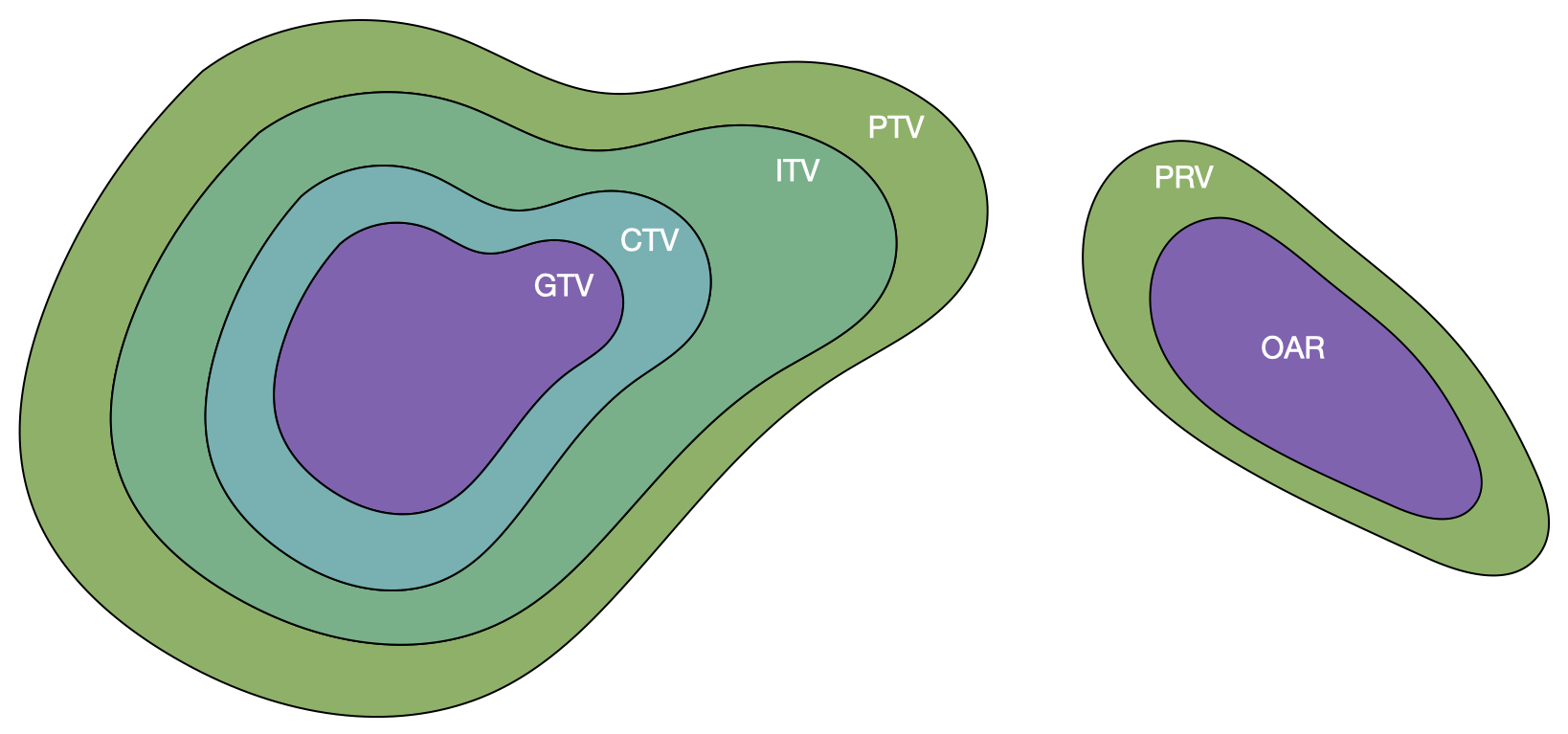}
  \caption{
  Margins added to gross tumor volumes (GTV) and organs at risk (OARs) to ensure that the treatment plan is robust against uncertainties. The latter means that even if the uncertainties materialize into errors, the structures will receive the planned dose. 
  }
  \label{fig:structures-margins}
\end{figure}

The standard way to deal with uncertainties is to add margins around critical anatomical structures to ensure they receive the planned dose \cite{simone_margins_2018}. Figure \ref{fig:structures-margins} presents and schematic of these margins for the tumor (left) and the organs at risk (right). As can be observed, in the case of the tumor, there can be three different contours in addition to the gross tumor volume (GTV), which is the one circumscribing the portion of the tumor visible in the image.  

The first tumor-related margin is the clinical target volume (CTV), which accounts for biological uncertainties. These arise because there might be some invisible spread of the tumor that goes beyond its boundaries. The extent of the CTV can vary depending on the location and type of tumor and is estimated by using the results of population-wise histopathological studies. The second and third tumor-related margins are the internal and planning target volumes, ITV and PTV, respectively, which account for uncertainties due to geometrical and dose-related factors. These are analogous to the planning organ at risk volume (PRV) margin in the case of OARs which ensures that important ones, like the brainstem, do not receive more dose than a predefined limit. 

The ITV, PTV, and PRV are computed using recipes that combine information of random and systematic errors in the dose delivery pipeline \cite{van_herk_errors_2004}. In practice, treatment centers quantify errors and uncertainties in their workflow and use them to calculate the margins using the recipes. Therefore, the extent of the margins might vary from one institution to the other.

\subsection{Reducing Geometric Uncertainty: Adaptation}
\label{sec:reduc-unc-adaptation}

Using margins for optimizing the treatment plan makes it robust because even if the uncertainties materialize (for instance, the tumor moves), they guarantee that the prescribed dose will be delivered to the structure. Furthermore, margins are a cost-effective strategy for dealing with uncertainty since they do not involve adding more time or resources to the clinical workflow. 

Nevertheless, gains in robustness when using large PTV margins come at the expense of preserving healthy tissue. With the previous generation of less conformal dose delivery technologies, the dose falloff was gradual, meaning that the dose distribution was fuzzy around the target volume. In this scenario, there is no incentive to put significant effort into reducing margins given that, even if they are removed, there is still going to be damage to surrounding healthy tissue due to the less conformal dose. 

Recent technologies such as VMAT and PT make it possible to create treatment plans with dose distributions that tightly circumscribe the tumor. In this case, a reduction of margins plays a significant role as it will result in less toxicity in the surrounding tissue and better patient outcomes. Therefore, for PT to realize its organ-sparing capabilities, it is necessary to reduce margins as much as possible, which entails eliminating sources of uncertainty from the dose delivery pipeline. 

An effective strategy to mitigate geometric uncertainties (due to setup variations, organ motion, and delineation variability) is to update the patient's treatment plan when their anatomy changes \cite{brock_adaptive_2019, albertini_online_2019}. To successfully implement APT schemes in the clinic, the first step is to identify the sources of uncertainty and their time frame. For instance, geometric variabilities due to organ motion can arise within some minutes if the patient moves their tongue while on the treatment couch; or in several days if they correspond to treatment-induced weight loss.

Having the sources of uncertainty and an estimate of their frequency, the next step is to implement monitoring mechanisms that can detect them and update the treatment plan accordingly. Section \ref{sec:workflow-implications-apt} provides a detailed explanation of different workflow-level strategies to deal with uncertainty at different time frames. To finalize, although adaptation schemes can help reduce margins, they are harder to implement in the clinic. The reason is that they are more time and resource-intensive. In practice, this factor often results in adaptive workflows that partially address uncertainty, requiring margins to account for the reminder.

\subsection{Other Ways of Mitigating Uncertainty}
\label{sec:other-ways-reduc-uncert}

Although this work focuses on adaptation, there are other mechanisms for eliminating or reducing sources of geometric uncertainty in the dose delivery pipeline \cite{vinod_review_2016}. The list below presents several strategies currently implemented in clinical practice, together with the type of uncertainty that they address. 
\begin{itemize}[noitemsep]
    \item (setup) Immobilization hardware coupled with lasers can be used both in the simulation and the dose delivery steps to ensure that there is as little as possible patient setup.
    \item (setup and delineation) Protocols can be used to ensure repeatability and consistency of the medical procedures, leading to less variation between and within observers. In the case of delineation, several guidelines have been proposed \cite{brouwer_ct-based_2015, gregoire_delineation_2014, lee_international_2018, gregoire_delineation_2018, gregoire_ct-based_2003, biau_practical_2019}.
    \item (setup, organ motion, and delineation) Complementing the planning CT scan with additional modalities like MR or PET can help make structures' boundaries clearer, which results in less uncertainty around their location and delineations.
    \item (delineation) Training of the ROs and RTTs ensures effective use of the guidelines. Training sessions are an opportunity to correct systematic errors caused by, for example, a lack of anatomical knowledge. 
    \item (delineation) Peer-reviewing makes catching and avoiding random and systematic delineation errors more likely. Also, discussing with peers helps clinicians become aware of their mistakes, making them less likely to happen in the future.
\end{itemize}

\section{Workflow Implications of Adaptation}
\label{sec:workflow-implications-apt}

Other things equal, switching from a photon to a proton-based treatment does not significantly change the clinical workflow. Nevertheless, as stated before, to realize the full potential of protons and their higher level of conformality, it is necessary to shrink the margins through adaptation. The latter does produce an effect on the clinical workflow, which magnitude varies depending on the granularity at which uncertainties will be addressed \cite{green_practical_2019, lim-reinders_online_2017}. This section presents the ingredients of adaptation and the resulting workflows.

% TODO: there is a citation that I would like to have but is missing. It corresponds to a paper that has multiple workflows for online, offline and real time adaptations.

\subsection{Image Guided Proton Therapy (IGPT)}

The most important enabler of adaptive workflows is the availability of images of the patient throughout the treatment. Image-Guided Proton Therapy (IGPT) is implemented by installing an image acquisition device in the treatment room, which can be used to get a snapshot of the patient's anatomy-of-the-day before the treatment starts \cite{jaffray_image-guided_2012, sterzing_image-guided_2011}. Ideally, it should be possible to re-optimize a treatment plan using the daily image, which would require a CT scan. Nevertheless, the extra radiation to the patient reduces the appeal of this alternative. Using an MR solves this problem but is more costly, and the technology that enables using MRIs for treatment planning is not yet ready for clinical use \cite{tetar_clinical_2019}. In general, a compromise has been reached of using CBCT scans which are cheaper to acquire and deliver small quantities of radiation to the patient, but which also are not suitable for treatment planning. Consequently, in current CBCT-based clinical workflows, if a change that requires replanning is detected, an offline adaptation session is triggered, which often entails running a time-constrained version of the treatment creation process.

\subsection{Adaptation Trigger and Schedules}
% discuss image guidance
% the adaptation trigger gives cadence/rhythm to the adaptive schedule

The first question that arises when implementing an adaptive workflow is when to perform the plan adaptation. The straightforward answer is that the plan should be updated every time the patient's anatomy changes to minimize the margin's extent. Nevertheless, this level of responsiveness is costly and, therefore, rarely implemented. In contrast, in clinical practice, adaptation sessions are usually scheduled based on estimates of the frequency of anatomical changes. Regardless if adaptations are triggered or scheduled it is possible to differentiate between three levels, each of which can be paired with different re-planning strategies \cite{green_practical_2019}. 

The first level, \textbf{offline adaptation}, deals with changes that become significant throughout multiple fractions. For instance, a tumor that gradually slides out of the radiation field or a shrinking parotid gland that receives a higher dose. The anatomical variations that this level addresses are more complex because they are related to the structures' shape and volume. Therefore, they are usually dealt with by re-executing the treatment creation process in-between fractions when the patient has left the treatment center, hence the offline in the name. This level of adaptation can be implemented via a trigger (decision made by the radiotherapy technologist after comparing the CBCT with the planning CT) or scheduling the plan updating sessions. 

The next level, \textbf{online adaptation}, addresses changes that occur in the time in-between fractions and can affect their effectiveness. A well-known example corresponds to the variations in the patient setup due to a failure to replicate it precisely. IGPT addresses this by adjusting the treatment machine parameters based on the patient's position shift. There are more complex changes that can also affect the delivered dose on a fraction basis. Examples include swelling of the oral cavity and filling of the sinuses. Detecting these variations entails delineating the daily CBCT, which is out of reach for most treatment centers. In these cases, margins or robustly optimized plans are required to guarantee tumor coverage \cite{cubillos-mesias_including_2019}. 

The third level, \textbf{real-time adaptation}, corresponds to changes that can happen multiple times within a fraction like peristaltic movements, heart beating, and breathing. In these cases, the plan needs to be updated multiple times within a fraction to eliminate the associated margin. This level of adaptation is characterized by predictable changes in the anatomy for which automatic replanning strategies have been devised. The context of real-time plan updating is extremely time-constrained, and, therefore, human intervention should be eliminated. Given the lack of human involvement and the fact that it is not used in practice for the HN (besides swallowing and positioning of the tongue, the HN area is a relatively static one in this time frame), this report does not further discuss real-time adaptation strategies.

\subsection{Adaptation-Induced Changes to Existing Processes}

There are two components that adaptive workflows introduce. First, the adaptation trigger produces an alert if there is an anatomical change that requires updating the plan. If only considering setup errors, then the current implementation of IGPT suffices. More complex organ-specific variations require comparing the daily contours to the original ones. If this change detection mechanism is implemented online, then the contouring time is restricted to a couple of minutes. Alternatively, the adaptation trigger could be performed in-between fractions, increasing the allowed time. Care must be taken if using a CBCT for the monitoring given that if there is a need for adaptation, a new image of the patient will be required. Ideally, the patient would not need to come twice to the treatment center for this. 

The second component of adaptive workflows is the re-execution of the treatment creation process, which Section \ref{sec:treatment-plan-creation} discussed. In current clinical practice, online plan adaptation is only feasible for simple errors such as those caused by changes in the setup. Therefore, complex adaptations that require plan re-optimization will likely happen offline in-between fractions. Despite the increased time with respect to within fraction updates, offline sessions, which need to fit in a day, are still time-critical compared to the initial treatment plan creation, which currently takes around three days in LUMC and HollandPTC. 

In addition to the reduced time, the nature of the task also changes, impacting resource availability. When creating the initial treatment plan, the goal is to find an optimal one regarding a set of dosimetric goals and constraints. In contrast, adaptation relies on comparing the current deviant plan with the original one, ensuring that their quality matches. The practical outcome of this is that only one image, a planning-quality CT, is taken. Then, in the contouring step, it is necessary to register this CT to the original set of images (for instance, the planning CT, MR, and PET), which introduces a new source of uncertainty.

\section{Head and Neck Considerations}
\label{sec:head-and-neck-considerations}

Due to its resource-intensive nature and limited availability, APT is commonly deployed for tumor sites that can benefit from its high level of conformality. Although most of the concepts in this document apply across tumor sites, the results of the study presented in Chapter \ref{ch:current-contouring-wf} and the discussion in Chapter \ref{ch:towards-apt} focus on the head and neck (HN) region.

\begin{figure}[h]
  \centering
  \includegraphics[width=\linewidth]{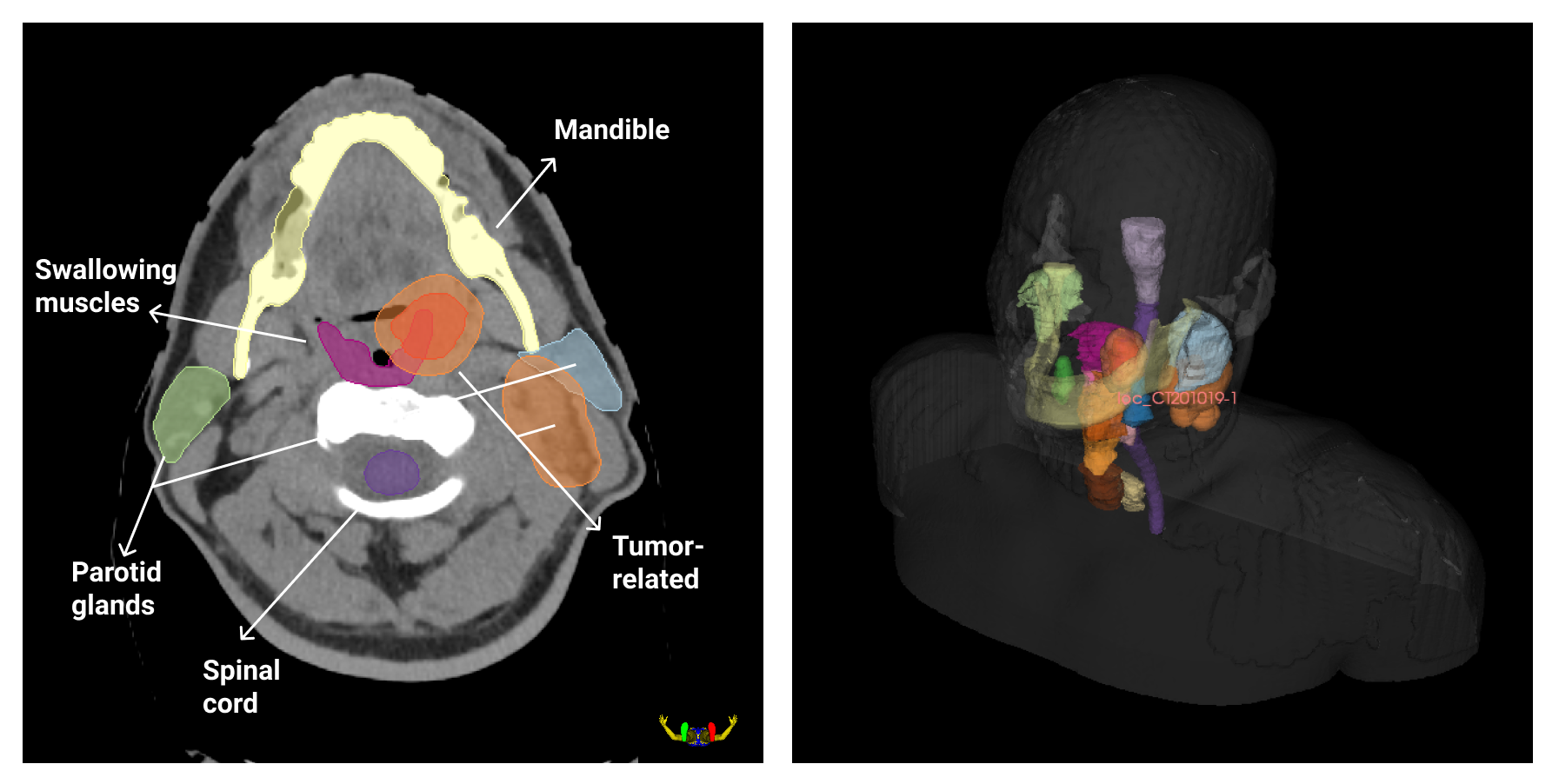}
  \caption{
  A subset of the contours of a HN patient at HollandPTC displayed on an axial slice (left) and in the 3D view (right). Due to the high number of critical structures, the HN is time consuming and resource intensive.
  }
  \label{fig:han-structure-density}
\end{figure}

In contrast with other areas such as the bladder and the lungs, the HN houses significantly more critical organs that, if damaged, can lead to severe side effects. For example, Figure \ref{fig:han-structure-density} presents the axial and sagittal slices of the CT of a patient with an oropharyngeal tumor. Note how in the vicinity of the tumor (in red), there are several critical structures like the spinal cord, the parotid glands, and the swallowing muscles. High radiation-induced toxicity in the salivary glands (parotids and submandibular glands) and the swallowing muscles can result in significant side effects like xerostomia (dry mouth) and dysphagia (difficulty swallowing).

% Figure: axial and sagittal slices of a patient with a nasopharyngeal tumor 

Table \ref{tab:dosimetric-info-hn} presents the principal structures in the context of HN cancers together with information about possible anatomical changes that they might suffer and their dosimetric goals and constraints. It is a subset of dose constraints presented by the Danish Head and Neck Cancer Group in their 2020 Radiotherapy Guidelines \cite{jensen_danish_2020}. Table \ref{tab:contouring-info-hn} in the next chapter complements it with information about the structure's delineation uncertainty, which affects the task difficulty. 

% Table that compiles information of all the structures that we consider in this document
\begin{table}[]
\begin{tabular}{l|p{2.5cm}|l|p{2.5cm}|p{2.5cm}|l}
\textbf{Group} & \textbf{Name}              & \textbf{Priority} & \textbf{Dosimetric goal or constraint} & \textbf{Side effect}                                       & \textbf{Tissue type} \\ \hline
OAR            & Mandible                   & medium            & $D_{max} \le 72 Gy$                    & Osteoradionecrosis                                         & bone                 \\[0.25cm]
OAR            & Parotid glands             & medium            & $D_{mean} \le 26 Gy$                   & Xerostomia                                                 & soft tissue          \\[0.25cm]
OAR            & Submandibular glands       & medium            & $D_{mean} \le 35 Gy$                   & Xerostomia                                                 & soft tissue          \\[0.25cm]
OAR            & Brainstem                  & maximum           & $D_{max} \le 54 Gy$                    & Neurological damage                                        & soft tissue          \\[0.25cm]
OAR            & Spinal cord                & maximum           & $D_{max} \le 45 Gy$                    & Neurological damage                                        & soft tissue          \\[0.25cm]
OAR            & Optic nerves               & high              & $D_{max} \le 54 Gy$                    & Visual disturbance                                         & soft tissue          \\[0.25cm]
OAR            & Optic chiasm               & high              & $D_{max} \le 54 Gy$                    & Visual disturbance                                         & soft tissue          \\[0.25cm]
OAR            & Eyes                       & high              & $D_{max} \le 45 Gy$                    & Retinopathy, conjunctivitis, dry eye syndrome and cataract & soft tissue          \\[0.25cm]
OAR            & Swallowing muscles         & medium            & $D_{mean} \le 55 Gy$                   & Dysphagia                                                  & soft tissue          \\[0.25cm]
OAR            & Oral cavity                & medium            & $D_{mean} \le 30 Gy$                   & Xerostomia and mucositis                                   &                      \\[0.25cm]
TV             & Primary GTV (GTVp)         & high              & Should receive full planned dose       & NA                                                         & various              \\[0.25cm]
TV             & Nodal GTV (GTVn)           & high              & Should receive full planned dose       & NA                                                         & soft tissue          \\[0.25cm]
ELN            & Elective Lymph Node Levels & high              & Should receive full planned dose       & NA                                                         & various              \\[0.25cm]
PS             & Implants                   & high              & NA                                     & NA                                                         & metal                \\[0.25cm]
PS             & Artifacts                  & high              & NA                                     & NA                                                         & various              \\[0.25cm]
PS             & Planning helpers           & high              & NA                                     & NA                                                         & various              \\ \hline
\end{tabular}
\caption{Example of clinical priorities, dose constraints, and side effects for a subset of the HN structures considered in PT \cite{jensen_danish_2020}. Note how these can be divided into organs at risk (OAR), target volumes (TV), elective lymph nodes fields (ELN), and planning structures (PS). While the structures in the first three groups can vary with the type of cancer and its location, the last one depends on the quality of the patient's planning CT. Artifacts or problems in the CT might make it unsuitable for optimizing PT treatment plans, given that protons are more sensitive to changes in image intensities. Planning structures help fix these issues by overriding image values with presets. For example, an artifact can be overridden with soft-tissue values, making it suitable for treatment planning. Planning structures are not further considered in this document.}
\label{tab:dosimetric-info-hn}
\end{table}

\section{Key Points}

The goal of this chapter was to understand the purpose of adaptation in PT. The following list presents a summary of what was discussed with links to the relevant sections. 

\begin{itemize}[noitemsep]
    \item PT enables more conformal treatments that destroy malignant cells \textbf{while preserving healthy surrounding tissue}. (Section \ref{sec:what-is-pt})
    \item The ultimate aim of PT is to deliver the prescribed dose to the patient's tumor. The dose delivery pipeline is designed to fulfill this goal. (Section \ref{sec:dose-delivery-pipeline})
    \item Uncertainties in the dose delivery pipeline can cause discrepancies in the delivered dose with respect to the planned one. (Section \ref{sec:geometric-uncertainties})
    \item To ensure that the tumor receives the prescribed dose, margins around the tumor are used. Nevertheless, these can result in additional damage to surrounding tissue. (Section \ref{sec:coping-with-unc-margins})
    \item By adapting the patient's treatment plan when anatomical changes are detected, geometrical uncertainties can be eliminated, which results in a reduction of the margins. (Section \ref{sec:reduc-unc-adaptation})
    \item Adaptation adds two new components to the dose delivery pipeline: adaptation trigger and plan re-optimization session. (Section \ref{sec:workflow-implications-apt})
    \item Adaptation is significantly more resource-intensive than margins, requiring extra images, contours, and, in some cases new treatment plans. For this reason, offline adaptation is the most common approach. (Section \ref{sec:workflow-implications-apt})
\end{itemize}

%% file: content/current-contouring-wf.tex
\chapter{Study of the Offline-APT Contouring Workflow}
\label{ch:current-contouring-wf}

This chapter presents the results of an observational study in Holland Proton Therapy Center (HollandPTC) and the Radiotherapy Department of Leiden University Medical Center (LUMC) in the Netherlands. The goal of the study was to understand how the contouring process fits in an offline-adaptive PT paradigm and which are bottlenecks that prevent it from scaling to online-adaptive PT. To achieve the goal, the results are structured into two levels. At the human-computer interaction level, the interactions between a single actor and the computer were recorded and analyzed. Then, the process level grounds these results in the clinical workflow discussing aspects like the distribution of labor and relationships with other processes. A detailed description of the methods used for gathering and analyzing the data of the study can be found in Appendix \ref{ch:methods-study-apt-workflow}. The last section of this chapter summarizes the key findings in terms of factors that affect contouring performance.

\section{Human-Computer Interaction Level}
% presents the detailed procedure of contouring a single anatomical structure. We focus on the actions that individuals perform in the software and the tools they use. We do this for multiple structures and multiple scenarios.

This section starts the description of the contouring process from the bottom: the interactions that individual users perform with contouring software to produce the delineations. 

\subsection{What is Contouring?}

\begin{figure}[h]
  \centering
  \includegraphics[width=\linewidth]{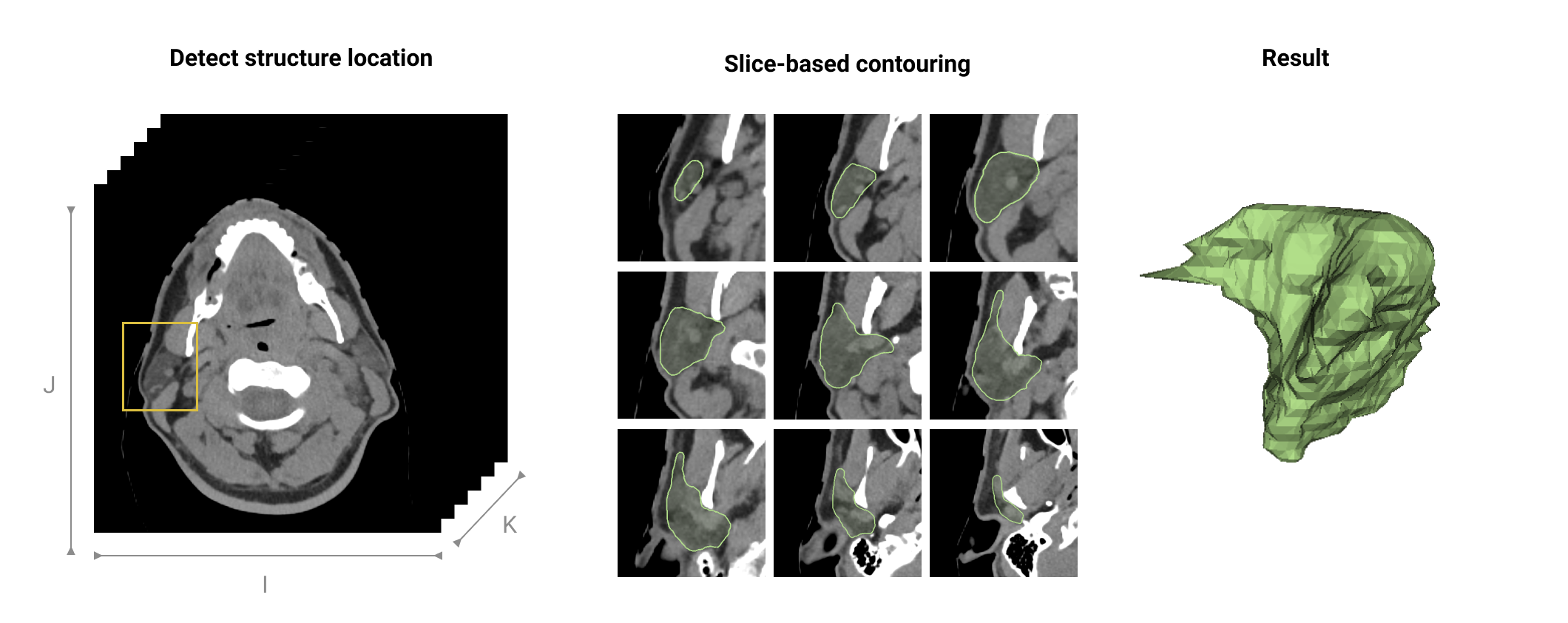}
  \caption{
  Process of contouring a right parotid gland. First (left), the structure needs to be located in the image volume, in this case, a CT. Then (middle), the clinician traverses the whole volume encircling those parts of the image that belong to the anatomical structure. This sequence of steps yields the three-dimensional object on the right, the patient's right parotid gland. Note how in clinical practice, there is no ground truth to assess the accuracy of the contour. Instead, the clinician must rely on the images, which, in several cases, provide incomplete or uncertain information. 
  }
  \label{fig:overview-contouring-process}
\end{figure}

Given a 3D image (for example, a CT or an MRI) of a patient with dimensions $I \times J \times K$ and a set of $S$ anatomies to delineate (which for HN is usually a superset of those presented in Table \ref{tab:dosimetric-info-hn}), the goal of contouring is to generate clinically-acceptable outlines for these structures in the image. Figure \ref{fig:overview-contouring-process} depicts an example of this process, where a slice of the right parotid gland has been delineated. In the figure, highlighted parts belong to the structure while the rest corresponds to the background (or to other structures).

When performing a delineation task, every user action (or lack of action) can be mapped to three general states: evaluation, execution, and miscellanea actions (like workspace setup). These coarse categories can be further subdivided into concrete actions, each of which can be carried out in different ways using contouring software (CSW) \cite{aselmaa_using_2017}. The list below presents these three levels of the action hierarchy. For the deepest level, the aim is not to be exhaustive with the features that CSW offers but rather to present an overview of the tools that have been adopted in clinical practice.

\begin{itemize}[noitemsep]
    \item Evaluation: the goal is to ensure that current delineations are correct and detect parts of the image volume where they are missing.
    \begin{itemize}[noitemsep]
        \item Analysis/visualization
        \begin{itemize}[noitemsep]
            \item Axis-aligned slice views (most used): three 2D scrollable views of the patient volume in the orthogonal, axial, sagittal, and coronal directions.
            \item 3D view (not used): a surface reconstruction of the contours that have been drawn. Enables inspecting global characteristics of the delineations such as their smoothness and inter-organ relationships.
        \end{itemize}
        \item Navigation
        \begin{itemize}[noitemsep]
            \item Consecutive slice switching (most used): every axis-aligned view shows only one slice of the image. To traverse the complete volume it is possible to scroll (or use the arrow keys) to move in a given direction.
            \item Point-and-click slice switching (rarely used): clicking in one point along the horizontal or vertical directions of an axis-aligned view will update the other two to match the selection. 
        \end{itemize}
    \end{itemize}
    \item Execution: the goal is to alter the contour in some way. It could be drawing it from scratch or editing an existing one.
    \begin{itemize}[noitemsep]
        \item Direct manipulation
        \begin{itemize}[noitemsep]
            \item Brush (most used): after setting the radius, it is possible to click and hover over any of the axis-aligned views to select voxels that should be part of the delineation. 
            \item Smart brush (rarely used): using contrast information in the image, snaps the brush to the borders of the structures, making it harder to overflow a contour.
            \item Mesh transformations (rarely used): makes it is possible to transform the whole contour by translating, rotating, or scaling the mesh.
            \item Mesh editor (not used): enables editing the contour in 3D by dragging the vertices that make up the mesh to the desired position. Depending on the value of the influence radius, several vertices adjacent to the one being dragged will follow, ensuring that the contour remains smooth. 
        \end{itemize}
        \item Indirect manipulation
        \begin{itemize}[noitemsep]
            \item Between slice interpolation (most used): by drawing the contour in only some slices across the axial direction, this tool permits completing the rest by interpolating the missing contours.
            \item Model-based contouring (rarely used): given a starting mesh, it attempts to fit it to the shape of the organ, considering image information. Usually, it does not produce satisfactory results in areas with low contrast in the structures' boundaries.
            \item Region growing (not used): given some seeds representing where the different structures are, it grows them into the contours. For a given organ, the growth stops when a sharp difference between voxels is detected (a boundary). Also, the algorithm can be steered by scribbling additional seeds.
            \item Atlas-based contouring (not available): given a curated dataset of images and their contours, attempts to deformable register the former to the input image and then propagates the contours. Finally, these are combined into the final ones.
            \item Deep learning-based contouring (not available): a model that has been trained on a large number of data (images and contours) is used to predict the structures' delineations in the input image. 
        \end{itemize}
    \end{itemize}
    \item Non-contouring interactions (NCI): user actions that do not result in a contour being affected but play a significant role in the delineation task. 
    \begin{itemize}[noitemsep]
        \item Layout change: enables switching from having the three axis-aligned views to having only one or comparing two images side-by-side.
        \item Window-level update: only works with CT images because of their standardized values. It limits the amount of information of the image that will be displayed, reducing the dynamic range and increasing the contrast of the structures whose values fall within. 
    \end{itemize}
\end{itemize}

The following sections use the taxonomy of actions presented above to understand which parts of the contouring task take the most time and why. The analysis uses coarse categories (visualization, non-contouring interaction, navigation, direct and indirect manipulation) to avoid yielding tool-dependent conclusions.

\subsection{Contouring From Scratch vs Refining Contours}
\label{sec:contouring-from-scratch-vs-refining}

There are two main approaches to contouring, which depend on the role of the user. On the one hand, the user can create the contours from scratch (approach 1), using tools like between slice interpolation to speed up the process. On the other hand, it is often possible to generate a starting point (for instance, by using atlas-based or deep learning-based methods or by propagating previous patient structures using rigid or deformable registration) that the user can then refine (approach 2).

% figure: approach 1 and approach 2 schematic

\begin{figure}[h]
  \centering
  \includegraphics[width=\linewidth]{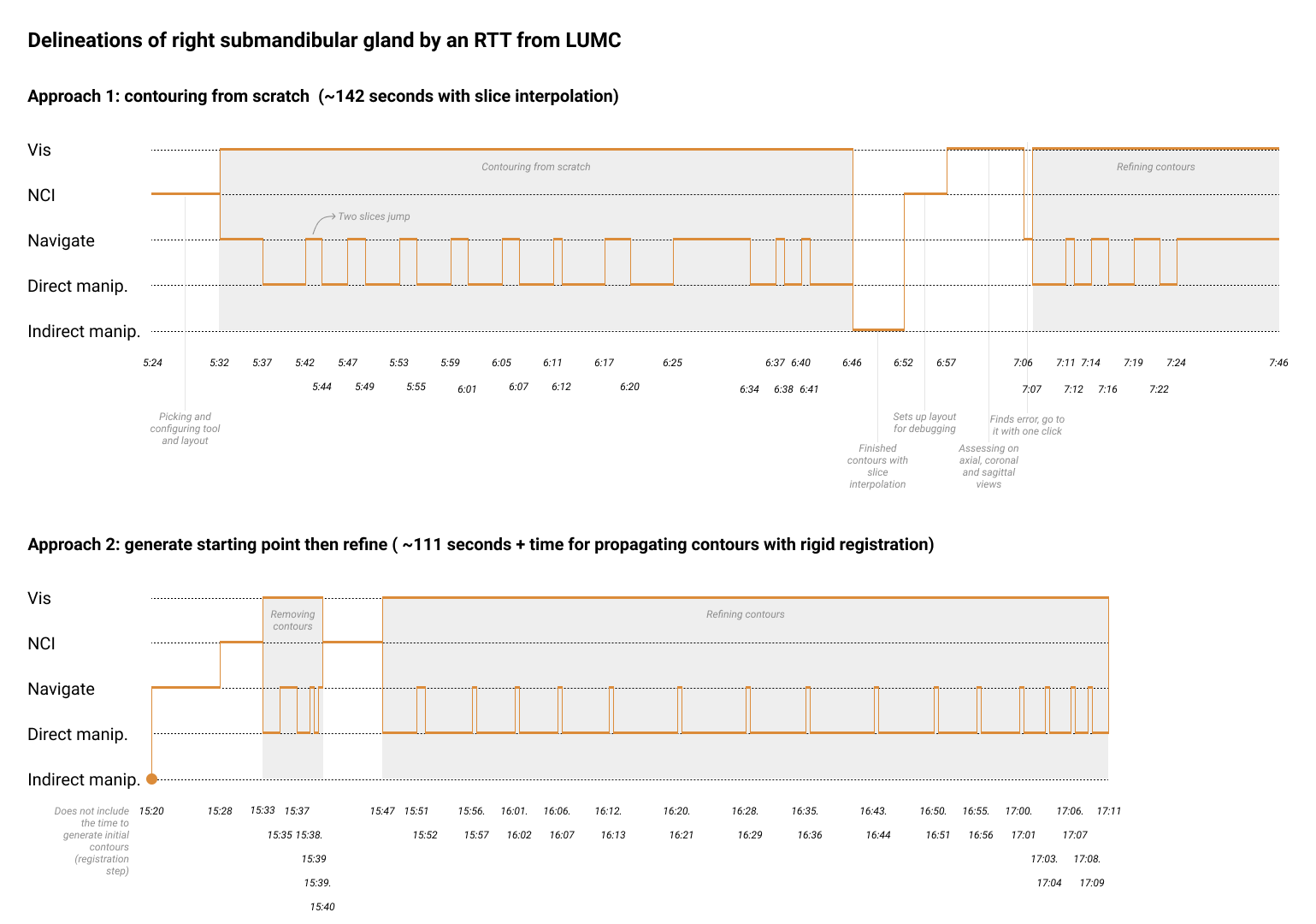}
  \caption{
  Comparison of the actions of the two main contouring approaches for the right submandibular gland of an HN cancer patient. In both cases, the delineations were performed by the same radiotherapy technologist (RTT) at LUMC, which has several years of experience. Note that the x-axis encodes time and makes both timelines comparable. For approach 1 (top), the RTT used the brush in combination with the between-slice interpolation tool. As can be observed, what characterizes this approach are long contour creation periods but short editing ones. For approach 2 (bottom), the RTT performed a rigid registration with another of the patient's scans and propagated the contour. In contrast to the previous approach, this one has a relatively small creation time (that all the structures share) but significant editing ones. Although, in this case, approach 2 was faster than 1, this depends heavily on the extent of the inaccuracies or the structures' extents, respectively. Therefore, in practice, clinicians tend to generate a contour and then, based on its quality, decide whether use and follow the second approach or not and default to the first one.
  }
  \label{fig:comparison-contouring-approaches}
\end{figure}

Figure \ref{fig:comparison-contouring-approaches} depicts a comparison of the two approaches performed by an experienced radiotherapy technologist for the right submandibular gland of a cancer patient. In this example, contouring from scratch took more time than refining contours (142 versus 111 seconds). Furthermore, if the timelines are divided into contour creation and contour editing stages (gray areas), it can be seen how for approach 1, the latter occupies a smaller portion of the overall time, suggesting that contouring from scratch yields fewer errors to address, at least from the perspective of the contours' author.

Although the example of the submandibular gland seems to indicate that refining automatically-generated contours is faster, this needs not always be the case and depends on the scaling behavior of each approach. Approach 1 scales with the structure's extent, meaning those that occupy more voxels will take longer to contour. In contrast, the second approach scales with the number of errors and their extent. Other things equal, if a structure's contour has inaccuracies in several slices, there is a possibility that erasing it and using approach 1 is the most efficient course of action. 

% Figure comparison of approaches

Finally, both contouring approaches discussed above are common in clinical practice, and it is left to the user to decide which one to choose depending on the contouring task at hand. The contouring from scratch approach seemed to be preferred when there were no preexisting contours of the patient (which could then be propagated) or when the automatically generated delineations were not good enough (often the case with methods like rigid registration or model-based contouring). In the latter case, the users often preferred to re-do the contours from scratch rather than correct them.

\subsection{Analysis of Contour Refinement}
\label{sec:contour-refinement-analysis}

Although contouring from scratch is prevalent in clinical practice, auto-contouring approaches that yield reasonable starting points are increasingly common in adaptive workflows due to their time-critical nature. In the clinic, commonly used methods include model \cite{acosta_multi-atlas-based_2014} and atlas-based segmentation \cite{pekar_automated_2004} and propagating contours using a rigid or deformable image registration \cite{kumarasiri_deformable_2014}. Additionally, vendors are starting to offer deep learning-based segmentation tools, which can reach expert-level performance in several organs \cite{nikolov_deep_2020}. The following paragraphs present the contour refining process for tumor-related structures, zooming into instances of error patterns that cause significant delays. This analysis complements and confirms the observations made for the parotid gland in the previous section.

\begin{figure}[h]
  \centering
  \includegraphics[width=\linewidth]{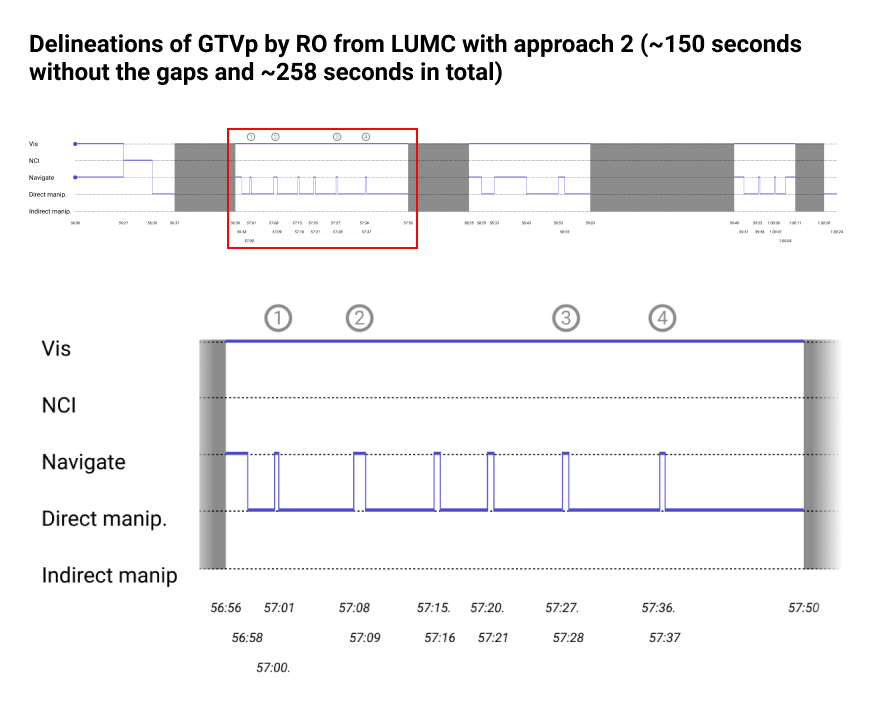}
  \caption{
  Interaction timeline for the refinement of the primary gross tumor volume (GTVp) after propagating it from a planning CT scan using rigid registration (contour creation time not considered). An experienced radiation oncologist (RO) from LUMC performed the edits. Dark gray areas correspond to the clinician being interrupted, which often occurs in clinical practice. Note how, in the zoomed-in area, there are a lot of redundant interactions, switching between navigation and direct manipulation. These interactions are presented in more detail in Figure \ref{fig:refinement-chunk-gtvp}.  
  }
  \label{fig:approach2-gtvp}
\end{figure}

To start, Figure \ref{fig:approach2-gtvp} presents the interaction flow for refining the propagated contours (via rigid registration) of the primary tumor volume (GTVp).  Similar to the case of the right submandibular gland in the previous section, the process can be subdivided into refinement chunks, which are areas in delineations that share an error pattern. The general journey of the user consists of finding a refinement chunk, determining the course of action, and executing it. For an anatomical structure, this process repeats multiple times until there are no unaddressed chunks. In the case of the GTVp, this takes around four minutes. 

% figure: whole interaction flow GTVp

\begin{figure}[h]
  \centering
  \includegraphics[width=\linewidth]{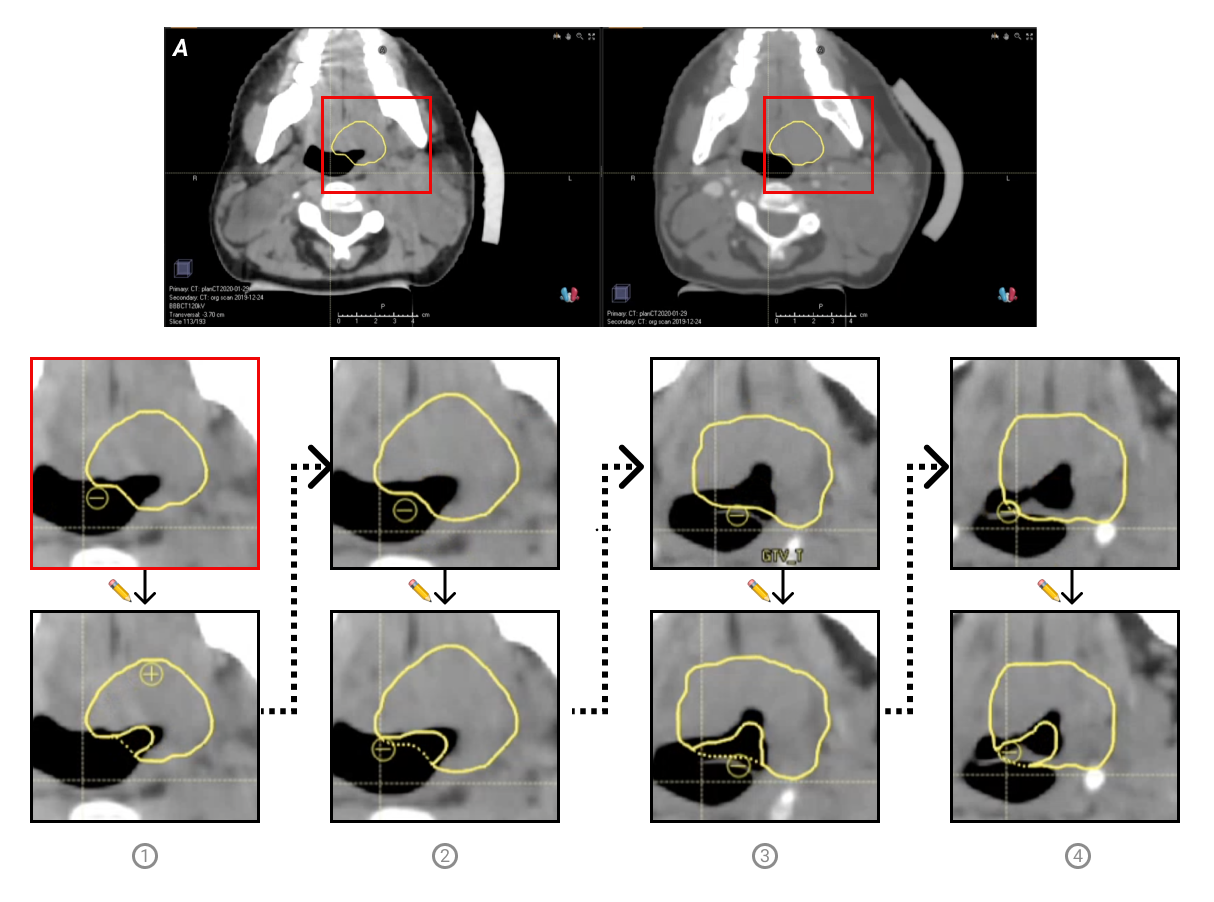}
  \caption{
  Refinement chunk from editing the delineation of a GTVp. The GTVp was propagated from the planning CT to the current one using rigid registration. After the propagation, the GTVp's contour includes one part of the trachea. It is necessary to fix inaccuracy to avoid surrounding organs receiving more radiation than necessary after adding the CTV and PTV margins. 
  }
  \label{fig:refinement-chunk-gtvp}
\end{figure}

Figure \ref{fig:refinement-chunk-gtvp} presents an example of a refinement chunk for the GTVp (removing the trachea's air from the GTVp contour), together with the sequence of actions that the user performed to address it. It is possible to observe how the process is highly redundant, with the user erasing the portion of the contour that does not correspond to GTVp across multiple slices. Even though one erasing interaction takes around 6 seconds, the whole chunk took approximately 1 minute because of the redundancy. 

\begin{figure}[h]
  \centering
  \includegraphics[width=\linewidth]{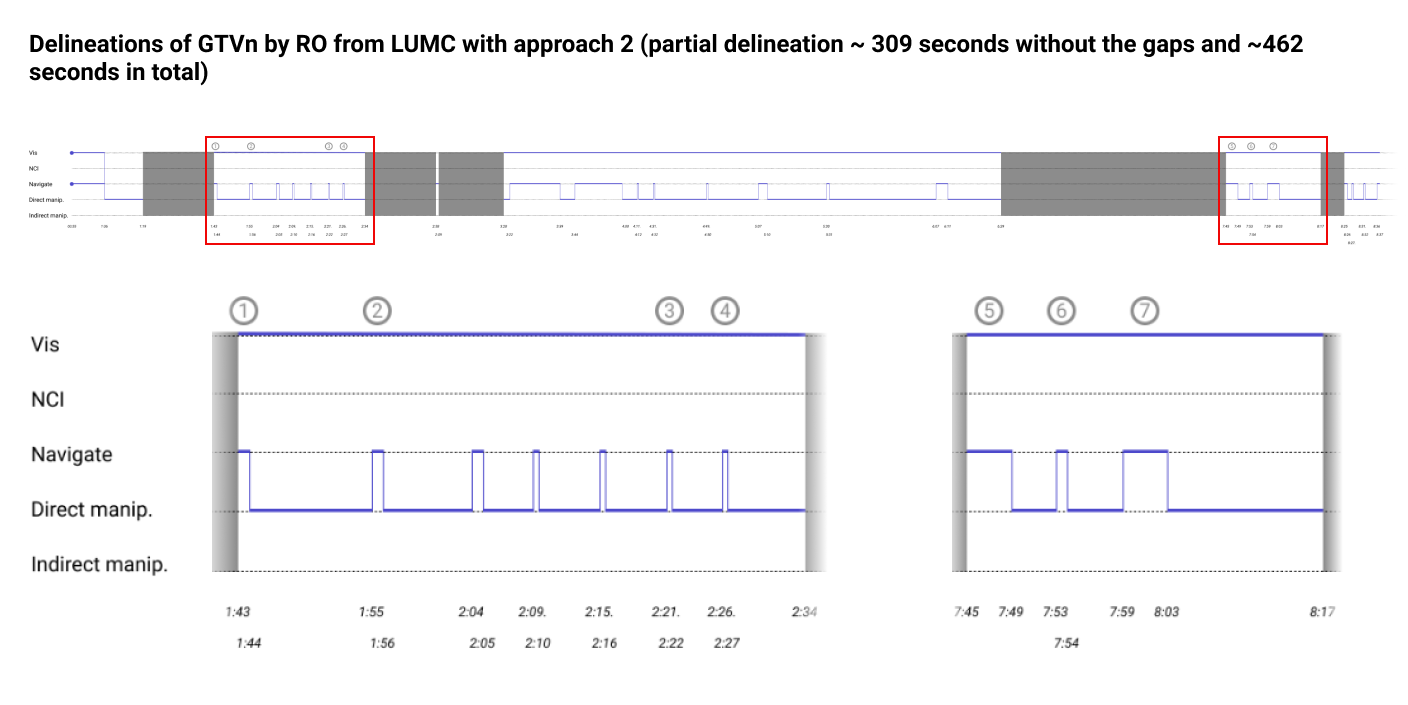}
  \caption{
  Interaction timeline for the refinement of the nodal gross tumor volume (GTVn) after propagating it from a planning CT scan using rigid registration (contour creation time not considered). An experienced radiation oncologist (RO) from LUMC performed the edits. Dark gray areas correspond to the clinician being interrupted, which often occurs in clinical practice. Note how, in the zoomed-in areas, there are a lot of redundant interactions, switching between navigation and direct manipulation. These interactions are presented in more detail in Figures \ref{fig:refinement-chunk1-gtvn} and \ref{fig:refinement-chunk2-gtvn}.  
  }
  \label{fig:approach2-gtvn}
\end{figure}

Moving to the contour refinement interaction flows of the nodal tumor volume (GTVn),  which Figure \ref{fig:approach2-gtvn} depicts, it is possible to observe a similar behavior. After an error or inaccuracy has been spotted, for instance at \textit{1:43} (the zoomed-in area in the left), there is a back and forth between direct manipulation and navigation interactions that follow. In contrast to the case of GTVp, though, the extent of the GTVn is significantly larger (spanning more voxels), which results in manipulation interactions that take longer. 

% figure  GTVn refinement

\begin{figure}[h]
  \centering
  \includegraphics[width=\linewidth]{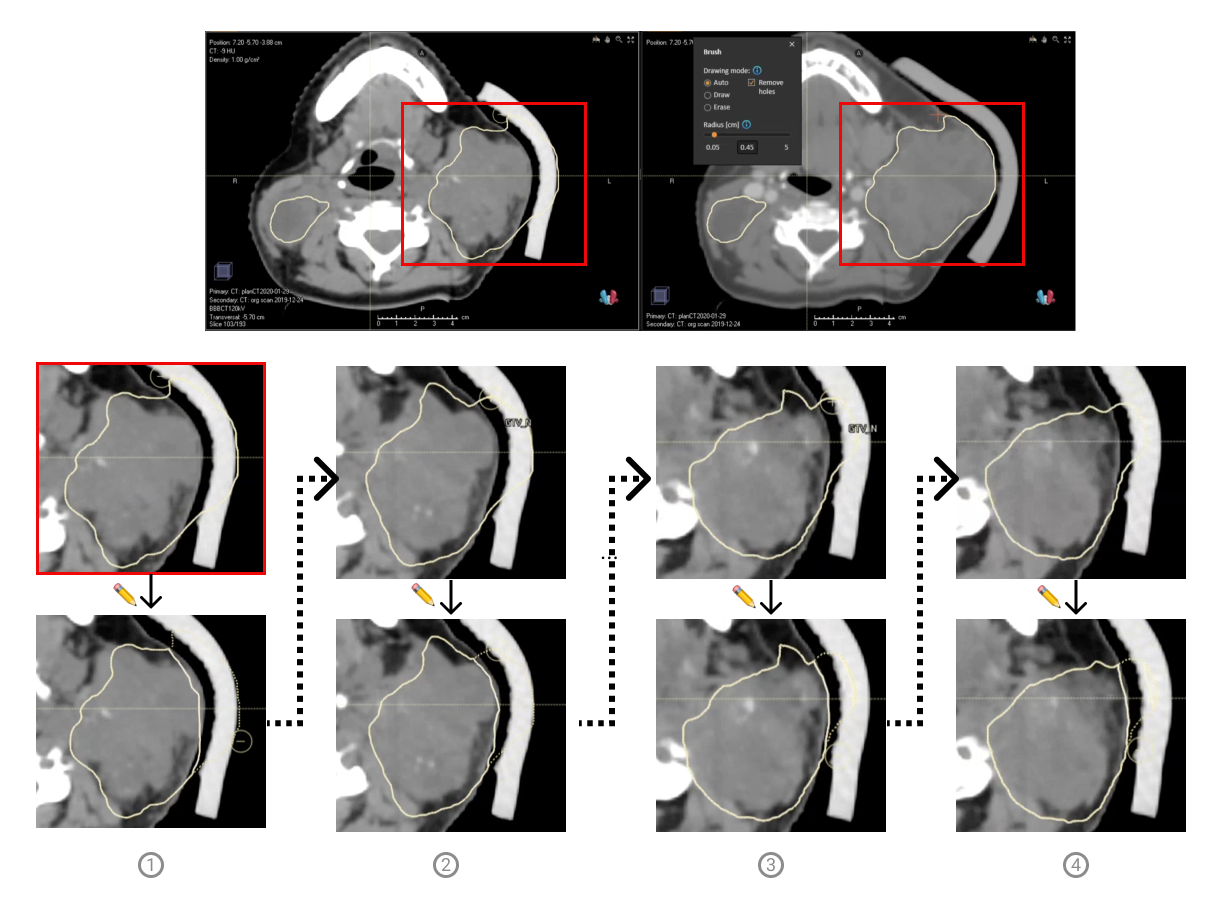}
  \caption{
  Refinement chunk from editing the delineation of a GTVn. The GTVn was propagated from the planning CT to the current one using rigid registration. After the propagation, the lateral side of the GTVn's includes air outside the patient's body. Usually, this inaccuracy can be quickly fixed by intersecting the GTVn contour with the contour of the patient's body. Nevertheless, the latter was not available in this case.
  }
  \label{fig:refinement-chunk1-gtvn}
\end{figure}

Zooming into the chunks that make up the GTVn refinement process permits corroborating the presence of redundancy. Figure \ref{fig:refinement-chunk1-gtvn} depicts a case in which a large area outside of the patient's body was mislabeled as GTVn. Although, in general, this can be quickly fixed by intersecting the GTVn contour with the body contour, in this case, the latter was not available, causing delays in the process. 

\begin{figure}[h]
  \centering
  \includegraphics[width=\linewidth]{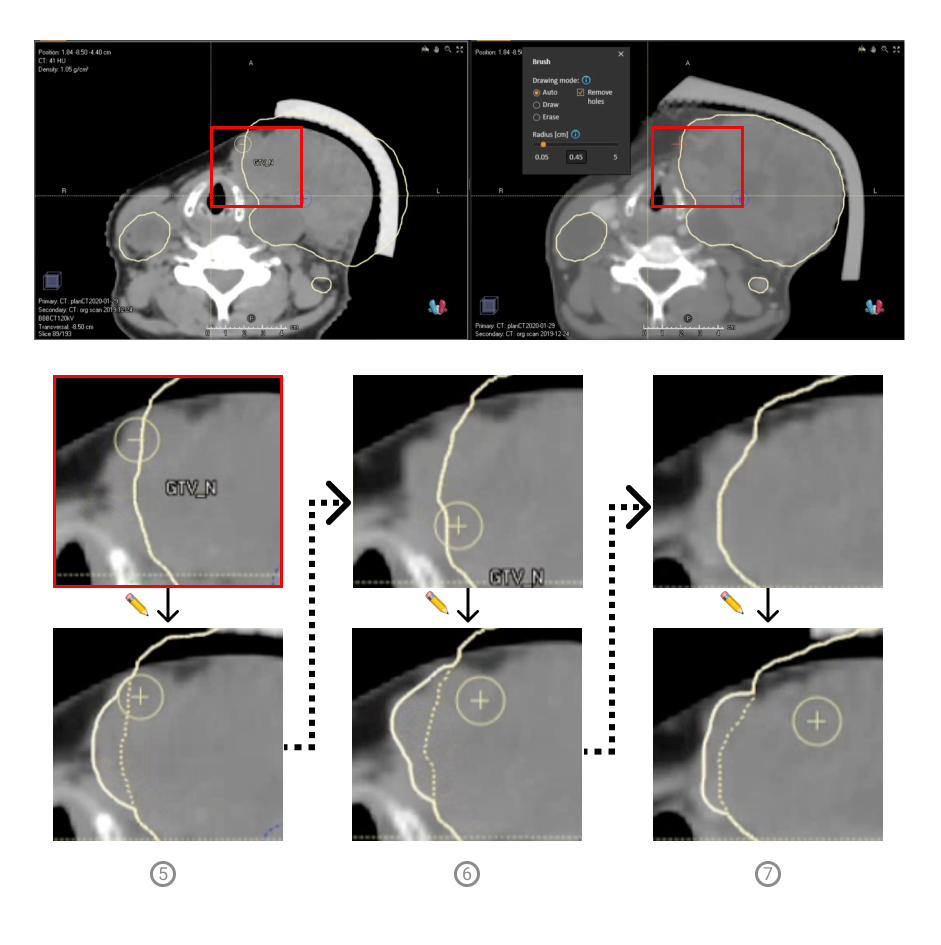}
  \caption{
  Refinement chunk from editing the delineation of a GTVn. The GTVn was propagated from the planning CT to the current one using rigid registration. After the propagation, the internal side of the delineation fails to include the whole structure. This error is challenging to fix because the precise location of the contour is uncertain. In these cases, the clinicians go back and forth the original delineations to ensure they match the propagated ones.
  }
  \label{fig:refinement-chunk2-gtvn}
\end{figure}

Figure \ref{fig:refinement-chunk2-gtvn} presents a more challenging error pattern. In this case, the rigid propagation failed to include an interior portion of the GTVn. In contrast with the case in the previous paragraph, this cannot be fixed by using existing automatic tools like auxiliary contours and algebra operations. Therefore, the user had to go over multiple slices extruding the delineation, which resulted in a refinement chunk that lasted around 30 seconds. 

Putting everything together, the performance of the contour refinement approach depends on several factors. On the one hand are objective attributes of the chunks, which depend on the choice of auto-contouring method, like the number of errors, their three-dimensional extent, and qualities like uncertainty in the image. On the other, subjective factors also seem to play a role. Based on a qualitative assessment of the automatically-generated contours, the users would often opt to do it again from scratch, an approach that, as the previous section showed, generally does not seem to yield performance gains. 

Finally, there seems to be a lack of tools tailored for contour refinement. While contouring from scratch can be significantly accelerated using tools like between slice interpolation \cite{aselmaa_influence_2017}, redundancy-aware tools like that do not exist for the contour refinement approach. Besides being able to translate or rotate a complete contour in 3D, more localized error patterns like those shown above are dealt with using the manual brush on a slice-by-slice basis.

% in discussion mention the fact that we spoke with users about other tools like scribble based contouring and they manifested an issue with those is the lack of control

\subsection{Refinement Chunks and Delineation Difficulty}
\label{sec:refinement-chunks}

% discuss what they are, their uncertainty/difficulty and clinical relevance

The previous sections used the term refinement chunk when discussing examples of interaction patterns in the contour refining approach. Formally, a refinement chunk is the unit of the contour refinement process and corresponds to a region of an automatically generated contour in which consecutive slices share features that might signal an error in the delineation. Figures \ref{fig:refinement-chunk-gtvp}, \ref{fig:refinement-chunk1-gtvn} and \ref{fig:refinement-chunk2-gtvn} are examples of refinement chunks that arise when using rigid registration to generate contours. Refinement chunks can also only involve one slice, as happens when the contours lack a slice or have an extra one in their cranial or caudal ends.

\begin{figure}[h]
  \centering
  \includegraphics[width=\linewidth]{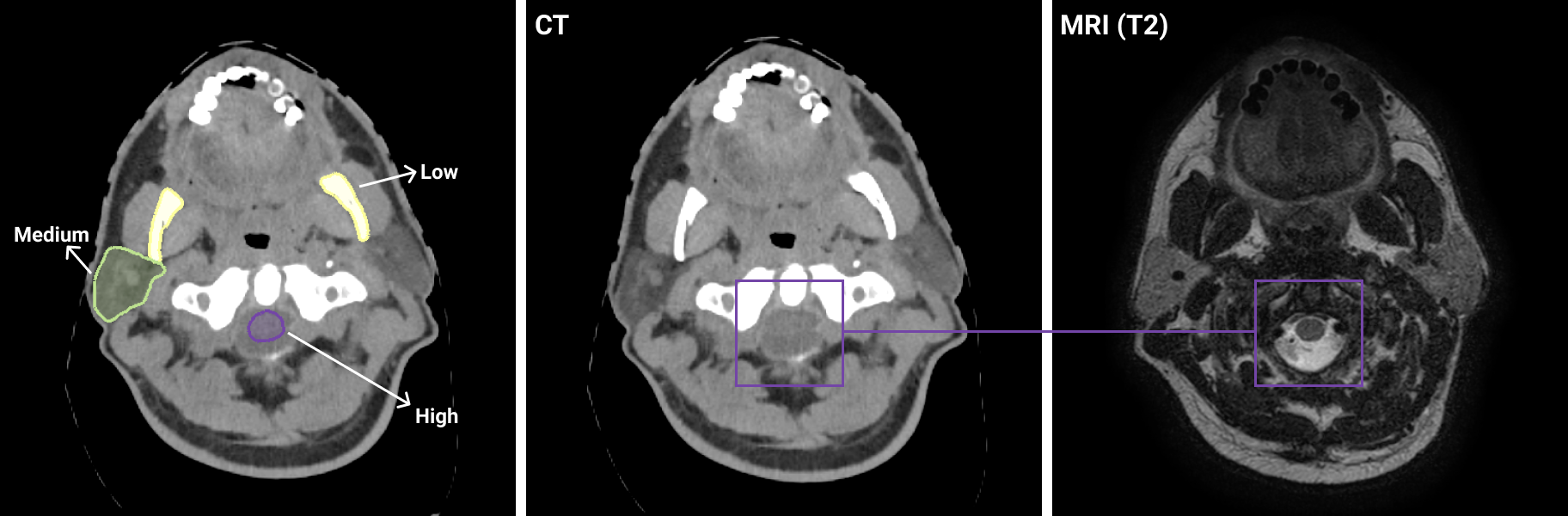}
  \caption{
  A slice of an image volume containing refinement chunks with varying uncertainty levels. The image on the left presents contours of the mandible, the right parotid gland, and the spinal cord, which have low, medium, and high uncertainty, respectively. The high degree of uncertainty when delineating the spinal cord with only a CT scan is due to the lack of contrast in the boundary (center image, purple square). An MRI can be used to supplement the CT, reducing the chunk's uncertainty (right image, purple square). Nevertheless, to profit from the additional information that the MRI provides, it is necessary to align or register it to the CT, which might introduce additional uncertainties.
  }
  \label{fig:refinement-chunk-uncertainty}
\end{figure}

In addition to the extent of the chunk, given by the number of slices and the number of voxels per slice that it spans, the time that it takes to resolve it depends on the image uncertainty. Figure \ref{fig:refinement-chunk-uncertainty} presents chunks with varying uncertainty levels. As can be observed, contours of low-uncertainty structures like the mandible correspond to a sharp intensity gradient in the image. In contrast, high-uncertainty ones like the spinal cord have no or little no correspondence to the CT's information. If there is an error in the latter, it will, similarly to the example presented in Figure \ref{fig:refinement-chunk2-gtvn}, take longer for the user to correct. 

% figure chunks with different levels of certainty (CT certain, CT uncertain + PET)

In practice, there are three ways of dealing with uncertain chunks. First, the planning image, usually a CT, can be complemented with others such as PET and MRI. For instance, by combining the CT scan on the center slice of Figure \ref{fig:refinement-chunk-uncertainty} with an MRI (right), it is possible to determine where the border of the spinal cord lies. Second, more certain adjacent slices can provide information about where the contour in the current one might lie. Third, when in doubt, clinicians are instructed to consult the guidelines, which indicate, based on landmarks, what to include in the contour. For example, the guidelines for organ at risk delineation in the HN recommend for the parotids to "include carotid artery, retromandibular vein and extracranial facial nerve" \cite{brouwer_ct-based_2015}.

All the mechanisms for dealing with uncertainty rely on the perception and interpretation of the user, which means that different clinicians might reach different conclusions as to where to draw the contours. Going back to Section \ref{sec:geometric-uncertainties} of the previous chapter, these variabilities, known as inter and intra-observer variabilities, add to other geometrical uncertainties like organ motion and have the potential to impact steps of the pipeline downstream. The goal of the processes presented in Section \ref{sec:supporting-processes} is to mitigate these diverging interpretations. 
 
%Finally, the characteristics of the chunks will vary depending on the contour-generating algorithm. Figure \ref{TODO}, which shows the contours of the medial part of the right parotid gland of a patient, exemplifies this. In this case, the gland had an inner lobe (blue circle left image) that was not detected by the DL-based and model-based contouring but was successfully propagated with deformable registration-based contouring.
% Figure: for the experiments/figures in this section: get a pCT - reCT pair that has planning contours. 

Finally, the chunk's characteristics can vary depending on the contour-generating algorithm. Algorithms that only rely on the CT information, for instance, model-based contouring, often have problems with structures that are not easy to see (like the mandible). Therefore, they will result in more and larger refinement chunks that require significant manual effort. In contrast, methods that use previous patient data, like deformable registration-based propagation, or from a database of patients, like deep learning or atlas-based auto-contouring, tend to perform better. Nevertheless, it is worth considering that even if they yield fewer refinement chunks, these will likely correspond to uncertain areas, making them harder to resolve.

\section{Process Level}

Having an in-depth understanding of the task of contouring an anatomical structure and how clinicians tackle it, this section grounds contouring in the clinical reality, describing how does it fit in HollandPTC's offline adaptive workflow (Figure \ref{fig:detailed-dose-delivery-pipeline-hptc} in Section \ref{sec:dose-delivery-pipeline} depicts what this type of workflow entails). 

\subsection{Anatomical Groups and Distribution of Labor}

The different anatomical structures involved in cancer treatment can be put into the four categories listed below. The groups are formed based on the structures' role during treatment planning. For instance, target volumes receive a therapeutic dose while the organs at risk should, if possible, receive no dose at all. The list below describes the different anatomical groups and their roles.

\begin{itemize}[noitemsep]
    \item Target volumes (TVs): areas of the patient's body that harbor malignant cells and therefore will receive a therapeutic radiation dose. This group includes the primary gross tumor volume (GTVp) and the nodal tumor volume (GTVc).
    \item Elective lymph node fields (ELFs): areas of the patient body, in the same region of the TVs, that might harbor malignant cells. These areas are larger and are defined based on landmarks. ELFs receive a prophylactic (which means lower) dose.
    \item Organs at risk (OARs): anatomical structures in the cancer region that should be spared, as much as possible, from being irradiated. Based on the type of tissue and severity of side effects, different OARs can have different priorities. Table \ref{tab:dosimetric-info-hn} exemplifies this for OARs in the head and neck. There, it can be observed how organs like the brainstem and the optic nerves are very high priority, using max dose constraints. In contrast, structures like the mandible have low priority given their higher resistance to radiation. 
    \item Planning structures (PS): additional patient-specific structures needed for creating their treatment plan. Examples of these are margins and tissue overrides to compensate for artifacts.
\end{itemize}

\begin{table}[]
\begin{tabular}{l|p{2.5cm}|l|l|p{2.5cm}}
Group & Name                       & Tissue type & Contouring difficulty & Reason difficulty                     \\ \hline
OAR   & Mandible                   & bone        & easy                  & high contrast                         \\[0.25cm]
OAR   & Parotid glands             & soft tissue & medium                & CT uncertainty                        \\[0.25cm]
OAR   & Submandibular glands       & soft tissue & medium                & CT uncertainty                        \\[0.25cm]
OAR   & Brainstem                  & soft tissue & hard                  & CT uncertainty                        \\[0.25cm]
OAR   & Spinal cord                & soft tissue & hard                  & CT uncertainty                        \\[0.25cm]
OAR   & Optic nerves               & soft tissue & medium                & CT uncertainty                        \\[0.25cm]
OAR   & Optic chiasm               & soft tissue & medium                & CT uncertainty                        \\[0.25cm]
OAR   & Eyes                       & soft tissue & easy                  & high contrast                         \\[0.25cm]
OAR   & Swallowing muscles         & soft tissue & hard                  & CT uncertainty                        \\[0.25cm]
OAR   & Oral cavity                & soft tissue & hard                  & CT uncertainty, landmark based        \\[0.25cm]
TV    & Primary GTV (GTVp)         & various     & medium                & CT uncertainty, heterogeneous         \\[0.25cm]
TV    & Nodal GTV (GTVn)           & soft tissue & medium                & CT uncertainty, extent, heterogeneous \\[0.25cm]
ELN   & Elective Lymph Node Levels & various     & hard                  & extent, landmark based                \\[0.25cm]
PS    & Implants                   & metal       & medium                & heterogeneous                         \\[0.25cm]
PS    & Artifacts                  & various     & medium                & heterogeneous                         \\[0.25cm]
PS    & Planning helpers           & various     & easy                  & heterogeneous                        
\end{tabular}
\caption{Contouring difficulties and factors that cause it for a subset of the HN structures considered in PT \cite{jensen_danish_2020}. The difficulty estimates in this table were derived after a brainstorming session with several clinicians (RTTs and ROs). This table complements Table \ref{tab:dosimetric-info-hn} in the previous chapter, which presents the priorities and dosimetric considerations for this subset of structures.}
\label{tab:contouring-info-hn}
\end{table}

The anatomical groups described above are used to distribute the contouring task among the radiation oncologist (RO) and the radiotherapy technologist (RTT), which are the two clinical actors that participate in the delineation process. In general, RO's are in charge of contouring high priority structures that are hard to see (like the swallowing muscles) and heterogeneous across the population (like GTVp). RTTs offload manual tasks from the ROs, generating starting contours of easier but time-consuming structures that the latter can then review. Interestingly, the introduction of auto-contouring methods creates a similar dynamic between the RTTs and these tools. The list below summarizes the capabilities of each actor and their role in the contouring workflow. Bayesian deep neural networks were used as an example of the auto-contouring capabilities, given that they constitute the state of the art \cite{mcclure_knowing_2019}. Section \ref{sec:fast-contouring-data-generation} provides detailed explanation of the capabilities and limitations of this technology.

% turn to table
\begin{itemize}[noitemsep]
    \item Radiation oncologist (RO)
    \begin{itemize}[noitemsep]
        \item Capabilities: knowledge of anatomy, cancer biology, and other aspects of the medical profession. In-the-job experience gained by a residency program where a senior RO mentors junior ones.
        \item Role: approving delineations of RTTs (OARs) and other ROs (TVs and ELFs). Making uncertain and high-risk clinical decisions (a clinical decision in this context means, for instance, delineating of TVs, ELFs, and critical OARs from scratch).
    \end{itemize}
    \item Radiotherapy technologist (RTT)
    \begin{itemize}[noitemsep]
        \item Capabilities: knowledge of the software and hardware used in different steps of the dose delivery pipeline. RTTs also have a basic knowledge of anatomy that improves with experience. Some RTTs can have dosimetric experience if they have been trained in treatment plan creation.
        \item Role: registering different information sources (for instance CT and MR images) and delineating the OARs. Depending on the RTT's experience level, they might choose to leave to the RO harder structures such as the brain stem and the swallowing muscles. 
    \end{itemize}
    \item DL-based auto-contouring
    \begin{itemize}[noitemsep]
        \item Capabilities: trained with a large number of curated contours of previous patients of the treatment center endows them with the capability to produce reasonable delineations of unseen images. Bayesian or probabilistic AIs can also point to parts of the delineations that are uncertain.
        \item Role: during the initial treatment plan creation, producing reasonable starting points of the OARs for which it was trained. During an adaptation, produce reasonable contours for all the delineations in the planning scan.
    \end{itemize}
\end{itemize}

\subsection{Contouring Contexts}

\begin{figure}[h]
  \centering
  \includegraphics[width=\linewidth]{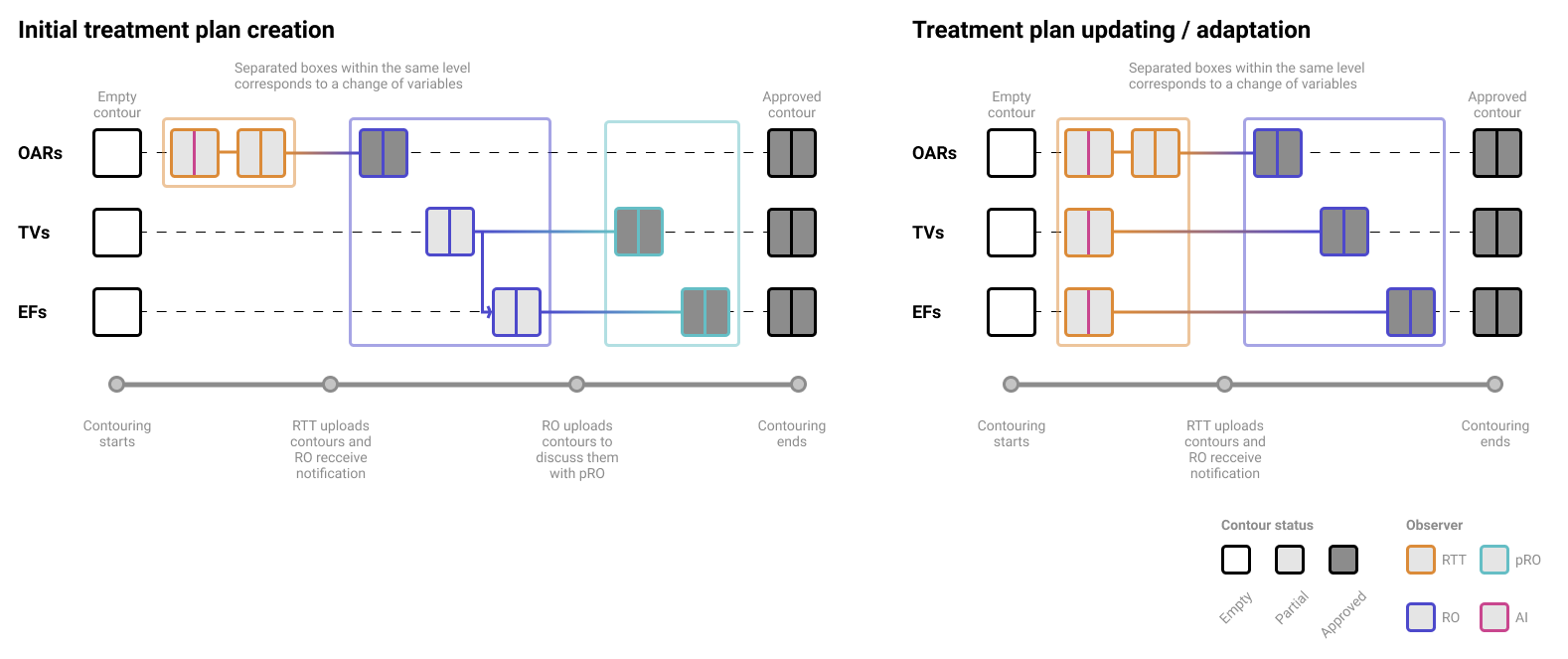}
  \caption{
  In the current workflows of HollandPTC and LUMC, contouring can happen in two contexts, which influence the process' execution flow. The first scenario, on the left, is contouring when creating the initial treatment plan. In this case, the goal is to produce as accurate as possible contours within a time frame of several days. Before a delineation is marked as approved (dark gray), at least two subjects, the last of which must be a radiation oncologist (RO), must have reviewed the partial delineations. Peer-reviewing can help reduce errors and slips caused by, for example, the difficulty of the task, the lack of experience, or tiredness. The second scenario, on the right, is contouring during an adaptation session. At this time, a treatment plan already exists and the goal is to update it for the current anatomy. As for contouring, the goal is to replicate or propagate the clinical decisions (contours) made when creating the initial treatment plan within a significantly shorter time frame (within a day or even within a fraction). The time-constrained nature of this scenario and the relatively simple nature of this task compared to the first scenario, permit relaxing the need for peer review.
  }
  \label{fig:overview-contouring-scenarios}
\end{figure}

There are two scenarios in the PT workflow that trigger the contouring process. The list below summarizes their differences. The first scenario is the initial patient treatment plan creation, which occurs in the days that follow the patient's arrival at the treatment center. In this context, the priority is to ensure that the contours are as accurate as possible. One strategy for this is to add additional images such as MR and PET, which can help with uncertain clinical decisions such as delineating the primary GTV. A second way to increase accuracy and consistency is by using peer-review. As the left diagram of Figure \ref{fig:overview-contouring-scenarios} shows, the different anatomical groups, which start being empty, need to be reviewed by at least two people, the last of which must be a RO. At the cost of longer execution times, peer-reviewing reduces the likelihood that errors and slips of RT team members propagate downstream, affecting the patient's treatment plan. In practice, contouring at planning can take several hours split over multiple days.% Appendix \ref{TODO} presents the workflow for all the anatomical structures in detail. 

% figure: two contouring contexts

The second scenario that requires contouring is when the initial treatment plan needs to be adapted to consider anatomical changes, which can happen several times during the treatment. In the current PT workflow, this change of context does not significantly alter the contouring process. Nevertheless, some differences should be highlighted. First, there is less available time for delineating given that the patient is undergoing treatment, and any delays can negatively affect tumor control and organ sparing. Second, when performing a plan adaptation, the original contours of the patient are available, which changes the contouring task's goal from creating the delineations to propagating them as accurately as possible. Finally, as can be observed in the diagram on the right part of Figure \ref{fig:overview-contouring-scenarios}, the peer-review requirements are lowered (there is no peer-RO anymore) because of the assumption that few or no new clinical decisions will need to be done.

Main differences between the two contouring scenarios: % convert to table
\begin{itemize}[noitemsep]
    \item Goal.
    \begin{itemize}[noitemsep]
        \item Planning: create contours, which involves, in some cases, making decisions under uncertainty.
        \item Adaptation: propagate decisions that were already made at planning. New decisions might need to be made if there are significant discrepancies between images.
    \end{itemize}
    \item Available time. 
    \begin{itemize}[noitemsep]
        \item Planning: several days because the patient has not started treatment yet.
        \item Adaptation: one or two days for the offline case and the duration of a fraction (usually around 20 min) for the online one. 
    \end{itemize}
    \item Available information. 
    \begin{itemize}[noitemsep]
        \item Planning: no pre-existing contours of the patient but other, registered, modalities such as MR, PET, and CT with and without contrast.
        \item Adaptation: contours from the planning phase. Other modalities from planning could be used, but usually, time does not allow for registering all of them. 
    \end{itemize}
    \item Distribution of labor.
    \begin{itemize}[noitemsep]
        \item Planning: every structure needs two checks, the second of which must be performed by a RO. 
        \item Adaptation: OARs are usually not checked by the RO, which is in charge of making sure that the TVs were correctly propagated and of updating the margins. No peer-reviewing of the TVs is needed.  
    \end{itemize}
\end{itemize}

\subsection{Relationships with Neighboring Processes}

In HollandPTC's dose delivery pipeline, which Figure \ref{fig:detailed-dose-delivery-pipeline-hptc} in the last chapter introduced, it is possible to observe the dependencies between the different processes (light gray arrows). These can be categorized into input and output relationships and vary depending on the context (for instance, treatment plan creation versus updating versus fraction delivery). The list below collects the relationships between processes in which contouring participates and highlights the implications of the relationship. For a complete description of the dose delivery pipeline and its components, refer to Section \ref{sec:dose-delivery-pipeline}. 

% process
% execution context
% relationship type: input/output
% time 
% subject

\begin{itemize}[noitemsep, nolistsep]
    \item Simulation (input):
    \begin{itemize}[noitemsep, nolistsep]
        \item Treatment plan creation: provides the planning CT to be contoured. Depending on the institution's resources, additional images such as MRIs and PET-CTs could be acquired too. The goal of providing multiple images is to reduce the difficulty of highly uncertain refinement chunks (see Section \ref{sec:refinement-chunks}). 
        \item Treatment plan update: only a CT scan is acquired due to resource limitations.
        \item Effect of the relationship with contouring: problems in the image such as noise, the partial volume effect, low resolution, or lack of information can increase the uncertainty associated with certain structures which will result in increased inter and intra-observer variability (see Section \ref{sec:geometric-uncertainties}). 
    \end{itemize}
    \item Registration (input):
    \begin{itemize}[noitemsep, nolistsep]
        \item Treatment plan creation: enables visual comparison of the planning image with the additional ones during contouring. To be able to compare, for instance, a CT and an MR side-by-side, it is necessary to map one image to the other so that they share the coordinate space. The registration or transformation can be rigid or deformable. The latter often yields a higher quality match but is harder to perform and to verify. 
        \item Treatment plan update: given that there are no additional images, the repeat CT is registered to the planning CT, which, in turn, enables comparison with additional planning images. Nevertheless, having two registrations introduces new sources of uncertainty. 
        \item Effect of the relationship with contouring: inaccurate registrations can lead to systematic errors in the contours if structures are not properly aligned.
    \end{itemize}
    \item Plan optimization (output): 
    \begin{itemize}[noitemsep, nolistsep]
        \item Treatment plan creation: uses the contours together with the dosimetric goals and objectives of the structures that they enclose to find machine parameters that can deliver the planned dose, optimally. 
        \item Treatment plan update: does not change significantly with respect to treatment plan creation. Some replanning strategies might use previous plans and contours as a starting point.
        \item Effect of the relationship with contouring: Changes in the contours' shapes can yield different plans or make it harder to find one that fulfills the criteria. Using a previous plan and contours might introduce additional uncertainties.
    \end{itemize}
    \item Plan evaluation (output):
    \begin{itemize}[noitemsep, nolistsep]
        \item Treatment plan creation: with the treatment plan and the contours it is possible to build the DVH, TCP, and NTCP (see Section \ref{sec:assessing-plan-quality}).  
        \item Treatment plan update: does not change significantly with respect to treatment plan creation.
        \item Effect of the relationship with contouring: changes in contours can alter the DVH, TCP, and NTCP which can, in turn, result in different outcomes of the decisions whether to accept the plan or not. At treatment plan updating, metrics of the update are compared against those of the approved treatment plan. This comparison introduces new sources of uncertainty in the evaluation process.
    \end{itemize}
\end{itemize}

\subsection{Supporting Processes}
\label{sec:supporting-processes}

In addition to the processes of the dose delivery pipeline described before, there are also supporting ones that aim at systematically reducing contouring errors and inter/intra-observer variabilities. Together with the strategies presented in Section \ref{sec:other-ways-reduc-uncert} in the previous chapter, these complement the APT framework helping to reduce uncertainty. The list below describes these processes and their goals. 

\begin{itemize}[noitemsep, nolistsep] % convert to table
    \item Commissioning: assesses the safety and performance of new technologies and processes before deploying them in the clinical workflow. An example of the role of commissioning is the current debate as to whether deep learning-based auto-contouring is ready for clinical practice. Several studies have pointed that using these technologies can lead to a reduction in contouring time while maintaining the quality of the delineations \cite{van_dijk_improving_2020, brouwer_assessment_2020}. Nevertheless, the rate of adoption of DLBS will vary between institutions based on the results from their commissioning processes.
    \item Quality assurance: assesses the safety and performance of the components of the clinical workflow periodically to ensure consistent outcomes. For instance, the auto-contouring algorithm might start to produce erroneous delineations after a change of the imaging protocol. The QA process should be able to detect and fix this issue to prevent introducing systematic errors in the dose delivery pipeline.
    \item Training: it can be with respect to the tools (for instance contouring or treatment planning software) or with institution-specific process' characteristics (for instance how to use the guidelines for contouring of HN cancer \cite{brouwer_ct-based_2015}). The goal is to establish a baseline in the contouring team to ensure consistent delineations that adhere to the deadlines and are produced in a short amount of time.
\end{itemize}

\section{Performance Factors}

Based on the results presented so far, this section compiles all the factors that affect the performance of the current offline-APT contouring workflows. These represent obstacles for implementing contouring in the gantry or offline but daily because of the heavy use of resources. The next chapter discusses several ways in which these could be addressed, yielding a fully adaptive PT workflow. 

\subsection{HCI-related}

\begin{itemize}[noitemsep]
    \item \textbf{Selection of contouring approach}. With reasonable starting contours, refining them is usually faster. On the other hand, contouring from scratch can be faster when the refinement chunks' extent is close to that of the organs. (Section \ref{sec:contouring-from-scratch-vs-refining})
    \item \textbf{Editing tool selection and availability}. Related to the previous point, tools tailored for a contouring scenario can significantly speed up the process. For instance, when contouring from scratch, between slice interpolation can cut the delineation time by more than half, making it the preferred tool. In contrast, there are no tools for the contour refinement approach, which forces the user to default to the manual brush, which is slow. (Sections \ref{sec:contouring-from-scratch-vs-refining} and \ref{sec:contour-refinement-analysis})
    \item \textbf{Extent of the organ or the refinement chunk}. Regardless of the contouring approach, the time that it takes to do so depends on 1) how many slices need to be inspected and 2) how much brushing will need to be done per slice. (Sections \ref{sec:contouring-from-scratch-vs-refining} and \ref{sec:contour-refinement-analysis})
    \item \textbf{Selection of algorithm to generate starting point}. There are multiple options to generate contours such as atlas-, model- and deep learning-based, or propagating pre-existing contours using rigid or deformable registration. These methods can produce refinement chunks with varying characteristics, which, in turn, can change how long they take to address. (Section \ref{sec:refinement-chunks})
    \item \textbf{Refinement chunk's uncertainty}. In addition to the extent, the time that it takes to address a chunk can increase if the user dwells on it due to uncertainty in the boundary's position. When there are multiple slices involved, it seemed like a decision made in one of them was then propagated to the rest, which suggests an overhead that does not scale with the number of uncertain slices in the chunk. (Section \ref{sec:refinement-chunks})
\end{itemize}

\subsection{Process-related}

\begin{itemize}[noitemsep]
    \item \textbf{Poor subject-task fit}. RTTs are tasked with contouring all OARs even if some of them will need to be skipped due to their difficulty or redrawn by the ROs. Also, is not clear whether certain tasks that are the exclusive domain of the RO such as tumor delineating could be within reach of RTTs with the right tools (like for example using the PET-CT).
    \item \textbf{Number of structures or refinement chunks to address}. One refinement chunk can take between a couple of seconds to a minute to address. If there are a couple of them, which is likely in areas with many contours like the HN, then the delineation refining process becomes a bottleneck.
    \item \textbf{Need of peer-review}. Peer-reviewing of contours is required during the creation of the initial treatment plan, which adds multiple hoops to the process, increasing the time it takes. This requirement is relaxed during adaptation where the goal is not to create the contours but to propagate them as faithfully as possible. 
    \item \textbf{Selection of auto-contouring method}. At adaptation, auto-contouring can either be registration-based or not.  Registration-based techniques also output a correspondence map between the images which makes the side-by-side comparison more straightforward. In contrast, with methods that do not return this mapping, it is necessary to register the images after the contouring which can be time-consuming and introduce new uncertainties. 
    \item \textbf{Information availability}. When creating the initial treatment plan CT scans are accompanied by MR and PET-CT, which reduce the delineation uncertainty. At adaptation, only a CT scan is available. If the planning images want to be used they need to be registered, which is a time-consuming process. 
\end{itemize}

%\subsection{User Requirements}

%% file: content/towards-apt-ready-contouring-wf.tex
\chapter{Towards Fast AI-Infused Contouring for Adaptive PT}
\label{ch:towards-apt}

The previous chapters presented the concept of adaptation in proton therapy, highlighting its importance for realizing the organ-sparing benefits of protons through margin reduction. They showed how several factors, such as organ motion and delineation variability, can lead to uncertainties in the dose delivery pipeline, resulting in deviations of the planned dose from the delivered one. To mitigate the uncertainties that arise due to geometric changes, it is necessary to update the patient's treatment plan as these are detected. Nevertheless, in the current APT workflow, this would significantly increase the demand for resources because several steps of the treatment creation process will need to be constantly re-executed. 

\begin{figure}[h]
  \centering
  \includegraphics[width=\linewidth]{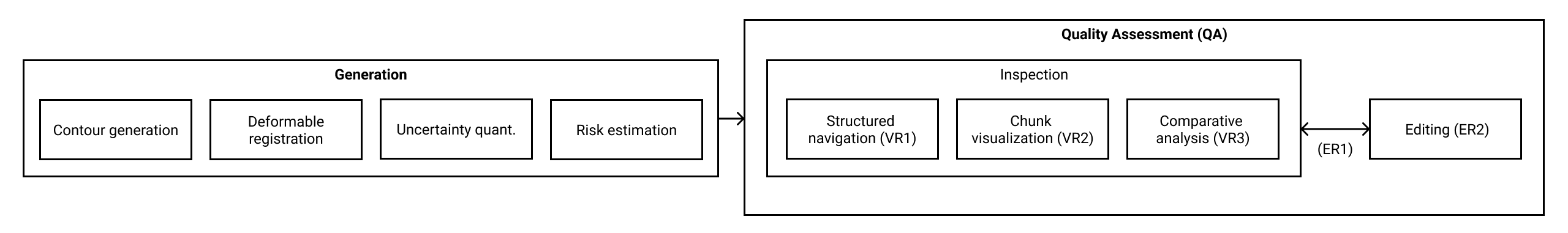}
  \caption{Components of fast contouring workflow (FCW). The FCW can be divided into an automatic data generation (Section \ref{sec:fast-contouring-data-generation} and a user-centric quality assessment (QA) Section \ref{sec:fast-contouring-user-qa} one. The main output of the former is the contours of the anatomical structures. Nevertheless, other pieces of information can be computed to support the QA process. Given state-of-the-art deep learning methods and clinical data sources, these are the uncertainty, the risk, and the deformable registration between the reference and daily images. In APT, QA's time budget is limited. Therefore, there is a need for visualization and interaction methods that: narrow search areas, reduce the time that the user spends analyzing the data, and reduce the number and length of interactions. These principles can be further separated into inspection (VR1-3) and edition-specific (ER1-2) ones.}
  \label{fig:components-fast-contouring-workflow}
\end{figure}

The goal of this chapter is to understand and characterize the role that the contouring process will play in future APT workflows, which account for daily anatomical variations. Concretely, it provides a detailed walkthrough of a concept of a fast contouring process tailored for time-critical scenarios. The concept description is structured in its three key components, which Figure \ref{fig:components-fast-contouring-workflow} depicts: fast generation, targeted inspection, and redundancy-aware editing. Although the advent of deep learning-based auto-contouring tools, which can in most cases achieve expert-level performance, has significantly sped up the first of these, contouring remains a bottleneck due to the time that it takes to inspect and edit faulty delineations, a process often referred to as quality assessment (QA). 

% Figure (three components of contouring workflows)

\begin{figure}[h]
  \centering
  \includegraphics[width=8cm]{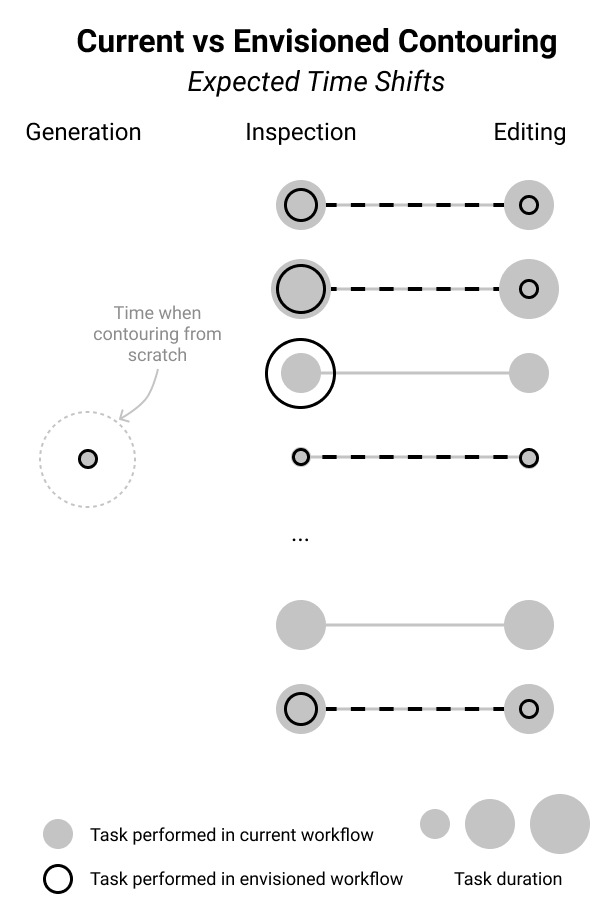}
  \caption{
  Expected shifts in tasks duration with the proposed concept. As can be observed, the contour creation step is significantly faster compared to contouring from scratch due to the usage of AI-based auto-contouring technology. In the regular workflow, after the contours have been generated, the clinician has to go through all the refinement chunks (inaccurate areas) and edit them. The time is shared between assessment and editing given that these tasks are tightly coupled. Also, the time of assessment and editing in the current approach depends on the difficulty and extent of the chunk. Moving to the proposed concept, the first thing to notice, is that the notions of uncertainty, risk and process-specific requirements will be leveraged to avoid needing to check all the chunks (Section \ref{sec:targeted-inspection}). Second, for clinically significant inaccuracies, the redundancy aware tool have the potential to significantly reduce the editing time by minimizing the required interactions \ref{sec:redundancy-aware-editing}). 
  }
  \label{fig:current-vs-envisioned-contour-creation-detection-editing}
\end{figure}

The proposed workflows aim at mitigating the QA bottleneck by relying on modern AI capabilities and clinical insights extracted from the literature review and the interviews and observation sessions of experienced clinicians. Figure \ref{fig:current-vs-envisioned-contour-creation-detection-editing} presents core components of fast contouring, which the following sections elaborate on in more detail. Before delving into these, Section \ref{sec:future-apt-workflows} presents a clinically-plausible dose delivery pipeline for online APT. Grounding the discussion in this pipeline validates this report's proposals, strengthening the case for implementing them in the clinic. Finally, Section \ref{sec:fast-contouring-wf-aspects} discusses workflow-level aspects of fast contouring like distribution of labor and constraints.

% Figure (three components of contouring workflows - time displacement)

\section{Future APT Workflows}
\label{sec:future-apt-workflows}

\begin{figure}[h!]
  \centering
  \includegraphics[width=\linewidth]{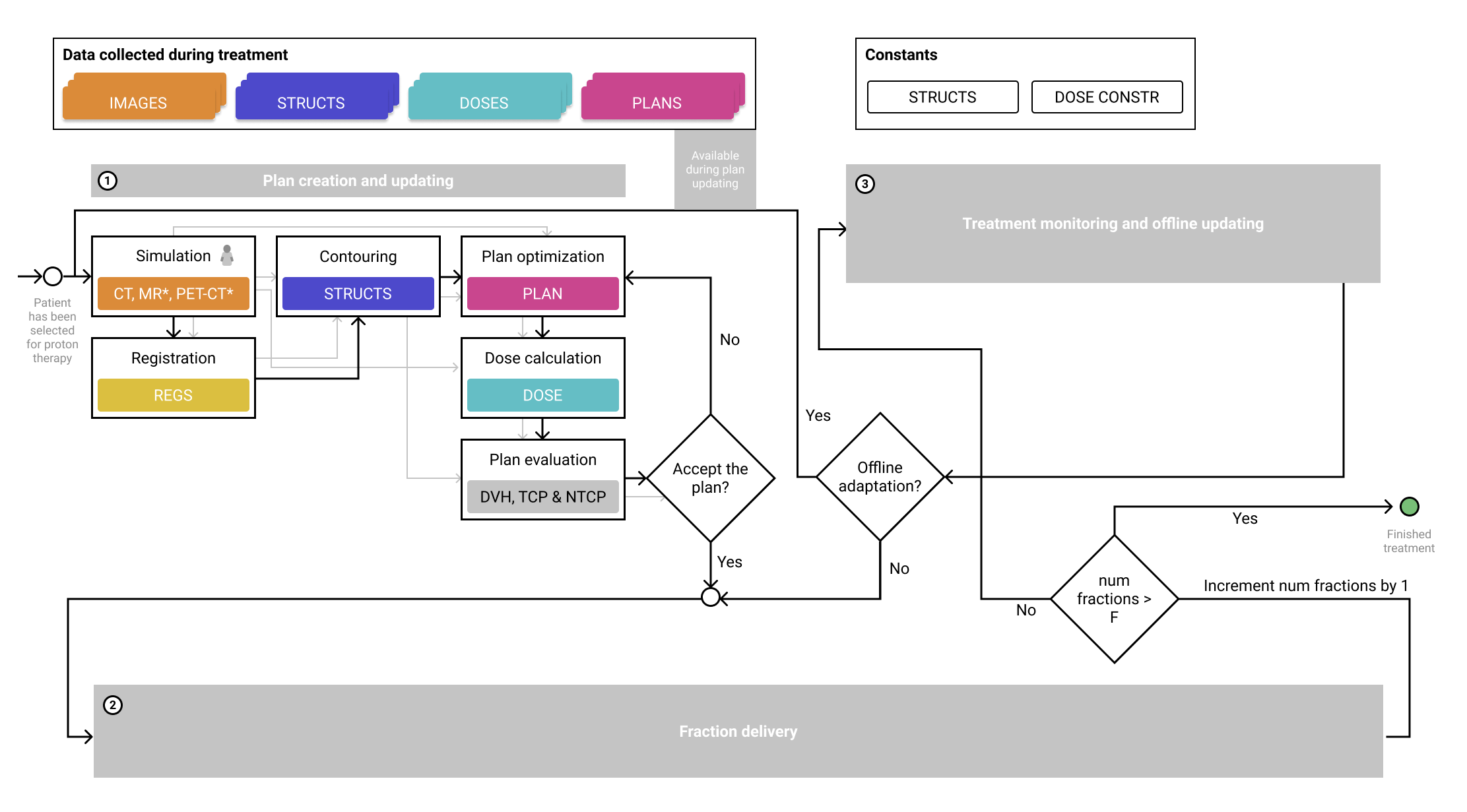}
  \caption{
  Template for defining APT-suitable dose delivery pipelines. Solid gray boxes denote general phases, boxes with solid black lines represent processes and diamonds decisions. Processes in the plan creation and updating phase change depending on whether it is the first time they are executed (treatment plan creation) or not (treatment plan updating). For the processes in the plan creation and updating phase, the diagram presents the relationships between inputs and outputs (light gray arrows). In general, the dose delivery pipeline is made of the plan creation and updating (1), fraction delivery (2), and treatment monitoring, and offline updating (3) phases. While the first one largely remains constant across institutions, the last two change depending on the adaptation paradigm. Figure \ref{fig:comparison-daily-workflows} presents examples of different implementations of the fraction delivery and treatment monitoring and offline updating stages.
  }
  \label{fig:online-adaptive-template}
\end{figure}

Figure \ref{fig:online-adaptive-template} presents a generic template of a dose delivery pipeline for adaptive radiotherapy with three main components. First is the treatment creation process, which runs when the patient arrives and, like is the case at HollandPTC and LUMC, every time their plan needs updating. Second, after creating the patient's initial treatment plan, the fractionated dose delivery process starts. The latter includes the activities that occur between the patient arriving at the treatment center and the irradiation of the tumor. The third component, which presence depends on the adaptation scheme, corresponds to an offline treatment monitoring and updating process that ensures clinically-acceptable treatment quality without using gantry time. In general, the last two processes occur multiple times, depending on the fractionation protocol. For instance, in HN PT the patient receives 70 Gy of radiation in 35 daily fractions and their treatment is monitored every week.

\begin{figure}[p!]
  \centering
  \includegraphics[width=\linewidth]{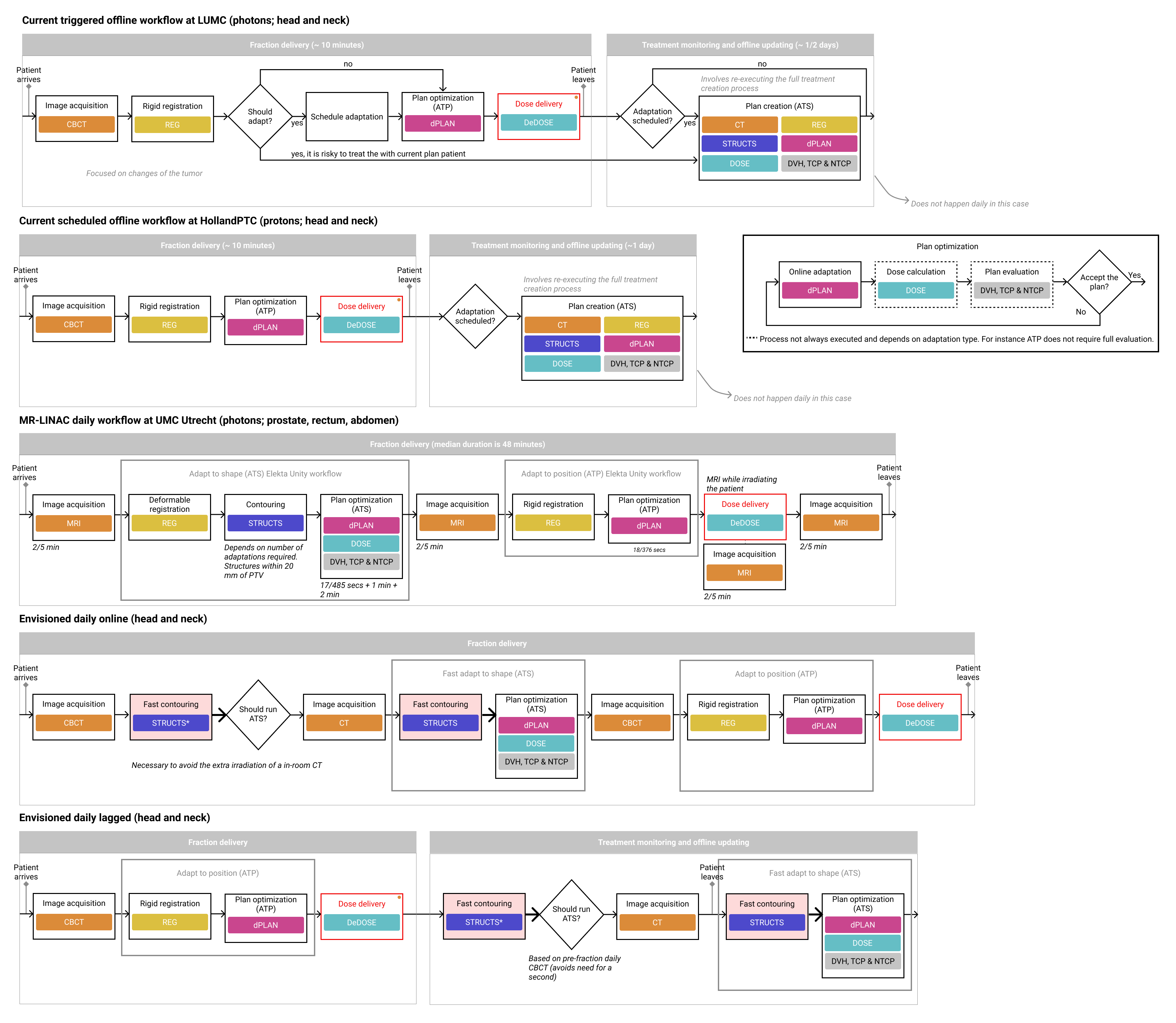}
  \caption{ Side-by-side comparison of different dose delivery pipelines. The first three rows show the clinical implementations at LUMC, HollandPTC, and UMC Utrecht, respectively. These include estimates of the times the main stages take for comparison. Note how LUMC's and HollandPTC's workflows are offline adaptive, but the former has a protocol for triggering adaptations while the latter does it weekly. In contrast, UMC Utrecht's MRI-LINAC workflow can perform daily online adaptation. The tradeoff is that the latter has a significantly longer fraction time. The last two rows present two alternatives of daily online adaptive workflows that could be implemented at HollandPTC, given its current capabilities. The Envisioned Daily workflow (EDW) is similar to UMC Utrecht's, allowing adaptation within the fraction. Given that the patient is laying on the treatment couch the whole time, in EDW, the ATS block, which is in charge of online geometry-related adaptations, is time-critical. The Envisioned Daily-Lagged Workflow (EDLW) removes the time pressure from the ATS block of EDW. This is done by performing ATS after the dose delivery and before the patient leaves the treatment center. The one-day adaptation lag is problematic for dynamic areas like the prostate and bladder. Nevertheless, HN is a more static region, with changes occurring daily. Given that HollandPTC uses CT-based imaging, in both EDW and EDLW, care must be taken with the number and type of the images taken to avoid giving more radiation to the patient than is safe. For this reason, both workflows include CBCT-based checks for deciding whether to trigger or not the CT-based ATS process. 
  }
  \label{fig:comparison-daily-workflows}
\end{figure}

As mentioned before, HollandPTC’s weekly-scheduled offline adaptation approach might not be able to account for all uncertainties arising from geometrical variations. Therefore, to leverage the tissue-sparing capabilities, the goal is to move towards an online daily adaptive paradigm. The following paragraphs describe three of the four versions of the dose delivery template that Figure \ref{fig:comparison-daily-workflows} summarizes. Namely, they start first presenting HollandPTC's current scheduled offline-adaptive protocol. Then, they introduce an example of a successful implementation of an online adaptive dose delivery pipeline at University Medical Center (UMC) Utrecht. Finally, this section describes two online-adaptive dose delivery pipelines for HN that are tailored to institutions with capabilities similar to those of HollandPTC's. Fast contouring, which the following sections describe, is a core building block of these workflows.

The second row of Figure \ref{fig:comparison-daily-workflows} recapitulates the current workflow at HollandPTC, which Section \ref{sec:dose-delivery-pipeline} described. As can be observed, during the delivery of the dose fraction, it only considers uncertainties related to deviations of the patient's setup. Because of the reduced number of processes to execute (relative to the other versions), a fraction at HollandPTC is relatively fast, lasting on average ten minutes. As for the geometry-related adaptations, the workflow addresses them weekly, repeating the plan creation process with some modifications. From the logistical perspective, scheduling adaptation sessions can be beneficial, allowing for efficient resource allocation. Nevertheless, from the point of view of APT, this workflow has a weakness: it addresses geometrical uncertainties periodically and not responsively. Given that a clinically-significant shape change could occur in between two adaptation sessions (for instance, filling of the bladder or the nasal cavities), it is then necessary to robustly optimize the treatment plan \cite{cubillos-mesias_including_2019}, which means more margins and, consequently, increased damage to the OARs.

Online adaptation is already possible in clinical practice. An example of this is the photon-based MR-linac used at University Medical Center Utrecht for treating prostate and rectal cancers \cite{winkel_adaptive_2019}. The MR-linac enables acquiring MR images and delivering the dose without switching rooms or changing the patient's position. Furthermore, because it uses radiation-free MR, this setup allows taking multiple images during the fraction, ensuring precise delivery of the planned dose. As can be seen in the third row of Figure \ref{fig:comparison-daily-workflows}, UMC Utrecht's fraction delivery workflow has two coarse steps before dose delivery. The second one is common to the previously described HollandPTC workflow and consists of correcting the patient positioning with simple transformations (translations and rotations), which they refer to as Adapt to Position (ATP). The novelty of this workflow is the first step, which they call Adapt to Shape (ATS). ATS detects anatomical changes using updated contours (propagated via deformable registration) and re-optimizes the treatment plan to match the clinical objectives set during the treatment creation process \cite{intven_online_2021, werensteijn-honingh_feasibility_2019}.

Compared to HollandPTC's current fraction delivery process, UMC Utrecht one takes significantly more time, with a median fraction duration of 48 minutes. The time increase can be traced back to two factors. First, several MRIs are acquired, each taking between 2 and 5 min. At HollandPTC, a single CBCT is acquired per fraction. Second, the ATS workflow requires re-contouring several of the patient's anatomical structures. Although in practice, clinicians speed up this process by using deformable registration to propagate the contours and then only considering those within a certain radius from the PTV for QA, it can still be a bottleneck if the patient requires several or significant adaptations. It is because of these time delays introduced by ATS that, at UMC Utrecht, it happens before ATP \cite{intven_online_2021}. Using this execution order ensures that deviations in the patient positioning that might have occurred during ATS do not result in a deviation of the dose distribution. The latter is more likely to be a factor in areas like the abdomen and pelvis where, for instance, the bladder can fill in-between ATS and dose delivery.

The clinical success that the MR-linac-based online adaptive workflow has shown suggests that it could be implemented at HollandPTC to endow its dose delivery pipeline with online adaptive capabilities. Nevertheless, currently, it is not possible to use an MR in proton therapy, with PT-suitable MR-linacs still five to ten years off from arriving in the clinic \cite{hoffmann_mr-guided_2020}. Therefore, future online APT workflows at HollandPTC, and also other cancer treatment centers that for several reasons cannot access an MR-linac, will need to rely on CBCT and CT, the predominant imaging technologies. Compared to MR-based workflows, a significant challenge of CT-based ones is the extra radiation that these imaging technologies deliver to the patient. Because of this, current practice consists of acquiring daily CBCTs, which have a lower dose, and only getting a new CT scan when adaptations are required. Future APT workflows will need to consider this compromise, using as much information as possible from the CBCTs and minimizing the use of the more radiation-intensive CT scans.
% is the extra radiation the problem or also logistical aspects?

Given the previous discussion, the last rows of Figure \ref{fig:comparison-daily-workflows} propose a couple of possible APT workflows that would be feasible to implement in treatment centers with similar capabilities to those of HollandPTC. As can be observed, similarly to the MR-linac, they employ ATS and ATP building blocks. Nevertheless, the imaging constraint described before demands changes in the execution order. The first workflow, Envisioned Daily (EDW), mirrors UMC Utrecht's, having the ATS block first and then the ATP one. In this case, the CT scan that ATS requires will only be taken if clinically significant geometrical changes are detected in the initial CBCT. After plan adjustments are done, a second CBCT will be taken to correct the patient's position before delivering the dose. It is worth noting that, with current HollandPTC's capabilities, this workflow would entail moving the patient between rooms, which is impractical, introduces additional uncertainties, and demands significantly more resources. Nevertheless, in the short term, HollandPTC plans to install a CT scanner in the gantry, which would make the setup similar to the one at UMC Utrecht’s albeit with different imaging techniques. Given its similarities with the MR-linac workflow, this one can deal with changes that occur in-between fractions. Nevertheless, it is expected to take more time to complete and will, at the minimum, impart double the current amount of CBCT-related radiation. Furthermore, the gantry operator must be familiar with contour delineation of OARs and tumor-related structures, which could put an additional burden on the treatment center's resource pool.

Although areas like the prostate, the rectum, and the abdomen can vary significantly between and even within fractions, this is usually not the case for the head and neck area, which has more stable anatomy \cite{morgan_adaptive_2020}. The second workflow, Envisioned Daily-Lagged (EDLW), in the last row of Figure \ref{fig:comparison-daily-workflows} builds on this assumption. Instead of performing ATS before the dose delivery process, it does it afterward, allowing to perform most of the ATS process without using gantry time. Although this workflow does not address anatomical changes that might have occurred between two consecutive fractions, it does consider the type of systematic changes that build up over several days. This characteristic makes it potentially more useful for HN than more dynamic areas like the prostate. Compared to the first envisioned workflow, this approach has three benefits. First, it does not require an extra CBCT for ATP, reducing irradiation. Second, the time in the gantry will be less, with ATS happening afterward. Third, decoupling ATS from the daily fraction process makes it less time-critical and allows for a decision-maker like a RO or an RTT certified for dosimetry to perform actions like re-contouring and re-planning without needing to be in the gantry at a given time. %The lists below summarize different aspects of the envisioned workflows.

\subsection{The Role of Fast Contouring in the Envisioned APT Workflows}

This report has two main goals. First, to understand and define the role of contouring in the daily online APT paradigms. Second, to propose protocols and tools that can realize the envisioned process. The dose delivery pipelines proposed in the previous paragraphs answered the first of these, making evident the importance of fast contouring as the enabler of ATS. As for the second goal, thinking about the implementation aspects of fast contouring requires a clear definition of the part of the workflow that it will occupy. Different contexts can have differing goals, requirements, and constraints, which will have an impact on the design process.

Considering Figure \ref{fig:online-adaptive-template} and the two last rows of Figure \ref{fig:comparison-daily-workflows}, it is possible to identify a pair of scenarios that use the delineation process. First, during plan creation to produce the set of delineations to optimize their initial treatment plan. Second, during fraction delivery and treatment monitoring for performing ATS. Recent works have focused extensively on the former, proposing general strategies to accelerate contouring that include allowing users to quickly shift between datasets \cite{aselmaa_using_2017}, using alternative input mechanisms like a stylus \cite{ramkumar_exploring_2013, multi-institutional_target_delineation_in_oncology_group_human-computer_2011}, or increasing the level of task automation \cite{van_dijk_improving_2020, aselmaa_influence_2017}. Nevertheless, relatively little attention has been paid to how to make contouring suitable for ATS, which makes it an obstacle for implementing ATP in the clinic. 

The proposals of this report build on the assumption that contouring can be further accelerated by tailoring it to ATS's requirements and available information. Concretely, out of the several ways in which the two scenarios differ, two main aspects impact the design of fast contouring tools for ATS. First, this scenario is time-critical. Although, in general, radiotherapy is time-constrained, the initial treatment plan creation process can spawn multiple days while adaptations should be done as soon as possible (often within a fraction) to prevent large systematic errors in the patient's treatment. Second, the available information changes, which, in turn, morphs the task's goal. At the fraction delivery state, information about the originally planned patient's treatment is available, which includes access to the initial contours, treatment plan, dose distribution, and images. Therefore, instead of aiming to recalculate them from scratch, which would involve repeating the treatment creation process, the goal is to use them as a reference. In practice this translates to, for example, not needing to decide where to draw the tumor's contour in an uncertain area and, instead, replicating a previously made delineation decision on the planning scan. Also, it means that the quality of the updated treatment plan will be judged based on how well it matches the original one as opposed to solely using dosimetric criteria.

The following sections describe how to implement the building blocks of fast contouring, which Figure \ref{fig:components-fast-contouring-workflow} at the beginning of this chapter depicts. The discussion is structured into the three core components of the process which are fast contour generation, a targeted inspection of generated contours, and redundancy-aware editing of faulty ones.

\section{Data Generation in Fast Contouring}
\label{sec:fast-contouring-data-generation}

The proposed contouring workflow has two phases, which are the consequence of increasingly reliable auto-conturing technologies that still require human supervision. The first stage consists of generating the delineations of the anatomical structures and additional data that can support the user-driven quality assessment process, which corresponds to the second stage. This section delves deeper into the generation process. Concretely, it presents the three main data types that can be gathered: contours, their uncertainty and information related to downstream tasks. Furthermore, for each of data types, this section presents time estimates for their computation based on state-of-the-art technologies. The latter being key for being able to use them in the time-constrained APT dose delivery pipeline. 

\subsection{Contour Generation and Deformable Image Registration}

The contour refinement approach introduced in the previous chapter relies on the assumption that initial contours have sufficient quality. Otherwise, users will opt to start from scratch, regardless of the algorithm or individual that created them. Currently, there are two categories of methods that consistently yield high-quality contours. These are:
\begin{itemize}[noitemsep]
    \item Deformable Image Registration-Based Segmentation (DIRS): given an image with approved contours and a new scan to delineate, the software first finds an optimal non-linear mapping, known as deformable registration, between the two. When the user agrees with the mapping, meaning that the images are correctly aligned, they can use it to propagate the original contours to the new image. In APT, DIRS can be used to propagate the patient’s contours from the initial planning scan to the daily or replanning one. If the former is not available, DIRS also works with atlases, yielding a family of methods known as (Multi-)Atlas-based Segmentation (MAS) in the literature.
    \item Deep Learning-Based Segmentation (DLS): deep learning is a type of artificial intelligence that can learn to perform a task by consuming large amounts of data. In the case of contouring, first, the deep neural network (DNN) is presented with tens or hundreds of image-contours pairs \cite{nikolov_deep_2020}. This training dataset should be representative of the contouring task in clinical practice to ensure that the DNN will be able to handle a wide array of patient cases. The training and validation \cite{vandewinckele_overview_2020} of the DNN happen offline. In clinical practice, the AI generates the contours minutes later after being presented with the patient image. 
\end{itemize}

Both DIRS and DLS can yield contours that are close to being clinically significant. Nevertheless, they differ in several ways. First, DLS methods can only segment anatomical structures that they have “seen” before. In contrast, DIRS can propagate all original structures with the computed mapping. This flexibility can accelerate the contouring process if there are patient-specific characteristics that the dataset used to train the DLS network did not contain. For instance, if the tumor is too big, it can deform surrounding organs. In this case, DLS will likely fail to produce suitable contours for the structures adjacent to the malignancy. In contrast, if those changes were present in the initial planning scan, DIRS will likely be able to replicate them in the new one, producing better quality contours.

A second difference between DIRS and DLBS is the type of errors they can yield. Given that DIRS uses an image with approved contours as a reference, it can propagate potential inaccuracies in those delineations to the new scans. These could cause systematic errors in the delivery of the dose. In contrast, DLBS-produced contours are a non-linear combination (or, roughly speaking, an average) of tens or hundreds of approved ones. Therefore, potential inaccuracies in the delineations will lead to random and not systematic errors in the dose delivery. Nevertheless, it is important to mention that using DLBS could also lead to systematic errors. For instance, if the contouring guidelines used in the clinic differ from those used for training the AI.

A third differentiator between DIRS and DLBS is that the former yields an extra output that could be useful for the quality assessment of the delineations. In an APT adaptation session, the goal of contouring is not so much to create the delineations from scratch. Instead, it is to make sure that those in the daily image match those in the initial one. The deformable mapping of DIRS enables comparing the two images and their contours, making it suitable for satisfying this requirement. Currently, if using DLBS, the deformation field needs to be generated ad-hoc via DIR. Several works are investigating DLBS methods that can also produce a deformation map together with the contours, which would alleviate this weakness \cite{elmahdy_joint_2021}.

In terms of efficiency, DLBS techniques can generate a set of contours in several seconds. Preliminary experiments show that delineating nine HN organs takes around a minute. This value is subject to factors such as the computer hardware, number of structures, the resolution of the image to contour, and the number of sample contours to generate. The latter, as the next section describes, can be used to quantify the AI’s uncertainty. As for DIRS, the bottleneck is generating the deformable mapping between images, which needs to be approved by the user before propagating the contours. The latter only takes a couple of seconds and depends on the number of structures and the resolution of the image grid.

To conclude, this section presented DLBS and DIRS the state-of-the-art auto-contouring methods for generating contours in APT. It showed how both have their unique strengths and weaknesses \cite{polikar_ensemble_2006}. Therefore, it makes sense to consider combining their outputs to improve delineations. This idea came up multiple times in discussions with clinicians. In practice, this means that the output of this step of the fast contouring workflow does not need to be a single set of contours. In contrast, it can be an ensemble of potential delineations for each structure. This ensemble can be formed with the contours of DLBS and DIRS or, as the next section shows, can be the output of a DLBS method.

To summarize:
\begin{itemize}[noitemsep,nolistsep]
    \item DLS can produce high-quality segmentations of anatomical structures on which it was trained on. DIRS can propagate all the structures in a reference image with approved contours. 
    \item DBLS delineation errors are random while DIRs ones can be systematic, propagating inaccuracies from the reference segmentations. 
    \item In addition to the contours DIRS also yields a deformation map that enables comparison of the two images. With when using DLS auto-contouring, the deformation map usually needs to be computed ad-hoc.
    \item Both DLBS and DIRS can generate contours in, at most, a minute. Nevertheless, DIRS requires the user to verify the deformable mapping before propagating the contours.
    \item Working with ensembles of contours from different auto-contouring methods can help mitigate their weaknesses.
\end{itemize}

\subsection{Uncertainty Quantification}

Uncertain contours are those that lie on image regions where the boundary line is not clear and therefore could vary between observers, causing inter and intra-observer variability (see Section \ref{sec:geometric-uncertainties}). When asked which areas of the contours they are unsure of, clinicians would likely point to those with two characteristics. First are areas where the image has noise, artifacts, or multiple structures with similar tissue types. Second are structures that do not match their prior knowledge and cannot be resolved only by looking at the image. For instance, CT scans often provide insufficient contrast in the inner part of the parotid gland. In these cases, clinicians use landmark-based guidelines to draw the boundary. 

\begin{figure}[h]
  \centering
  \includegraphics[width=\linewidth]{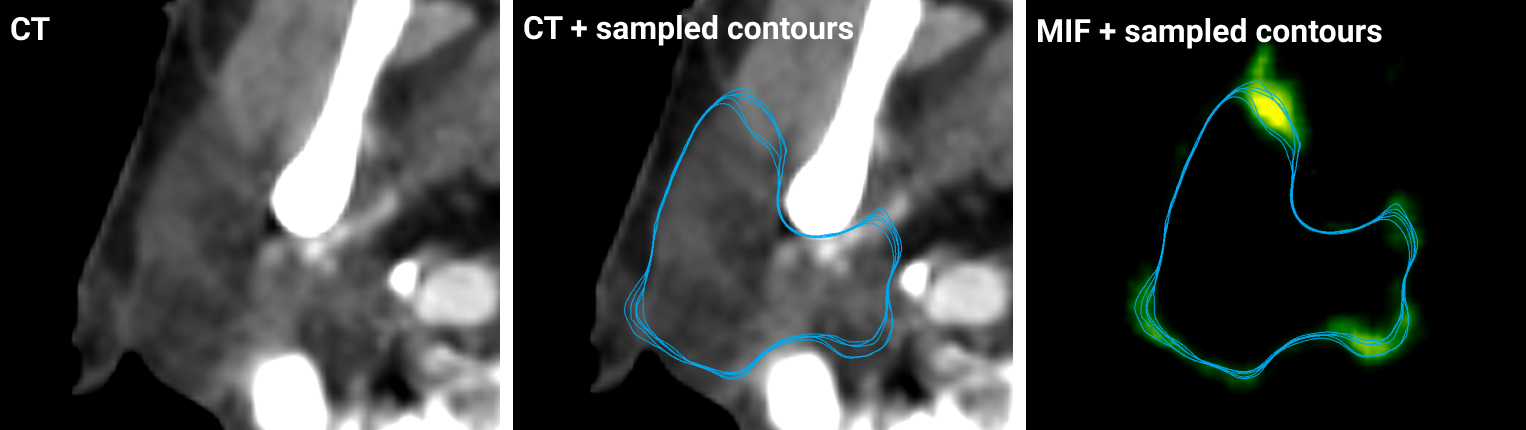}
  \caption{
  Contour and uncertainty generation capabilities of state-of-the-art Bayesian neural networks. The left panel shows a slice of the right parotid gland of an HN cancer patient. Given the lack of contrast in the image, some of the boundaries are hard to detect. These regions are referred to as uncertain ones. As the center and right panels illustrate, Bayesian NNs can produce multiple contours, which can then be transformed into uncertainty using mutual information (right panel) or entropy. These uncertainty maps could help the users quickly locate potential inaccuracies \cite{hui_quality_2018} or could be used as input to active learning methods that use uncertainty to guide the querying process \cite{top_active_2011}.
  }
  \label{fig:dl-uncertainty-parotid}
\end{figure}

In recent years, the field of Bayesian Deep Learning has appeared, allowing DLS methods to quantify the uncertainty of the predicted contours \cite{mody_comparing_2021, mukhoti_evaluating_2019}. Generally speaking, a Bayesian DLS generates an ensemble of possible delineations. The left pane of Figure \ref{fig:dl-uncertainty-parotid} shows a slice of a parotid gland that has some hard-to-contour areas. The center pane presents several of the AI-generated contours. These can be aggregated into a per-pixel measurement of uncertainty using metrics like entropy or mutual information. The image in the right pane displays the resulting heatmap.

% figure: uncertain slice | multiple contours | mif mask

Uncertainty can be a valuable signal for guiding the contour QA process. For instance, in the current clinical workflow, when RTTs are unsure about a delineation, they often let the RO in charge of reviewing them know. ROs can then use this information to focus on parts of the contours that have a higher chance of being wrong. In a similar vein, uncertainty from the DLS could be used to help direct the clinicians' attention to areas of the delineations with potential errors. Nevertheless, infusing uncertainty in the contouring workflow can be challenging.

In the perfect scenario, uncertain contours should be the only ones with potential inaccuracies. In other words, uncertainty should correlate with hard-to-resolve parts of the image. If this is true, AI's uncertainty could be used for speeding up the contouring process similarly as in the case of the RTT/RO interaction sketched above. Nevertheless, preliminary experiments indicate that the AI can be uncertain in easy-to-resolve areas (false positives) and confident in hard-to-resolve, and likely incorrect, ones (false negatives). When relying on AI uncertainty to guide the QA process, these false positives and negatives undermine users' trust, forcing them to review all the contours. Furthermore, false negatives can affect patients' outcomes if the clinicians rely too heavily on the AI and its uncertainty.

More research is needed to improve uncertainty quantification and to understand its role in the contouring workflow and its impact on users' trust. One promising direction is to train AI models that are optimized for generating uncertainty that is more likely to correlate with inaccurate regions. Another one would be to augment AI's uncertainty with others like clinicians or uncertainty derived from image features \cite{al-taie_uncertainty_2015, saad_exploration_2010}.

In terms of efficiency, Bayesian DLS methods are more computationally demanding than their uncertainty-free counterparts. The reason is that, instead of a single contour per structure, Bayesian DLS generate an ensemble of them. Preliminary experiments indicate that generating 15 samples for 9 HN OARs takes around a minute. Then, computing the per-pixel uncertainty based on entropy or mutual information will take some additional seconds.

To summarize:

\begin{itemize}[noitemsep,nolistsep]
    \item Bayesian DLS methods can produce a per-pixel estimate of uncertainty, which denotes how certain the network is about a given contour.
    \item The efficiency of Bayesian DLS depends on how large the ensemble of possible contours is. 
    \item The main challenge with current uncertainty quantification techniques is the presence of false positives and false negatives which could undermine users trust and affect patients’ outcomes.
    \item AI uncertainty could be complemented with other sources of uncertainty like image information or domain knowledge. An example of the latter are delineation guidelines, which list areas for which delineations are ambiguous.
\end{itemize}

% SOMETHING ABOUT UNCERTAINTY IN DIR?

\subsection{Information From Neighboring Processes}

Contouring is not an isolated process. Instead, other steps of the dose delivery pipeline use delineations as inputs. Information about how changes in the contours affect downstream processes could help guide their quality assessment. For instance, in a time-constrained scenario, it would make sense to prioritize delineations that pose a greater risk on treatment plan optimization and other decision outcomes. In practice, computing the contours' risk would entail obtaining multiple contours samples and estimating how derived quantities change. The uncertainty-generating capabilities of Bayesian DLS make this sampling possible.

Augmenting uncertainty with its risk can help further restrict the user's search area in the quality assessment process. A concrete example is to use a DVH to determine whether the initial treatment plan still satisfies the goals and constraints in the anatomy of the day. The first step would be to generate an ensemble of contours and their uncertainty using a Bayesian DLS. Then, multiple DVHs based on the possible delineation variations could be computed using the planning dose distribution. In the context of this example, risky uncertain areas would be those where metrics derived from the DVH of the delineation samples deviate significantly from the planning ones, exhibit a substantial variance, or go from acceptable to not, or vice versa.

The concept of the risk that uncertain contours pose is a multi-dimensional one that depends on the downstream tasks. The list below presents the two scenarios in the proposed APT workflow in which performing a sensitivity analysis to inform the contouring task makes sense. It also presents details of what the computational pipeline looks like and how long computations would take \footnote{The exact times still need to be corroborated with clinical partners.}, approximately, with state-of-the-art hardware and algorithms.

% add literature to timings
\begin{itemize}[noitemsep]
    \item To decide whether the current treatment plan is clinically acceptable for daily anatomy. Clinicians use contours to determine whether there are changes in the structures' shapes that grant updating the treatment plan. Furthermore, they often use dose metrics to ensure that plan is still clinically acceptable.
    \begin{itemize}[noitemsep]
        \item Pipeline: contour sample -> DVH based on planning dose -> TCP and NTCP models
        \item Timings: generating 15 contour samples takes around 1 minute for 9 HN OARs, computing a DVH depends on the number of structures and takes between 30 seconds and 1 minute per contour sample. 
    \end{itemize}
    \item To find a new optimal treatment plan. Variations in the contours can lead to different configurations in the optimization process. In turn, each of these will produce a different dose distribution.
    \begin{itemize}[noitemsep]
        \item Pipeline: contour sample -> plan re-optimization -> dose calculation -> DVH based on recalculated dose -> TCP and NTCP models
        \item Timings: generating 15 contour samples takes around 1 minute for 9 HN OARs, re-calculating the treatment plan can range from several minutes to seconds depending on the type of optimization (for example, full optimization versus using a library of plans), dose calculation can take several minutes or a couple of seconds depending on the algorithm (Montecarlo vs analytical), and computing a DVH depends on the number of structures and takes between 30 seconds and 1 minute. This means that, with current technologies, one execution of this pipeline will take at least a minute. 
    \end{itemize}
\end{itemize}

The timings sketched above show that computing risk metrics can be time-intensive, depending on the number of contour samples considered. It is possible to narrow down the structures and variations used for these analyses based on their relevance. This information could come from a retrospective study of a sample of the patient population or clinical experience. Also, from information that the clinicians have of the patient. For instance, during the interviews, and also in the literature \cite{werensteijn-honingh_feasibility_2019}, clinicians commonly limited the inspection area based on the distance from the PTV. Also, they often use a reduced subset of structures for plan adaptation sessions. Building these heuristics in the risk analyses could expedite the process without significantly affecting the treatment outcome.

To conclude, it is necessary to note that establishing feedback loops between contouring and neighboring processes is not always a good idea and must be done with care. Concretely, this type of analysis only makes sense if the patient already has an optimal treatment plan and the goal is to update it to the daily anatomy. The goal of having information about the variations' risk is not to tell the users which contour to pick but to indicate which parts of the contours have a higher weight on the patient's outcome. Clinicians can use this information to focus on specific areas of the volume. Then, comparing the daily contours with the reference ones, clinicians can pick or draw the most suitable ones.

\section{Quality Assessment of Contours in APT}
\label{sec:fast-contouring-user-qa}

The second block of the fast contouring workflow is assessing the quality (QA) of the generated contours. In contrast with the preceding data generation part, QA is user-driven. Therefore, its efficiency depends on how fast the users can perform the activity. Concretely, the general task that the system should allow the users to complete is:

\begin{itemize}[noitemsep]
    \item To produce delineations of the patient's daily anatomy that match those in the reference treatment plan. 
\end{itemize}

As Figure \ref{fig:components-fast-contouring-workflow} shows, the QA task has two interlinked components: inspection of the delineations and editing of inaccurate ones. In the following, a concept of a possible user interface will be used to describe a set of principles that can be used to design fast contouring tools. These principles resulted from the interviews and observation sessions with clinicians and the literature review that previous chapters of this report have presented.

\subsection{Targeted Inspection}
\label{sec:targeted-inspection}

The goal of the inspection task consists of being able to detect parts of the contours that might be inaccurate. A tool that accelerates inspection must provide the following functionality:
\begin{itemize}[noitemsep]
    \item \textbf{Provide semantics-based mechanisms to navigate through the contours (VR1)}. Currently, the user navigates through slices in multiple orientations, correcting inaccuracies as they appear. A top-down view in which users drill down from anatomical structures to interesting areas of the contours (see the concept of refinement chunks in Section \ref{sec:refinement-chunks}) would enable avoiding checking areas of the volume that are not important for the task at hand. 
    \item \textbf{Provide multiple views of the current area of interest (VR2)}. Clinicians usually scroll through multiple slices and use planes in varying orientations to get a sense of the problem. For fast contouring, this should be possible with fewer interactions.  
    \item \textbf{Allow effective comparison of daily and reference information (VR3)}. An inaccuracy in a daily contour can be found or confirmed by looking into the reference one. Furthermore, comparing derived information like uncertainty or DVH can help decide whether an inaccuracy needs to be fixed or not.
\end{itemize}

% figure: a concept of a risk-aware tool
\begin{figure}[h]
  \centering
  \includegraphics[width=\linewidth]{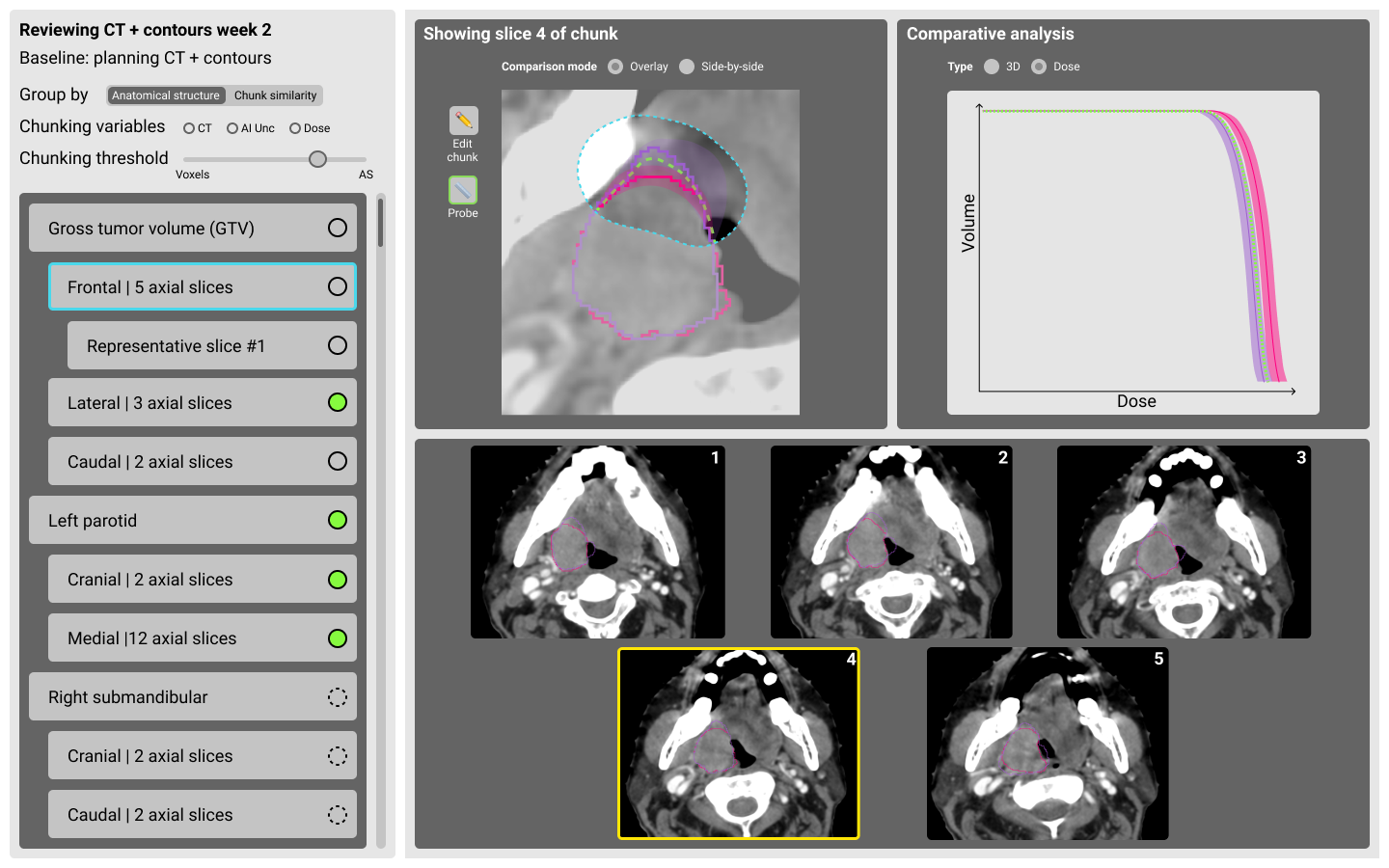}
  \caption{
  Concept of an uncertainty and risk-aware tool for efficiently determining which refinement chunks to edit in a treatment plan update session. The inputs of this system are ensembles of contours generated by an AI for the planning and adaptation images, their uncertainty, and clinical data from neighboring tasks (dose distribution in this case). With these ensembles, it is possible to quantify the delineation's uncertainty and use it to perform a sensitivity analysis on clinical endpoints such as the dose-volume histogram (DVH). The left panel of the wireframe lists the refinement chunks, which are areas of the contours with spatially correlated error patterns. The user has control over the generation and listing of these chunks, which can be handy for prioritizing structures in time-critical scenarios (VR1). Once the user selects one of the chunks, in this case, a subset of 5 slices in the frontal part of the gross tumor volume (light blue border), the panel on the right of the UI updates with a rich and information-dense display (VR2). This panel has two main components. The widgets in the first row enable a direct visual comparison between the planning and the adaptation images, their contours, and potentially, risk metrics (VR3). The second row permits a detailed view of the area that the chunk spans. In this case, through a grid of slices. The DVH on the top right portion of the right panel permits the user to understand how uncertainty impacts conclusions regarding the dose the delineated structure will receive. If the edits are not clinically significant, meaning in this case that there is not a lot of variation in the DVH metrics, then the user can avoid performing the edit and continue to the next chunk. If unsure about the analysis, the user can probe it by drawing a contour and seeing its implications on the DVH in real-time (dashed green line). Finally, depending on the context in which the tool is being used, its filtering and ordering capabilities can be adapted (VR1). For instance, only certain changes such as filling of the sinuses are expected to occur in-between fractions. The system could include this information to allow the user to focus on potential issues in these areas.
  }
  \label{fig:wireframe-risk-aware-chunk-selection}
\end{figure}

Figure \ref{fig:wireframe-risk-aware-chunk-selection} presents the concept of a tool that follows the three requirements presented above. Regarding VR1, the left pane lists multiple areas of interest, refinement chunks, computed based on characteristics of the image, the contours, and their uncertainty \cite{saad_probexplorer_2010, saad_exploration_2010, al-taie_uncertainty_2014, prassni_uncertainty-aware_2010, hui_quality_2018, chen_automated_2015, altman_framework_2015}. Also, these can be augmented with downstream information like the dose distribution. The user can choose the grouping scheme of the refinement chunks, in this case by the organ to which they belong, and also the list order based on clinical criteria like the chunks’ distance to the PTV and their priority in the plan optimization process.

Once the user has identified a chunk to inspect based on the filtering criteria, the right pane shows an information-rich chunk-dependent display (VR2). For instance, for chunks that span multiple slices, these can be shown, sparing the user the need to scroll back and forth. The slice view can be complemented with a 3D view that affords other mesh-based editing mechanisms. Finally, if the user is interested in the downstream effect of the contours, they can toggle views of, for example, the sensitivity analysis performed on the DVH.

For the third requirement (VR3), the tool provides mechanisms for comparing a chunk of the daily anatomy with the reference one. If a DIR is available, this will be used to align the volumes and the contours. Otherwise, the tool allows comparing two organs’ contours and their surroundings. This is done by generating a DIR per organ which is an easier task than doing so for the whole image. As for the quantitative displays, like for instance DVH, multiple visual channels such as line color and texture will be leveraged to help the user differentiate between scenarios.

Finally, it is important to note that there is uncertainty in several parts of this analysis pipeline. For instance mapping a scan with AI-generated contours to the reference one alters the uncertainties from the AI which those steeming from the propagation procedure. Therefore, in all the information displays that the tool offers, uncertainty can be toggled and visualized in different ways. Being able to switch it off is an important user-facing requirement as clinicians said multiple times during the interviews that they want as little clutter in the contours as possible. Also, uncertainty might produce false negatives, which could reduce the users’ trust in the provided chunks. Therefore, the tool also supports using slices as refinement chunks, which amounts to falling back to the current way of performing QA.

\subsection{Redundancy-Aware Editing}
\label{sec:redundancy-aware-editing}

After detecting an inaccurate contour, the next step is to fix it. Based on this report’s findings, editing tools should adhere to two requirements:

\begin{itemize}[noitemsep]
    \item \textbf{Editing setup should adapt to the problem at hand (ER1)}. For instance, an inaccurate contour might span several slices. Currently, editing of this chunk happens on a slice-per-slice basis drawing on a 2D canvas. Alternatively, this chunk could be inspected using a grid of slices or a 3D view. Editing mechanisms should be tailored to the unique affordances of these visualizations to spare the users from needing to switch between visualizations to perform the edits.
    \item \textbf{Minimize the number of redundant interactions (ER2)}. When the user edits a contour, they provide extra information that the system should be able to use for similar cases. For instance, neighboring inaccurate slices often share error patterns. After editing one of these, the user should not need to repeat the interaction in the others. 
\end{itemize}

% figure: concept TODO: update
\begin{figure}[h]
  \centering
  \includegraphics[width=\linewidth]{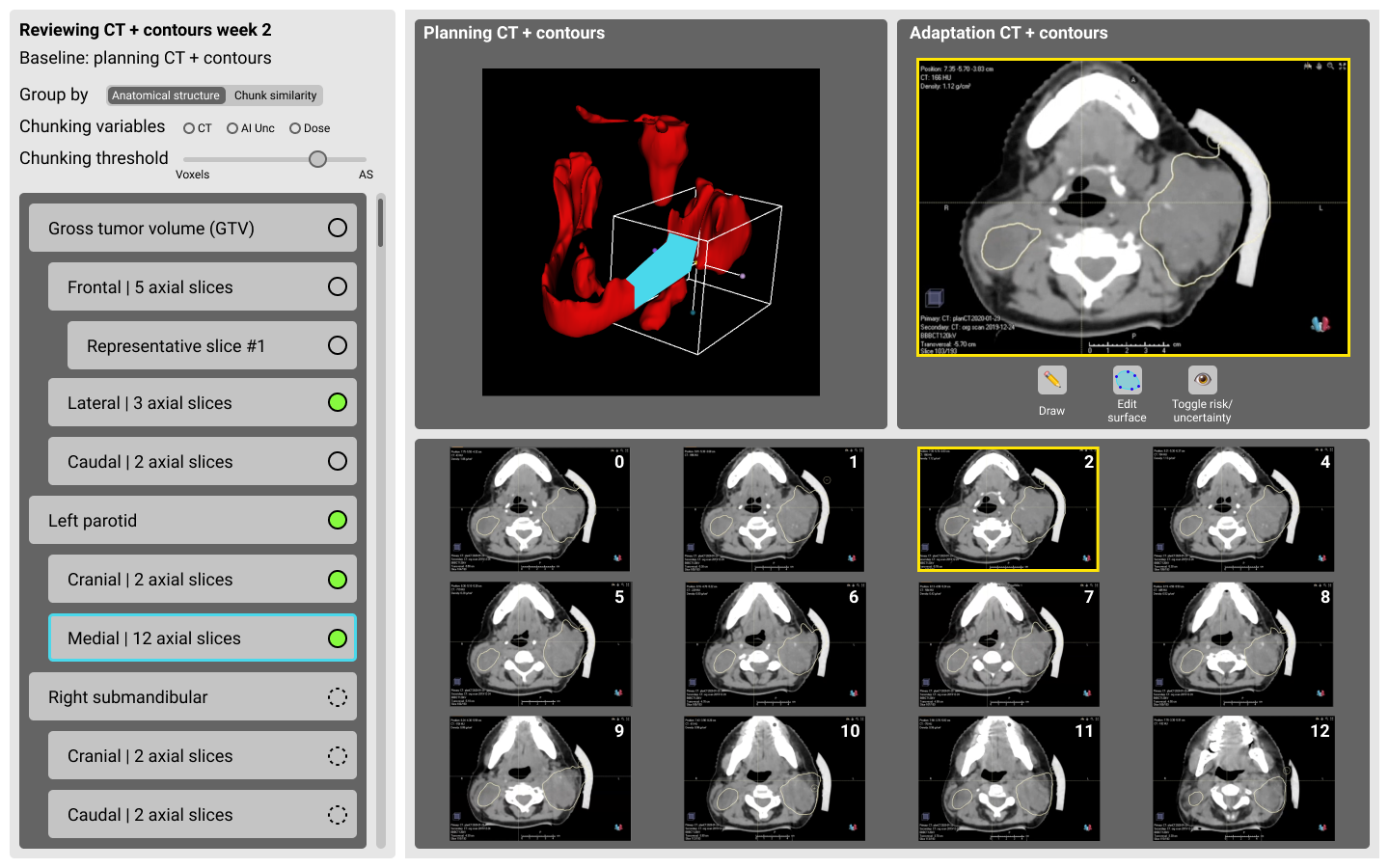}
  \caption{
  Concept of a redundancy-aware tool for efficient editing of inaccurate contours. If, after using the tool from Figure \ref{fig:wireframe-risk-aware-chunk-selection}, the user decides to edit the chunk, they can do it right away, without switching visualizations (ER1). The goal of the redundancy-aware editing system is to spare the user from multiple interactions (ER2). In the example, the refinement chunk involves adjusting the contour of the left parotid in 12 slices. Suppose the user takes around four seconds to edit one. Then, it would take them almost a minute to do the whole chunk. Considering that, in clinical practice, there will be more than one of these, editing becomes an obstacle towards the implementation of time-critical APT. The proposed concept addresses this by offering the user a complete view of the editing area, in this case, the 12 slices. Furthermore, it highlights and enlarges a representative one, which the user can edit. Once they have updated the boundary in the enlarged slice, the interactive AI propagates it across the other 11. If the user is not satisfied with the propagation, they can pick the slice they are not satisfied with and impose new constraints. The expected effect of this tool on the workflow is to reduce the number of interactions required to address chunks that spawn multiple slices. 
  }
  \label{fig:wireframe-redundancy-aware-editing}
\end{figure}

Figure \ref{fig:wireframe-redundancy-aware-editing} presents a concept of an implementation of an editing tool that follows these requirements. Using the visualization capabilities of the concept from the previous section, the user found out that a chunk of 12 slices of the medial region of the left parotid needs fixing. Instead of switching to the 2D slice view, the system lets the user perform the edits in the image grid where they found the issue (ER1). As for ER2, the user does not need to fix the contours on the 12 slices. Instead, they provide their input in a representative one, highlighted at the top of the right pane. With this information, an interactive DLS system (similar to those presented in \cite{lin_scribblesup_2016}) propagates the decision to the other slices, sparing the user several time-consuming interactions. Alternatively, the user could have also used the 3D view to perform the edits. In this case, mesh-based uncertainty-aware Laplacian editing tools \cite{sorkine_laplacian_2004} let the user fix the problem in fewer interactions.

Two aspects reinforce focusing on having coherent visualization and editing views (ER1) and on minimizing the number of user interactions (ER2). On the one hand, one of the preferred tools in the clinic, between-slice interpolation, already exploits redundancy in the contour creation phase. Its wide adoption and positive impact on task performance \cite{aselmaa_influence_2017} prove the benefits of eliminating redundant interactions and provide a baseline to measure the proposed concept. Second, of all the proposed interactive contouring tools, few, to none, have found their way into clinical practice. The main issue with these tools, which enable interaction through mechanisms like scribbles \cite{lin_scribblesup_2016} or fiducials \cite{maninis_deep_2018, dai_boxsup_2015}, is the lack of coherence between the input of the user and what happens in the viewport, which results in perceived reduced control and predictability. The proposed system would alleviate this by tightly coupling editing mechanisms with the visualizations where users found the contouring issues.

Similar to the contour inspection case, during the observation sessions, the users expressed several preferences regarding features of contour editing tools. Mainly, they remarked how important control and predictability are. In general, they distrust tools that are too proactive based on simple interaction mechanisms like scribbles as they fear the resulting contours will still require extensive edits. Editing tools that adhere to ER1 and ER2 should mitigate these concerns.

\section{Fast Contouring Workflow Considerations}
\label{sec:fast-contouring-wf-aspects}

After describing the components of the fast contouring workflow, it remains to discuss how these fit in the dose delivery pipelines for APT proposed at the start of this chapter. The paragraphs below structure this discussion into two main points: distribution of labor and orchestration of tasks. In summary:
\begin{itemize}[noitemsep, nolistsep]
    \item RTTs should be able to execute the fast contouring workflow without RO assistance. The latter would only intervene in exceptional cases.
    \item For the Envisioned HollandPTC workflow 1, fast contouring has a time budget of around 5 minutes. Given this time criticality, not all the data will be available. At the minimum, the contours and the uncertainty should be. 
    \item For the Envisioned HollandPTC workflow 2, fast contouring can occur anywhere in-between fractions. Nevertheless, to free clinical resources, it should still take significantly less time than the 1/2 hours that the current workflow takes. Given the more relaxed time constraints, additional data will be available, which can further speed up the QA process. 
    \item For Envisioned HollandPTC workflow 1 fast contouring needs to be executed as a single block. In contrast, for Envisioned HollandPTC workflow 2 the generation and QA parts can be split, allowing more time for the former and also better task allocation. 
\end{itemize}

Currently both at HollandPTC and LUMC, RTTs are in charge of the fraction delivery stage. ROs only participate when something out of the ordinary things happens that halts the fraction. For example, when the tumor presents sudden large changes in shape or volume. Nevertheless, this is possible because no contouring or plan updating, both of which require RO's approval, happen during the fraction. One of the goals of the presented fast contouring tools is to eliminate the need for peer review in the adaptation process. Two mechanisms are responsible for this. First, producing contours that match a reference is a less ambiguous task than creating them from scratch. The former is the scenario that RTTs would face during an adaptation and effective comparison tools would help them accurately propagate the delineations. Second, fast contouring would have safeguards that alert ROs in case of potential issues. This is similar to adaptation triggers currently used at LUMC to decide whether the plan should be adapted or not. 

Regarding the orchestration of the fast contouring tasks, it varies depending on the chosen dose delivery pipeline. In Envisioned HollandPTC workflow 1, ATS occurs with the patient on the treatment couch. In this context, fast contouring is a time-critical activity. Currently, HollandPTC's and LUMC's fraction delivery processes take around 20 minutes, without including ATS. In contrast, UMC Utrecht's workflow, which includes ATS takes a median of 48 minutes. Although the permissible gantry time is institution-dependent, for the ensuing analyses, this report considers this duration an upper limit of what is possible. In the MRI linac workflow, the ATS block takes around 10 minutes. Assuming that re-planning can be done in less than 5 minutes, this would leave a 5-minute time slot for the fast contouring workflow. Given that, at the minimum, the generation stage takes around 2 minutes, there are around 2 minutes available for QA. 

Moving to Envisioned HollandPTC workflow 2, it imposes less tight time constraints on fast contouring given that it happens after the fraction delivery has finished and the patient has left the treatment center. Nevertheless, to enable daily APT of all HN patients, fast contouring should still be executable in a reduced time frame. Ideally, less than the hour that the current contouring workflow takes. A benefit of this workflow is having the time in between fractions. This allows decoupling of the generation and QA processes. Allowing more time for the former, which does not rely on user intervention, would result in richer datasets that can further speed up the QA process.

%% file: content/conclusion.tex
\chapter{Conclusion}

% 1) affirm that you delivered on the claims made in the introduction
% 2) summarize the contributions of the work
% 3) make any key points that you would like to leave to the reader
% 4) point to a bright future a better world

Realizing the benefits of proton therapy requires reducing the extent of safety margins around critical structures. A significant component of these margins is geometric uncertainties which can be mitigated by re-optimizing or adapting the patient's treatment plan. In adaptive workflows, the delineation process plays a key role, allowing for the detection of geometric changes and providing the inputs for the treatment plan creation process. Nevertheless, presently it also constitutes the main obstacle for implementing the most responsive daily-adaptive PT. The reason for this is that, despite advances in auto-contouring technology, clinicians still need to verify and correct the generated contours to prevent downstream dose deviations.  

This report aimed to understand how to bring the current contouring workflows closer to what is required for the future, more responsive, APT dose delivery pipelines. It started by providing a context as to why adaptation is necessary to realize the benefits of PT and presented the implications of APT in the design of the dose delivery pipeline. Then, it presented the results of the Study of the Offline-APT Contouring Workflow. The study's goal was to characterize contouring as it is currently implemented in an offline-adaptive framework, identifying potential bottlenecks both at the process and human-computer interaction levels. For this, several clinicians at two Dutch cancer treatment centers, Holland Proton Therapy Center and the radiotherapy department of Leiden University Medical Center, were interviewed and observed while they performed their routine delineation tasks.

The study yielded two main observations that could be leveraged to accelerate contouring in the context of an adaptive framework. On the one hand, neighboring processes can provide information that is useful for guiding the delineation process. On the other hand, the contouring refinement task requires a lot of redundant user interactions. Developing controllable and predictable tools that leverage this could significantly shrink the delineating time. The report concluded by presenting a proof of concept of a fast and robust contouring workflow that proposed a way to implement these observations in clinical practice. The thought experiment demonstrated that online-adaptive-ready contouring workflows are within reach if they are supported by scenario-depend user interfaces that leverage the affordances of state-of-the-art Bayesian deep neural networks.

%% file: content/appendix-process-specification.tex
\chapter{PT Workflow Process Specification}
\label{ch:appendix-process-specification}

\begin{table}[h]
\begin{tabular}{p{0.35\linewidth} | p{0.6\linewidth}}
\multicolumn{2}{c}{\textbf{Image Acquisition or Simulation}}                                                                                                                                                                                                                                                                                  \\ \hline
Context          & Initial treatment plan creation                                                                                                                                                                                                                                                                                            \\ \hline
Duration         & 20 minutes per modality                                                                                                                                                                                                                                                                                                    \\ \hline
Patient required & Yes                                                                                                                                                                                                                                                                                                                        \\ \hline
Inputs           & None                                                                                                                                                                                                                                                                                                                       \\ \hline
Outputs          & A set of three-dimensional images                                                                                                                                                                                                                                                                                          \\ \hline
Executed after   & None                                                                                                                                                                                                                                                                                                                       \\ \hline
Description      & Acquire 3D images of the patient. Usually a Computarized Tomography (CT) is taken, but additional modalities such as Magnetic Resonance (MR) and Positron Emission Tomography (PET) can be added to reduce uncertainty in the delineation step. Which extra modalities to take depend on the available time and resources.
\end{tabular}
\label{table:ps-simulation}
\end{table}

\begin{table}[h]
\begin{tabular}{p{0.35\linewidth} | p{0.6\linewidth}}
\multicolumn{2}{c}{\textbf{Registration}}                                                                                                                                                                                   \\ \hline
Context          & Initial treatment plan creation                                                                                                                                                                          \\ \hline
Duration         & 30 minutes                                                                                                                                                                                               \\ \hline
Patient required & No                                                                                                                                                                                                       \\ \hline
Inputs           & Reference (CT) and other modalities to register (MR, PET-CT)                                                                                                                                             \\ \hline
Outputs          & For each pair of reference image-additional modality, a transformation that maps their coordinate spaces                                                                                                 \\ \hline
Executed after   & Simulation                                                                                                                                                                                               \\ \hline
Description      & Given a reference image, which will be used to optimize the plan, align other modalities to its coordinate system to enable comparison in the contouring step.
\end{tabular}
\label{table:ps-registration}
\end{table}

\begin{table}[h]
\begin{tabular}{p{0.35\linewidth} | p{0.6\linewidth}}
\multicolumn{2}{c}{\textbf{Contouring}}                                                                                                                                                                                                                                       \\ \hline
Context          & Initial treatment plan creation                                                                                                                                                                                                                            \\ \hline
Duration         & Between 70 minutes and 6 hours                                                                                                                                                                                                                             \\ \hline
Patient required & No                                                                                                                                                                                                                                                         \\ \hline
Inputs           & Reference (CT), other modalities to register (MR, PET-CT), and their registrations. Also, a list of structures to contour                                                                                                                                  \\ \hline
Outputs          & A set of contours for anatomical structures relevant to the patient's cancer, based on the tumor's type and location                                                                                                                                       \\ \hline
Executed after   & Registration or simulation if using a single CT image for contouring                                                                                                                                                                                       \\ \hline
Description      & Given a list of anatomical structures that will be considered in plan optimization, delineate them in the reference image (usually a CT scan), using other -registered- images or extra information when needed.
\end{tabular}
\label{table:ps-contouring}
\end{table}

%\begin{figure}[h]
%  \centering
%  \includegraphics[width=10cm]{images/tables/ps-contouring.png}
%  \label{table:ps-contouring}
%\end{figure}

\begin{table}[h]
\begin{tabular}{p{0.35\linewidth} | p{0.6\linewidth}}
\multicolumn{2}{c}{\textbf{Plan Optimization}}                                                                                                                                                                                                                                                                                                                                                                                                                        \\ \hline
Context          & Initial treatment plan creation                                                                                                                                                                                                                                                                                                                                                                                                                    \\ \hline
Duration         & Several hours                                                                                                                                                                                                                                                                                                                                                                                                                                      \\ \hline
Patient required & No                                                                                                                                                                                                                                                                                                                                                                                                                                                 \\ \hline
Inputs           & CT image, contours, dosimetric goals and objectives of anatomical structures                                                                                                                                                                                                                                                                                                                                                                       \\ \hline
Outputs          & Machine parameters to deliver an optimal dose distribution                                                                                                                                                                                                                                                                                                                                                                                         \\ \hline
Executed after   & Contouring                                                                                                                                                                                                                                                                                                                                                                                                                                         \\ \hline
Description      & Find a configuration of the treatment machine's that can deliver an optimal dose distribution to the patient. An optimal dose distribution is the one that satisfies dosimetric goals and constraints set beforehand. An example of such optimization goals for the head and neck area could be to deliver 70 Gy of radiation to the tumor while keeping the dose to the parotid glands below 26 Gy.
\end{tabular}
\label{table:ps-plan-optimization}
\end{table}

\begin{table}[h]
\begin{tabular}{p{0.35\linewidth} | p{0.6\linewidth}}
\multicolumn{2}{c}{\textbf{Dose Calculation}}                                                                                                                                                                                                                                                                                                    \\ \hline
Context          & Initial treatment plan creation                                                                                                                                                                                                                                                                                               \\ \hline
Duration         & 1 to 2 minutes                                                                                                                                                                                                                                                                                                                \\ \hline
Patient required & No                                                                                                                                                                                                                                                                                                                            \\ \hline
Inputs           & CT image, contours, treatment plan                                                                                                                                                                                                                                                                                            \\ \hline
Outputs          & Simulated dose volume                                                                                                                                                                                                                                                                                                         \\ \hline
Executed after   & Plan optimization                                                                                                                                                                                                                                                                                                             \\ \hline
Description      & Given a treatment plan, simulates the delivery of the dose to the patient's anatomy based on the CT image and the contours. This results in a dose volume, which is of the same shape of the image and, for each voxel, provides information about how many Gy will be delivered.
\end{tabular}
\label{table:ps-dose-calculation}
\end{table}

\begin{table}[h]
\begin{tabular}{p{0.35\linewidth} | p{0.6\linewidth}}
\multicolumn{2}{c}{\textbf{Plan Evaluation}}                                                                                                                                                                                                                 \\ \hline
Context          & Initial treatment plan creation                                                                                                                                                                                                           \\ \hline
Duration         & Several minutes                                                                                                                                                                                                                           \\ \hline
Patient required & No                                                                                                                                                                                                                                        \\ \hline
Inputs           & CT image, contours, treatment plan, simulated dose volume                                                                                                                                                                                 \\ \hline
Outputs          & Dose volume histogram, tumor control probability (TCP) and normal tissue complication probability (NTCP) models                                                                                                                           \\ \hline
Executed after   & Dose calculation                                                                                                                                                                                                                          \\ \hline
Description      & Produces the quantitative information that relates the dose with the patient's anatomy and the outcome. This information can be used to decide whether the plan's quality is sufficient or not.
\end{tabular}
\label{table:ps-plan-evaluation}
\end{table}

%% file: content/appendix-study-apt-methods.tex
\chapter{Methods Study of the Offline-APT Contouring Workflow}
\label{ch:methods-study-apt-workflow}

% Method in one line: literature survey -> structured interviews -> thematic analysis [transcribing -> define codes -> check codes and recode -> define themes -> check codes and themes and recode and retheme] -> shadowing sessions -> activity analysis -> connect to automation theory -> yield opportunities for the usage of AI in this expert domain

\begin{figure}[h]
  \centering
  \includegraphics[width=\linewidth]{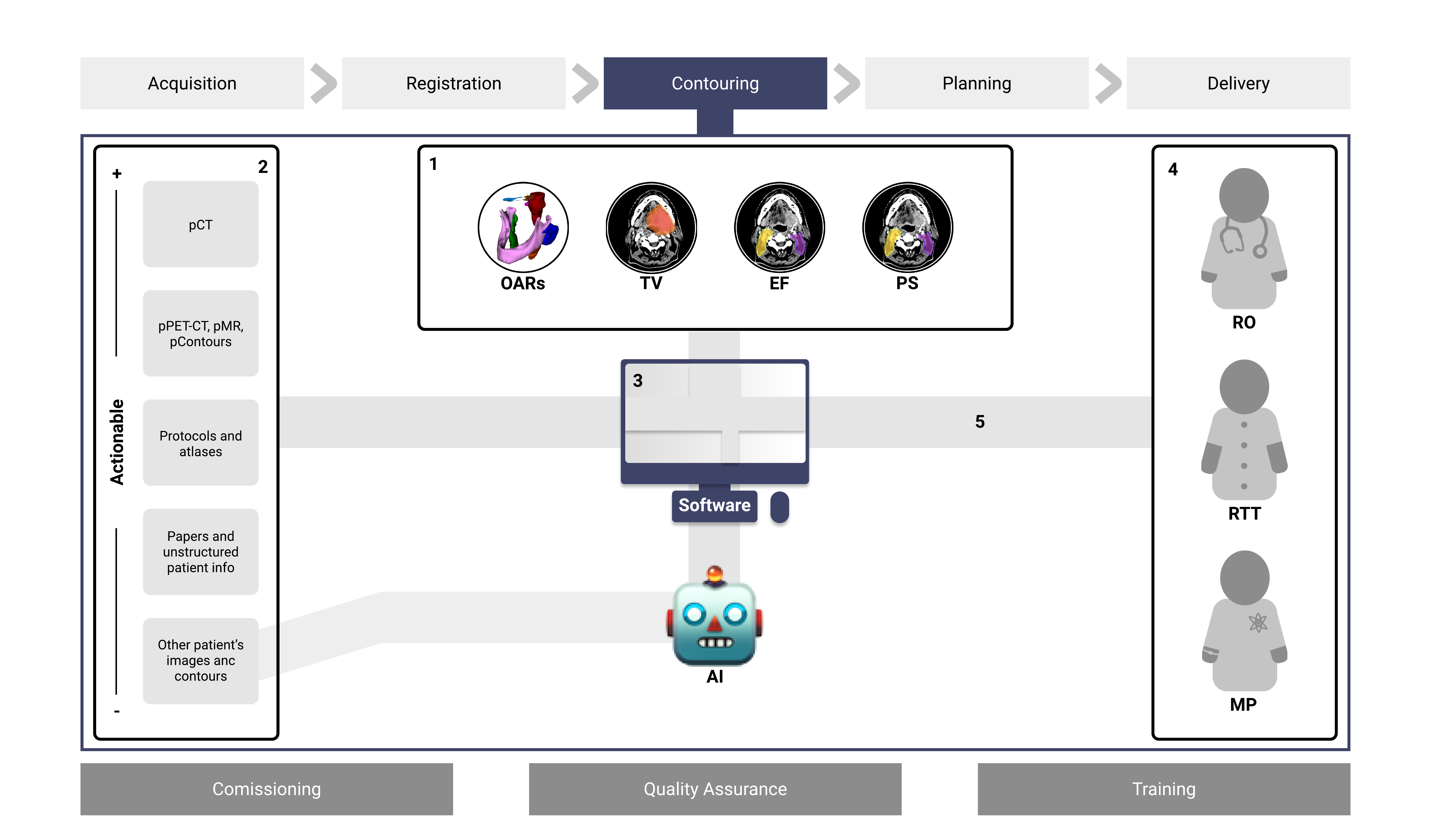}
  \caption{Performing an Activity Theory analysis of the interview and observational data yielded five aspects that play a central role in contouring: the object of the activity, the data sources, the software, the RT practitioners, and the hierarchical structure of the activity and performance factors.}
  \label{fig:results-contouring-activity}
\end{figure}

The main goal of this work is to understand the challenges and opportunities for implementing DL-based auto-contouring AI in the clinical setting. To achieve meaningful conclusions that directly relate to the clinical practice, a user-centered approach was used, which enabled getting a complete picture of the contouring workflow. Then, the results were grounded into the context of the APT and AI technology by obtaining a detailed the required concepts through a review of the literature. Below we describe the data sources and data gathering and analysis methodologies that were used in this work.

\section{Data Gathering}

This research was approved by the Institutional Review Board at Delft University of Technology and informed consent was obtained from each participant. The majority of the data used in this work was gathered by interviewing and observing the participants of the contouring activity. Concretely, two radiation oncologists (RO) and two radiotherapy technologists (RTT) from two medical centers in the Netherlands. On the one hand was Leiden University Medical Center (LUMC), which has a large, established, department dedicated to photon-based RT. On the other hand was the recently opened Holland Proton Therapy Center (HollandPTC), which specializes in proton therapy (PT), a newer RT paradigm that leverages protons to spare the organs around the tumor. Table \ref{tab:participants-qualitative-sessions} condenses the information of the qualitative sessions, which we describe in more detail below.  

These qualitative sessions were supplemented with other sources of information that included a review of the literature and institution-specific documents. The former, together with AI-based delineation prototypes developed by the team in the past months, formed our understanding of the contouring activity, how it fits in the RT workflow, and the potential challenges and opportunities to integrate AI-based contouring methods. The rest of this section details the methodology for gathering these data. 

\subsection{Interviews} 

To get an understanding of the contouring activity and how it fits in the RT workflow, three sessions based on semi-structured interviews were devised \footnote{To ensure the safety of the participants and the researchers during the COVID pandemic, we conducted all the sessions using an online video meeting service.}. The first one focused on obtaining a general understanding of the contouring activity and how it fits in the RT workflow. This session was also used to build rapport with the participants. For this, a draft of the RT workflow based on the literature was presented to the participants. At multiple times in the presentation, they were asked questions to elicit a discussion that improved our understanding of the workflow. Given that the sessions took place asynchronously, the workflow draft and related questions were constantly updated, based on the learnings.

\begin{figure}[h]
  \centering
  \includegraphics[width=\linewidth]{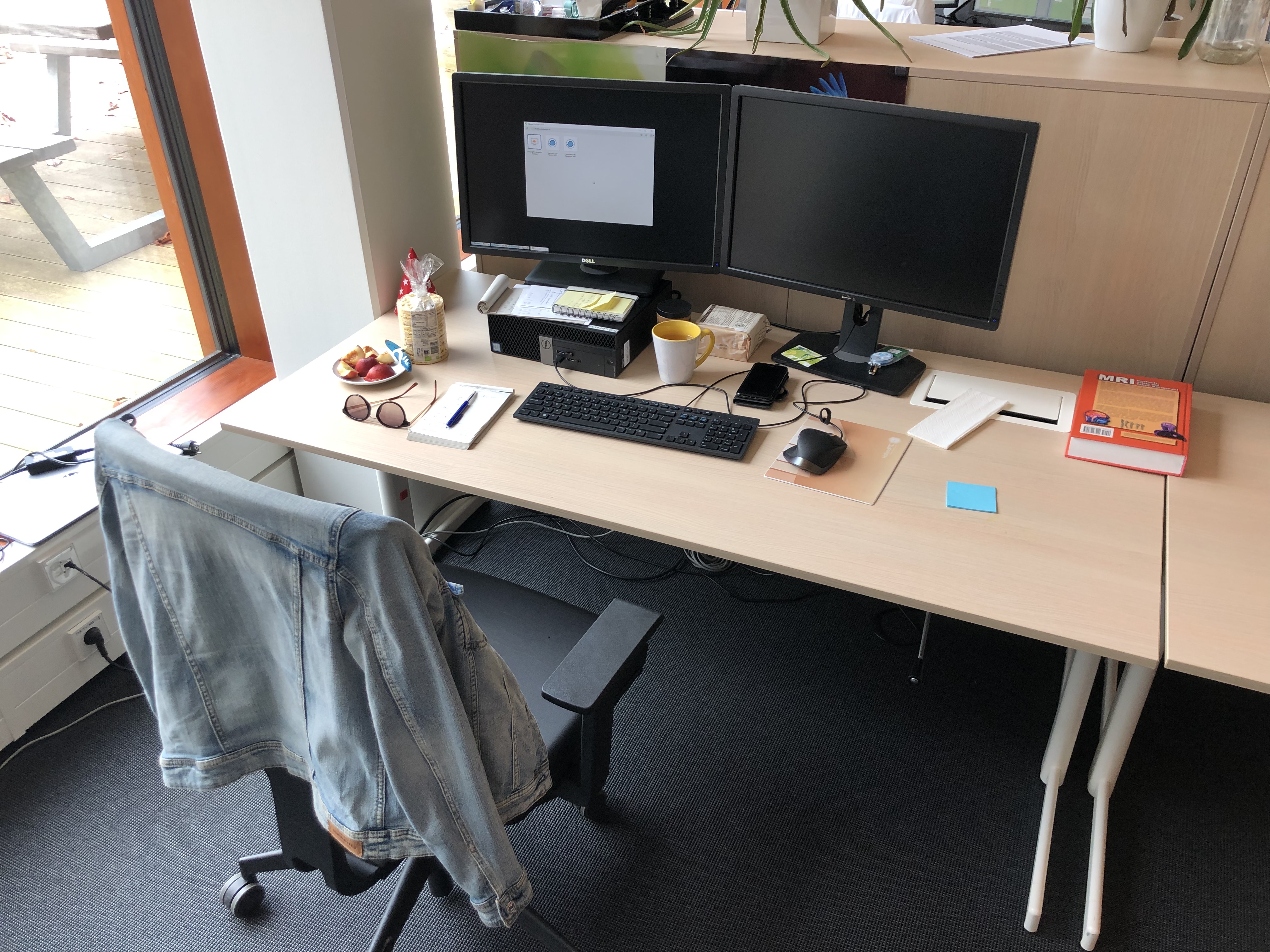}
  \caption{Workspace of radiotherapy technologist (RTT) at HollandPTC. It is common to have two large monitors and a high-quality mouse or pen tablet to ease contouring. The offices are open, which promotes collaboration but also interruptions, which might affect task performance.}
  \label{fig:hptc-rtt-workspace}
\end{figure}

The following two sessions were used to understand how RT practitioners contour in clinical practice. For this, participants were asked to contour a patient with their usual tools and software. Figure \ref{fig:hptc-rtt-workspace} shows examples of the workspaces at LUMC and HollandPTC. The first sessions focused on the scenario in which the patient needs to be contoured from scratch after being accepted into the clinic. In the second, the case where the patient's treatment plan needs to be adapted to anatomical changes during the treatment was simulated. To ensure that sessions could be recorded, an anonymized dataset of several patients was prepared in collaboration with the medical centers. For each patient, we had the planning and adaptation images (pCT, pMR, pPET-CT, reCT) and the planning and adaptation contours that were used to plan that patient's treatment. For all the sessions, the audio, video, and screen of the participants were recorded, which totaled 14 hours of unstructured data.

\subsection{Naturalistic Observation} 

In addition to the semi-structured interviews, naturalistic observation sessions were conducted with most of the participants. The goal of these sessions was to obtain a situated picture of the contouring activity, avoiding potential biases due to the participants preparing the data of the interviews. For the sessions, the participants were accompanied for one of their scheduled clinical contouring sessions, which lasted between one and two hours. The participants were not bothered during their work. Additionally, these sessions were only recorded in the form of field notes that were coded along with the rest of the qualitative data in the analysis phase.

\subsection{Literature} 

In this work, research literature was used with different goals. Initially, to gather an initial understanding of the contouring activity and its role in the RT workflow. Using keywords like "radiotherapy", "proton therapy", "radiation oncology", "workflow", "contouring" and "delineation" in the principal radiation oncology venues yielded a pool of papers used to build the draft of the RT workflow used in the first interview session. An important starting point was \cite{castadot_adaptive_2010}, which provided an initial seed from which to start. Later on, the search was narrowed to the contouring activity and, in particular, to the use of AI-based automation. This knowledge was used to prepare the targeted questions of the second and third sessions of the interviews. To reduce the amount of variability and make it easier to compare different sources, the papers were filtered to only include those related to the head and neck area.

\subsection{Clinical Documents and Other Sources of Information} 

In addition to the qualitative studies and the literature, documents that RT practitioners at LUMC and HollandPTC use when contouring were collected and analyzed. These included consensus papers, guidelines, protocols, flowcharts, and delineation atlases. Additionally, informal discussions with other staff members, such as medical physicists and RTTs specialized in treatment planning, were carried out. These helped refine the analyses. For instance, after the semi-structured interviews, an informal focus group meeting was conducted at HollandPTC to discuss the difficulty of contouring different anatomical structures. This meeting provided an understanding of how subjective and objective factors can influence the contouring activity's performance.

\begin{table}[]
\begin{tabular}{ccccc}
\textbf{ID} & \textbf{Institution} & \textbf{Role} & \textbf{\begin{tabular}[c]{@{}c@{}}Interview\end{tabular}} & \textbf{\begin{tabular}[c]{@{}c@{}}Naturalistic \\ Observation\end{tabular}} \\ \hline
P1                      & LUMC                 & RO            & 1, 2, 3                                                                & Yes                                                                          \\
P2                      & LUMC                 & RTT           & 2, 3                                                                   & No                                                                           \\
P3                      & HollandPTC                 & RO            & 1, 2                                                                   & Yes                                                                          \\
P4                      & HollandPTC                 & RTT           & 1, 2, 3                                                                & Yes                                                                         
\end{tabular}
\caption{Participants of the qualitative sessions. There were two radiation oncologists (RO) and two radiotherapy technologists (RTT) from two institutions in the Netherlands. Due to their tight schedules, not all of them could participate in all the sessions.}
\label{tab:participants-qualitative-sessions}
\end{table}

\section{Data Analysis}

The data-gathering phase yielded a large corpus of unstructured data that included audio transcripts, field notes, and literature. Thematic Analysis (TA), a method for analyzing qualitative data, was used to make sense of this information \cite{braun_2006, braun_2012}. To start, possible patterns in the data were identified by reading the complete data set multiple times. Recent literature in Human-AI interaction places a strong emphasis on the tasks without paying attention to the socio-technical system that surrounds them \cite{amershi_guidelines_2019, cai_hello_2019, beede_human-centered_2020, yang_re-examining_2020}. In these initial passes of the data, a more holistic understanding of the contouring process was necessary to integrate auto-contouring technologies effectively. 

Activity Theory (AT) provides a framework for understanding the tasks in their context by looking at the object, the subject, the task structure, and the technical and cultural factors \cite{nardi_1996, kaptelinin_1999, kaptelinin_2012}. AT was used as the theoretical base for the next step of the TA, which is the coding process. Based on this theory, codes were assigned to extracts of the data set, according to the component of the contouring activity to which they relate. At this stage, fine-grained codes such as "personal preferences", "task experience", "usage of tools" and "usage of medical images" were used. Also, issues such as potential overlap or relationships between codes were not yet considered. The coding process yielded more than twenty codes that were reviewed by the coauthors of this work. 

\begin{figure}[h]
  \centering
  \includegraphics[width=\linewidth]{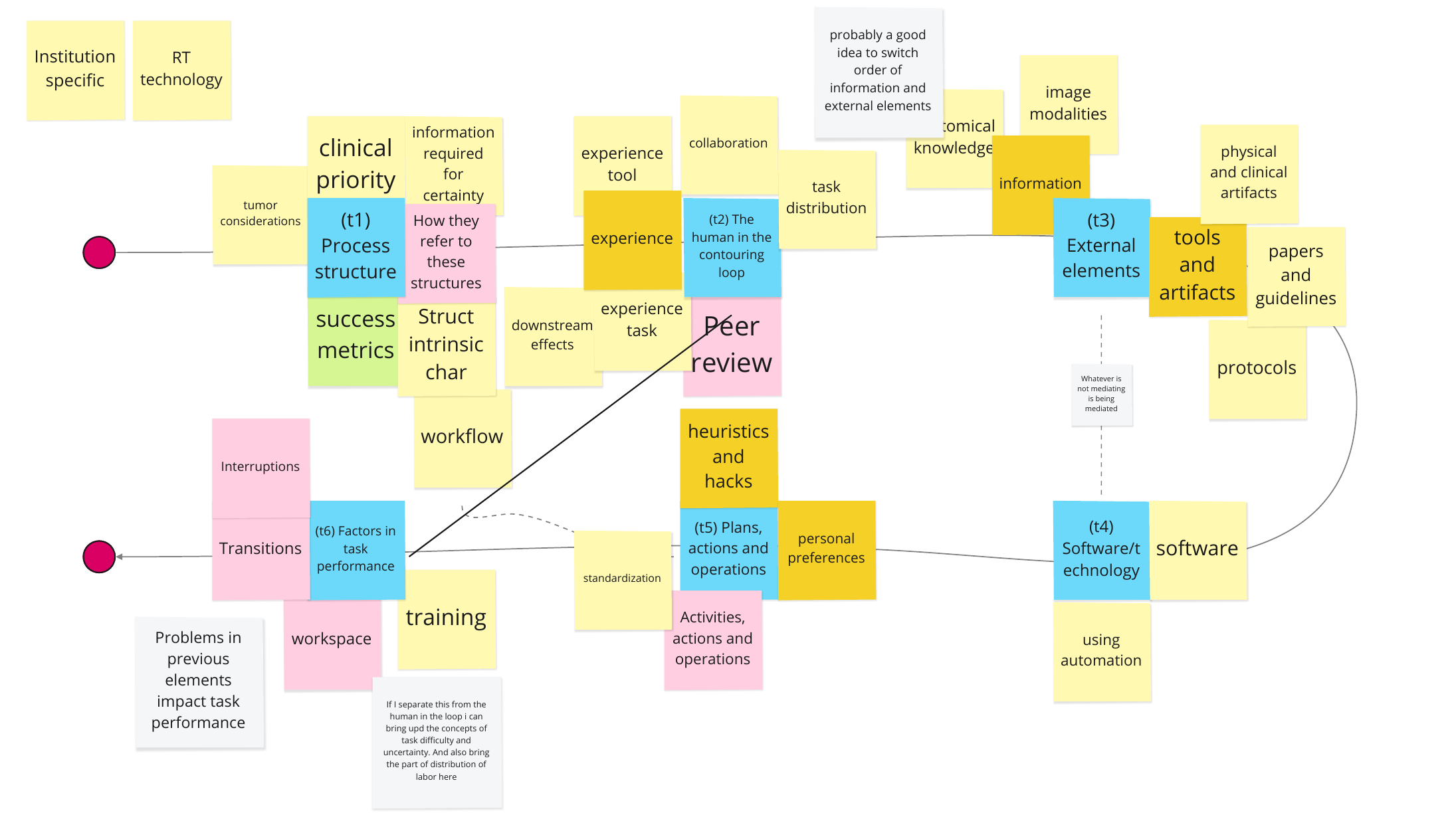}
  \caption{Intermediate iteration of the theme definition process in the Thematic Analysis. At this point, we had grouped the codes according to the different components of the Activity Theory Framework.}
  \label{fig:iteration-thematic-analysis}
\end{figure}

The next step of TA is to produce themes that organize data and provide a rich answer to the original question. For this, multiple rounds of arranging and grouping the codes according to the different axis of AT were performed. Figure \ref{fig:iteration-thematic-analysis} depicts a post-it board of an intermediate iteration of the theme definition process. At the beginning of the theme definition process, there were around ten themes, which were reduced to the five elements in Figure \ref{fig:results-contouring-activity}: the object of the activity, the data sources, the software, the RT practitioners, and the hierarchical structure of the activity and performance factors. 

